\documentclass[12pt,epsf,amssymb]{article}

\usepackage{graphicx}
\usepackage{amsfonts}
\usepackage{amsmath}
\usepackage[usenames]{color}
\usepackage{amssymb}

\def\Z{\mathbb{Z}}
\def\R{\mathbb{R}}

\def\P{\mathbb{P}}
\def\Im{\mathrm{Im}}

\begin{document}
\baselineskip 0.6cm
\newcommand{\gsim}{ \mathop{}_{\textstyle \sim}^{\textstyle >} }
\newcommand{\lsim}{ \mathop{}_{\textstyle \sim}^{\textstyle <} }
\newcommand{\vev}[1]{ \left\langle {#1} \right\rangle }
\newcommand{\bra}[1]{ \langle {#1} | }
\newcommand{\ket}[1]{ | {#1} \rangle }
\newcommand{\Dsl}{\mbox{\ooalign{\hfil/\hfil\crcr$D$}}}
\newcommand{\nequiv}{\mbox{\ooalign{\hfil/\hfil\crcr$\equiv$}}}
\newcommand{\nsupset}{\mbox{\ooalign{\hfil/\hfil\crcr$\supset$}}}
\newcommand{\nni}{\mbox{\ooalign{\hfil/\hfil\crcr$\ni$}}}
\newcommand{\EV}{ {\rm eV} }
\newcommand{\KEV}{ {\rm keV} }
\newcommand{\MEV}{ {\rm MeV} }
\newcommand{\GEV}{ {\rm GeV} }
\newcommand{\TEV}{ {\rm TeV} }

\def\diag{\mathop{\rm diag}\nolimits}
\def\tr{\mathop{\rm tr}}

\def\Spin{\mathop{\rm Spin}}
\def\SO{\mathop{\rm SO}}
\def\O{\mathop{\rm O}}
\def\SU{\mathop{\rm SU}}
\def\U{\mathop{\rm U}}
\def\Sp{\mathop{\rm Sp}}
\def\SL{\mathop{\rm SL}}

\def\change#1#2{{\color{blue}#1}{\color{red} #2}\color{black}\hbox{}}


\begin{titlepage}

\begin{flushright}
LTH/689 \\
UCB-PTH-06/02 \\
LBNL-59632 \\
\end{flushright}

\vskip 2cm
\begin{center}
{\large \bf Proton Decay, Yukawa Couplings and \\ Underlying Gauge Symmetry 
in String Theory} 

\vskip 1.2cm
${}^a$Radu Tatar and ${}^b$Taizan Watari

\vskip 0.4cm

${}^a${\it Division of Theoretical Physics, Department of Mathematical Sciences

The University of Liverpool,
Liverpool,L69 3BX, England, U.K.

rtatar@liverpool.ac.uk}

${}^b${\it Department of Physics and Lawrence Berkeley National 
Laboratory,

University of California, Berkeley, CA 94720, USA

TWatari@lbl.gov} \\

\vskip 1.5cm

\abstract{In string theory, massless particles often originate from 
a symmetry breaking of a large gauge symmetry $G$ to its subgroup $H$. 
The absence of dimension-4 proton decay in supersymmetric theories 
suggests that $(\bar{D},L)$ are different from $\bar{H}(\bar{\bf 5})$ 
in their origins. In this article, we consider a possibility that 
they come from different irreducible components 
in $\mathfrak{g}/\mathfrak{h}$. 
Requiring that all the Yukawa coupling constants of quarks and leptons 
be generated from the super Yang--Mills interactions of $G$, 
we found in the context of Georgi--Glashow $H=\SU(5)$ unification 
that the minimal choice of $G$ is $E_7$ and $E_8$ is the only 
alternative. 
This idea is systematically implemented in Heterotic String, M theory 
and F theory, confirming the absence of dimension 4 proton decay 
operators. 
Not only $H=\SU(5)$ but also $G$ constrain operators of effective field 
theories, providing non-trivial information.
} 

\end{center}
\end{titlepage}


\section{Introduction}

String theory is a well-formulated theory of quantum gravity, and 
it is also known to be able to (almost) realize the (supersymmetric) 
standard model of particle physics at low-energies 
\cite{a1,a2,cv0,Penn4,cv1,DESY,dona1,dona2}. 
Since 1990's, the understanding of string theory has improved 
dramatically, 
and an enormous variety of string vacua have been discovered.
But, too much variety in string vacua and too much success in 
describing the standard model with string theory might also imply that 
there is no hope to obtain testable predictions from string theory 
in a near future.

Although the current string theory has a plethora of vacua, there is 
one thing in common among them. One of the most intriguing features 
of string theory is that it supports higher-dimensional gauge theories 
with 16 supersymmetry charges where scalar, fermion and vector fields 
are unified. Thus, at least at high energies, the interactions of 
quarks, leptons and Higgs fields are under the constraint 
of the gauge principle. String theory can realize a large gauge group 
$G$ and this symmetry can be broken down spontaneously to a subgroup 
$H$ such as SU(5), either by vector bundles in Heterotic theory 
or by resolution of singularities in M / F theory. 
Particles in low-energy spectrum charged under $H$ arise from 
$\mathfrak{g}/\mathfrak{h}$ part of the high-energy supersymmetric 
vector multiplet. The old ideas of Higgs multiplets \cite{gauge-higgs} 
or quarks and leptons \cite{coset-ql} coming out of a coset space 
$\mathfrak{g}/\mathfrak{h}$ or its complexification are realized 
naturally in string theory.
 
In the present work, we focus on the underlying large gauge symmetry 
$G$ and 16 supercharges. The usage of the non-linearly realized 
symmetry $G$ as a constraint of low-energy interactions is an idea 
that dates back to 1960's. When the symmetry $G$ is a gauge symmetry, 
and it is combined with the underlying 16-charge supersymmetry, 
{\it all} the interactions of quarks and leptons are constrained 
by the non-linearly realized gauge symmetry. We are not obtaining 
numerical predictions, as there is a lot of room for corrections 
and arbitrariness to come in. 
Our approach is to use the underlying symmetries as a powerful tool 
at the level of knowing whether certain interactions exist or not. 

We start from an assumption that the Yukawa couplings originate from 
supersymmetric Yang--Mills interactions of a large symmetry $G$ that 
contains $H=\SU(5)$. Theories with low-energy supersymmetry should 
not have dimension-4 proton decay operators, which means that there 
must be some difference in the origins of $\bar{\bf 5}=(\bar{D},L)$ 
and $\bar{H}(\bar{\bf 5})$ multiplets, although their representations 
under the unified SU(5)$_{\rm GUT}$ gauge group are the same. 
Thus, we introduce another assumption that they have different origins in 
$\mathfrak{g}/\mathfrak{h}$, so that these two multiplets have 
different interactions. Then we prove in section \ref{sec:BottomUp} 
that $G$ has to be at least $E_7$. The argument that leads to this 
statement does not depend on any implementation in string theory 
and it is only based on 
commutation relations and irreducible decomposition of Lie algebra. 
String theory supports $E_7$ symmetry in Heterotic, M- and F-theory. 
$G=E_8$ contains the minimal 
$E_7$ gauge symmetry and is also supported by string theory.
They are the only possibilities for the underlying gauge symmetry 
in String Theory.
The only way to get these groups other than in Heterotic theory is to go 
to the nonperturbative limits of Type IIA or Type IIB, which are 
M theory and F theory, respectively.
If one works in the perturbative type IIA or type IIB, the vacua with 
intersecting D-branes cannot be used for realising these symmetries. 
We show that there are overall seven 
possibilities in the way low-energy particles are identified in the coset 
space $\mathfrak{e}_7/\mathfrak{su}(5)$ or $\mathfrak{e}_8/\mathfrak{su}(5)$.

The idea presented in section \ref{sec:BottomUp} is realized explicitly 
in Calabi--Yau compactification of Heterotic theory in section 
\ref{sec:Het}, in $G_2$-holonomy compactification of eleven-dimensional 
supergravity in section \ref{sec:M} and finally in elliptic Calabi--Yau 
compactification of F-theory in section \ref{sec:F}. 
For instance, in Heterotic compactification, we see that the absence of 
dimension-4 proton decay is guaranteed (at the perturbative level) when 
a stable rank-5 vector bundle $V_5$ has a negative-slope sub-bundle 
with certain properties. Origins of low-energy multiplets are 
identified for each symmetry-breaking pattern, in each section. 
Discussion in section \ref{sec:BottomUp} guarantees that Yukawa couplings 
of charged leptons, down-type and up-type quarks are generated from 
the super Yang--Mills interactions, and in some cases, Dirac neutrino Yukawa 
couplings as well. 
Many yet-to-be answered questions are raised in this article, and 
at the same time, some predictions are also derived, using the constraint 
of the underlying gauge symmetry and supersymmetry. 

{\bf Reading guide}: The only prerequisite for section \ref{sec:BottomUp} is 
some knowledge in Lie algebra, and this section will be readable without 
details of string theory and it presents the key idea of this 
article. The later three sections for Heterotic (section \ref{sec:Het}) 
M- (section \ref{sec:M}) and F-theory (section \ref{sec:F}) are almost 
mutually independent, but assume the contents of section \ref{sec:BottomUp}.
For those who do not want to read all through the three sections, summary of 
physics is given in section \ref{sec:Sum}.

\section{Bottom-up Classification of the Origins of Matters}
\label{sec:BottomUp}

In this article, we only consider Georgi--Glashow SU(5)$_{\rm GUT}$ 
unification. A good bottom--tau Yukawa unification and a similar 
degree of hierarchy among Yukawa couplings of down-type quarks and 
charged leptons are consistent with this unification. The disparity 
between the large mixing angles in the neutrino oscillations and 
the small mixing angles in the quark sector can be attributed 
to the difference in the nature of 
SU(5)$_{\rm GUT}$-${\bf 10}=(Q,\bar{U},\bar{E})$ representations 
containing quark doublets $Q$ and 
SU(5)$_{\rm GUT}$-$\bar{\bf 5}=(\bar{D},L)$ representations containing 
lepton doublets $L$ \cite{non-parallel,anarchy}. 
All these are the motivations for us to consider SU(5)$_{\rm GUT}$ 
unification. Since no extra gauge symmetry prevents right-handed 
neutrinos from having large Majorana masses, tiny neutrino 
masses may be explained by the see-saw mechanism \cite{see-saw}, 
and this may be another good feature of the SU(5)$_{\rm GUT}$ 
unification. If the gauge group is larger, it has to be broken 
in a field-theoretical way. But, the analysis in this article 
can also be taken just as an example, and one can freely extend 
the analysis to other unification groups. 
Indeed, the above argument is not as rigorous as it might appear.

\subsection{Assumptions}

In order for us to be able to understand the Yukawa couplings better, 
we assume that all the Yukawa couplings of up-type and down-type 
quarks and charged leptons are generated from super Yang--Mills 
interactions of an underlying symmetry $G$, and all the low-energy 
multiplets are given their origins in the coset space 
$\mathfrak{g}/\mathfrak{su}(5)$.

Another assumption that we introduce in this article is that 
$\bar{\bf 5}=(\bar{D},L)$ and $\bar{H}(\bar{\bf 5})$ have distinct 
origins in the coset space $\mathfrak{g}/\mathfrak{su}(5)$. This is 
to come up with models without dimension-4 proton decay operators.
Indeed, we always need Yukawa couplings of down-type quarks and 
charged leptons 
\begin{equation}
 W \ni y \; \bar{\bf 5} \cdot {\bf 10} \cdot \bar{H}(\bar{\bf 5}),
\label{eq:dYukawa}
\end{equation}
while the dimension-4 proton decay operators should be absent\footnote{
If the supersymmetry breaking masses of the squark--slepton sector
are much higher than the electroweak scale, the constraints on those 
operators are weaker. But, even for the supersymmetry breaking masses 
of order $10^{10}$ GeV, the coefficients of the operators 
(\ref{eq:dim4}) should be much smaller than unity.}
\begin{equation}
 W \ni \bar{\bf 5} \cdot {\bf 10} \cdot \bar{\bf 5} 
  = \bar{D} \cdot \bar{U} \cdot \bar{D} 
  + \bar{D} \cdot Q \cdot L 
  + L \cdot \bar{E} \cdot L, 
\label{eq:dim4}
\end{equation}
or protons would not remain stable. The operators (\ref{eq:dYukawa}) 
and (\ref{eq:dim4}) differ only by $\bar{H}(\bar{\bf 5})$ in 
(\ref{eq:dYukawa}) and $\bar{\bf 5}$ in (\ref{eq:dim4}).
If they are identified with different parts of the coset space 
$\mathfrak{g}/\mathfrak{h}$, then the operators (\ref{eq:dYukawa}) and 
(\ref{eq:dim4}) transform differently under some symmetry in $G$.
Thus, such symmetry can kill the dangerous operators (\ref{eq:dim4}) 
while allowing the necessary Yukawa couplings 
(\ref{eq:dYukawa}).\footnote{
This assumption is to try to solve the dimension-4 proton decay 
problem by restricting the underlying symmetry $G$ and its breaking 
pattern. The other possibility that is not covered here is that 
those two multiplets have different properties under some symmetry 
of compactification geometry, although they are exactly in the same 
irreducible component in the coset space $\mathfrak{g}/\mathfrak{h}$. 
This possibility has to restrict the compactification geometry, 
instead of $\mathfrak{g}/\mathfrak{h}$ and its breaking pattern, 
in order to solve the problem of dimension-4 proton decay.} 
For the moment being, we keep an open mind to all kinds of possibility for the 
choice of such symmetry. We shall see in this section that there are strong restrictions 
on the symmetry. 

\subsection{Up-type Quark Yukawa Couplings from $E_6$ Symmetry}
  
In order to figure out the minimal underlying symmetry $G$, 
we proceed by enhancing $G$ a little by little on necessity basis. 
The enhanced symmetry provides both massless matters and 
their interactions. The 2 phenomenology assumptions constrain 
the origin of low-energy particles and the underlying symmetry $G$. 
The choice of $G$ and its breaking pattern can 
be determined purely in representation theory of Lie algebra, 
without paying much attention to the explicit implementation in 
string theories. The net chirality in the low-energy spectrum 
depends on topological aspects of geometry of compactification 
manifold, and  we discuss separately in a later section. 
For this reason, we do not distinguish $\mathfrak{g}/\mathfrak{h}$ 
from its complexification.

Let us start off with the {\bf 10} representation and its up-type 
quark Yukawa coupling. 
The origin of ${\bf 10}$ representation can be in 
$\mathfrak{so}(10)/\mathfrak{su}(5)$. If it is to be realised in 
Type IIA or Type IIB string theory, it is (1,0) (and (-1,0)) open 
string connecting 5 parallel D-branes and their orientifold mirror 
images. It is also easy to find this representation in Type I string 
theory or Heterotic theory. Thus, the existence of this representation 
alone does not tell us a lot.

The most intriguing feature of SU(5)$_{\rm GUT}$-unified theories 
is that there must be up-type quark Yukawa couplings 
\begin{equation}
 W \ni {\bf 10}^{ij} \; {\bf 10}^{kl} \; H({\bf 5})^m \epsilon_{ijklm}.
\label{eq:uYukawa}
\end{equation}
If the multiplets above were made of open strings in Type IIA or 
Type IIB string theories, SU(5)-indices could be contracted only 
between a pair of superscript and subscript, combining a $\sigma=0$ 
boundary and a $\sigma=\pi$ boundary of open strings. It is unlikely 
that those Yukawa couplings are generated in Type IIA or Type IIB 
string theories. But, we know that at least in Heterotic theories, 
compactification of the $E_8 \times E_8$ theory can provide such 
a Yukawa coupling, and it is also clear from the Heterotic--M-theory 
duality \cite{wittendual} and Heterotic---F-theory duality 
\cite{evidence,MV1,MV2,FMW,Het-F-4D} that both M-theory and F-theory 
can describe such Yukawa coupling, as well.
Although the description will not be available within the Type IIA or 
IIB limits, M-theory and F-theory can describe the Yukawa couplings. 
We cannot expect that all the low-energy particles are described 
by simple open strings ending on D-branes, but by membranes \cite{BSV}
or by $(p, q)$ strings \cite{GZ}, respectively, in M-theory or 
in F-theory. In both descriptions, an $E_r$-type ($r=6,7,8$) 
singularity has to be involved.

The origin of the $H({\bf 5})$ multiplet containing the up-type 
Higgs doublet $H_u$ can be identified in a coset space 
$\mathfrak{e}_6/\mathfrak{su}(5)$, rather than 
$\mathfrak{su}(N)/\mathfrak{su}(5)$ or 
$\mathfrak{so}(N)/\mathfrak{su}(5)$ with large $N$. The irreducible 
decomposition\footnote{Here, Res$^G_H$ means restriction of group 
from $G$ to $H$. This has nothing to do with residue.} 
of $\mathfrak{e}_6$-{\bf adj.} representation is given by 
\begin{eqnarray}
 {\rm Res}^{E_6}_{\SU(5)_{\rm GUT} \times \SU(2)_2 \times \U(1)_6} 
  \; \left(\mathfrak{e}_6\mbox{-}{\bf adj.} \right)
 & \simeq & 
    ({\bf adj.},{\bf 1})^0 \oplus ({\bf 1},{\bf adj.})^0 \oplus 
    ({\bf 1},{\bf 1})^0 \nonumber \\
 & &  \oplus \left[(\wedge^2 {\bf 5},{\bf 2})^1 \oplus 
               (\wedge^4 {\bf 5},\wedge^2 {\bf 2})^2
             \right] \oplus {\rm h.c.},
\label{eq:e6dcmp}
\end{eqnarray}
where a multiplet $(R_5,R_2)^{q_6}$ is in a representation $R_5$ 
of unbroken SU(5)$_{\rm GUT} \subset E_6$, in $R_2$ of underlying 
SU(2)$_2 \subset E_6$ that commutes with SU(5), and $q_6$ denotes 
the charge under the U(1)$_{q_6}$ symmetry.\footnote{This U(1)$_{q_6}$ 
symmetry and U(1)$_{q_7}$ that appears later are related to the two 
U(1) symmetries $E_6 \rightarrow \SO(10) \times \U(1)_\psi$ and 
$\SO(10) \rightarrow \SU(5)_{\rm GUT} \times \U(1)_\chi$ in \cite{PDG} 
as 
\begin{equation}
\left( \begin{array}{c} q_{\chi} \\ q_{\psi} \end{array} \right) \equiv
\left( \begin{array}{c} \sqrt{24}Q_\chi \\  \sqrt{\frac{72}{5}} Q_\psi 
       \end{array} \right) = 
 \frac{1}{3} \left(\begin{array}{cc} -5 & 1 \\ 1 & 1 \end{array} \right)
 \left(\begin{array}{c} q_6 \\ q_7 \end{array} \right).
\end{equation} }
This irreducible decomposition contains not only a multiplet 
in SU(5)$_{\rm GUT}$-$\wedge^2 {\bf 5}$ representation, a candidate for 
${\bf 10}=(Q,\bar{U},\bar{E})$ multiplets, but also one in 
SU(5)$_{\rm GUT}$-$\overline{\wedge^4 {\bf 5}}$, that is, 
in SU(5)$_{\rm GUT}$-${\bf 5}$ representation. 
This is a candidate of $H({\bf 5})$.

There is no problem in incorporating this coset space in Heterotic 
$E_8\times E_8$ theory, since it contains $E_6$ as a subgroup of 
$E_8$. The $E_6$ symmetry is broken down to SU(5)$_{\rm GUT}$ 
by turning on non-trivial vector bundle with the structure group 
SU(2)$_2 \times$U(1)$_{q_6}$ inside the $E_6$.
The net chirality at low energy can be obtained by the topological 
nature of the zero-modes of the vector bundle.
In M-theory and F-theory descriptions, the $E_6$ vector multiplet
does not come from strings, but from membranes \cite{BSV} 
(or $(p,q)$ open strings \cite{GZ}). Singular ALE fibration or 
elliptic fibration can give rise to $E_6$, which can be broken down 
to SU(5)$_{\rm GUT}$ by deforming the geometry in the fibre direction 
over base manifolds. Chiral matters arise at the loci of enhanced 
singularity. In any of Heterotic, M-theory and F-theory descriptions, 
$E_6$ symmetry is realised, is broken to SU(5)$_{\rm GUT}$ and 
chiral matters in the representations in the coset space of 
$\mathfrak{e_6}/\mathfrak{su}(5)$ are available in the low-energy 
effective theories. More details of the descriptions are explored 
in later sections. In this section, we further pursue 
general aspects that do not depend on the specific implementation 
in either one of Heterotic, M- or F-theory.

In Heterotic $E_8 \times E_8$ formulation, D = 10 ${\cal N} = 1$ 
super Yang--Mills interaction is described in terms of D=4 
superpotential as, e.g., \cite{AGW} 
\begin{equation}
 W = \int d^6 y 
  \tr{}_{E_6} \left( \Sigma^a(y,x) \left[(\partial^b - \Sigma^b(y,x)),
             \Sigma^c(y,x) \right]
      \right) \epsilon_{abc},
\label{eq:16SUSY}
\end{equation} 
where $a=1,2,3$ correspond to 3 complex dimensions of the compact 
Calabi--Yau space. This comes from the second term of the Heterotic
theory superpotential 
\begin{equation}
 W \ni \int_Z \Omega \wedge \left(dB - \omega_{YM} +\omega_{grav.}
 \right),
\label{eq:Het-super}
\end{equation}
where $\Omega$ is the global holomorphic 3-form of Calabi--Yau 3-fold, 
and $\omega_{YM}$ and $\omega_{grav.}$ are Chern--Simons 3-forms.
Chiral zero-modes live in some irreducible representations, with 
certain wave function depending on the internal coordinates $y$'s.
Thus, this superpotential gives rise to some interactions at 
low energy, and in particular, we expect that the Yukawa coupling 
is obtained in this way.
However, since the overall compactification geometry does not preserve 
all the 16 SUSY charges, the value of resulting coupling constants 
will not be equal to the gauge coupling constant of the unbroken 
SU(5)$_{\rm GUT}$. It is easy to see from the Lie algebra 
of $E_6$ that the up-type quark Yukawa coupling (\ref{eq:uYukawa}) 
is contained in (\ref{eq:16SUSY}), when 
$\overline{(\wedge^4 {\bf 5},\wedge^2 {\bf 2})}$ is 
identified with $H({\bf 5})$:
\begin{equation}
 W \ni (\wedge^2 {\bf 5},{\bf 2})\cdot (\wedge^2{\bf 5},{\bf 2})
   \cdot \overline{(\wedge^4 {\bf 5},\wedge^2 {\bf 2})} = 
  {\bf 10} \cdot {\bf 10} \cdot H({\bf 5}).
\label{eq:uYukawaE6}
\end{equation}
The existence of the up-type quark Yukawa coupling is not specific 
to the Heterotic theory. The only essences are the underlying $E_6$ symmetry, 
and the existence of Yang--Mills interactions that have 
16 supersymmetry charges locally. The presence or absence of 
interactions among chiral matters is tied to that of vector fields 
by the supersymmetry and is ultimately determined 
by the Lie algebra of $E_6$.
Since the geometric engineering of M-theory and singular elliptic 
fibre of F-theory give rise to $E_6$ vector multiplet with 
16 supersymmetry charges locally, the same results as 
in the Heterotic description follow in the M-theory and F-theory 
descriptions as well. 
For instance, the Figure 7 of \cite{GZ} can be interpreted as the 
$(p,q)$ string description of the up-type quark Yukawa coupling 
(\ref{eq:uYukawaE6}).

The coset space $\mathfrak{e}_6/\mathfrak{su}(5)$, however, offers 
only one type of multiplet SU(5)$_{\rm GUT}$-$\bar{\bf 5}$+h.c. 
representation. Since the minimal supersymmetric extension of the 
standard model contains at least two different multiplets 
in SU(5)$_{\rm GUT}$-$\bar{\bf 5}$ representation, namely
$\bar{\bf 5}$ and $\bar{H}(\bar{\bf 5})$, the coset space 
$\mathfrak{e}_6/\mathfrak{su}(5)$ has to be extended so that 
they are incorporated. 

\subsection{The Minimal $E_7$ Symmetry}
\label{ssec:bottomup}

A coset space $\mathfrak{e}_7/\mathfrak{su}(5)$ contains 
$\mathfrak{e}_6/\mathfrak{su}(5)$ and provides more variety 
in the multiplets. First, $E_7$ symmetry contains 
SU(6)$_1 \times$SU(2)$_2 \times$U(1)$_{q_6}$, and 
the $\mathfrak{e}_7$-{\bf adj.} representation has 
the irreducible decomposition 
\begin{eqnarray}
 {\rm Res}^{E_7}_{\SU(6)_1 \times \SU(2)_2 \times \U(1)_{q_6}} 
   \left( \mathfrak{e}_7\mbox{-}{\bf adj.} \right) & \simeq & 
     ({\bf adj.}={\bf 35},{\bf 1})^0 \oplus ({\bf 1},{\bf adj.})^0 
      \oplus ({\bf 1},{\bf 1})^0 \nonumber \\ &  \oplus &  
     \left[ (\wedge^2 {\bf 6},{\bf 2})^1 \oplus 
            (\wedge^4 {\bf 6},\wedge^2 {\bf 2})^2 \oplus 
            (\wedge^6 {\bf 6},{\bf 2})^3 
     \right] \oplus {\rm h.c.}.
\label{eq:e7dcmp}
\end{eqnarray}
The unification group SU(5)$_{\rm GUT}$ is identified with 
the subgroup of $\SU(5) \times \U(1)_{q_7} \subset \SU(6)_1$. 
Thus, when the $E_7$ symmetry is broken down to SU(5)$_{\rm GUT}$, 
we have the following irreducible components:
\begin{eqnarray}
 ({\bf adj.},{\bf 1})^{0} & \rightarrow & ({\bf adj.},{\bf 1})^{(0,0)} 
   \oplus \left[ ({\bf 5},{\bf 1})^{(0,6)} \oplus {\rm h.c.} \right],\\
 (\wedge^2 {\bf 6},{\bf 2})^1 & \rightarrow & 
    \left[ (\wedge^2 {\bf 5},{\bf 2})^{(1,2)} \oplus 
           ({\bf 5},{\bf 2})^{(1,-4)} \right] + {\rm h.c.}, \\
 (\wedge^4 {\bf 6},\wedge^2 {\bf 2})^2 & \rightarrow & 
   \left[ (\wedge^4 {\bf 5},\wedge^2 {\bf 2})^{(2,4)} \oplus 
          (\wedge^3 {\bf 5},\wedge^2 {\bf 2})^{(2,-2)} \right] 
     +{\rm h.c.},  
\end{eqnarray} 
where $(R_5,R_2)^{(q_6,q_7)}$ describe the representations under 
SU(5)$_{\rm GUT} \times$SU(2)$_2 \times$U(1)$_{q_6} \times$U(1)$_{q_7} 
\subset E_7$. In particular, there are three different multiplets in 
SU(5)-$\bar{\bf 5}$ representation, namely, 
$\overline{({\bf 5},{\bf 1})^{(0,6)}}$, 
$\overline{({\bf 5},{\bf 2})^{(1,-4)}}$ and 
$(\wedge^4 {\bf 5},\wedge^2 {\bf 2})^{(2,4)}$. 
Thus, the two low energy multiplets in the SU(5)-$\bar{\bf 5}$ 
representation, namely, $\bar{\bf 5}=(\bar{D},L)$ and $H(\bar{\bf 5})$,
can be identified with 2 different multiplets. 
As long as they are different, there will be no dimension-4 proton 
decay operators (\ref{eq:dim4}).

It is preferable if the 2 multiplets are identified so that 
the down-type quark Yukawa coupling is generated from the 16-SUSY 
Yang--Mills interaction. In order to identify properly, let us 
look at the $E_7$ version of the superpotential (\ref{eq:16SUSY}):
\begin{eqnarray}
 W & \ni & \left[
   (\wedge^2 {\bf 6},{\bf 2})^1 \cdot (\wedge^2 {\bf 6},{\bf 2})^1
    \cdot \overline{(\wedge^4 {\bf 6},\wedge^2 {\bf 2})^2} 
  + (\wedge^2 {\bf 6},{\bf 2})^1 \cdot 
    (\wedge^4 {\bf 6},\wedge^2 {\bf 2})^2 \cdot 
    \overline{({\bf 1},{\bf 2})^3} \right] + {\rm h.c.} \nonumber \\
  & + & \overline{(\wedge^2 {\bf 6},{\bf 2})^1} \cdot 
     ({\bf adj.},{\bf 1})^0 \cdot (\wedge^2 {\bf 6},{\bf 2})^1 
      + \overline{(\wedge^4 {\bf 6},\wedge^2 {\bf 2})^2} \cdot 
     ({\bf adj.},{\bf 1})^0 \cdot (\wedge^4 {\bf 6},\wedge^2 {\bf 2})^2
   + \cdots ,
\label{eq:poten}
\end{eqnarray}
which is, in terms of SU(5)$_{\rm GUT}$, 
\begin{eqnarray}
 W & \ni & 
  (\wedge^2 {\bf 5},{\bf 2})^{(1,2)} \! \!  \cdot \!\!
     (\wedge^2 {\bf 5},{\bf 2})^{(1,2)} \! \! \cdot \!\!
     \overline{(\wedge^4 {\bf 5},\wedge^2 {\bf 2})^{(2,4)}} + 
  ({\bf 5},{\bf 2})^{(1,-4)} \!\! \cdot \!\! 
     (\wedge^2 {\bf 5},{\bf 2})^{(1,2)} \!\! \cdot \!\! 
     \overline{(\wedge^3 {\bf 5},\wedge^2 {\bf 2})^{(2,-2)}} 
  \nonumber \\
 & + &   
   (\wedge^2 {\bf 5},{\bf 2})^{(1,2)} \!\! \cdot \! 
     (\wedge^3 {\bf 5},\wedge^2 {\bf 2})^{(2,-2)} \! \cdot \! 
     \overline{({\bf 1},{\bf 2})^{(3,0)}} + 
   ({\bf 5},{\bf 2})^{(1,-4)} \!\! \cdot \! 
     (\wedge^4{\bf 5},\wedge^2 {\bf 2})^{(2,4)} \! \cdot \! 
     \overline{({\bf 1},{\bf 2})^{(3,0)}} 
    \nonumber \\
 & + & 
    \overline{({\bf 5},{\bf 2})^{(1,-4)}} \!  \cdot \! 
      \overline{({\bf 5},{\bf 1})^{(0,6)}} \! \cdot \! 
      (\wedge^2{\bf 5},{\bf 2})^{(1,2)} + 
    (\wedge^4{\bf 5},\wedge^2 {\bf 2})^{(2,4)} \! \cdot \! 
        \overline{({\bf 5},{\bf 1})^{(0,6)}} \! \cdot \!
        \overline{(\wedge^3 {\bf 5},\wedge^2{\bf 2})^{(2,-2)}} 
     \nonumber \\  
  & &   + {\rm h.c.} + \cdots.
\end{eqnarray}
Therefore, for instance under the 2 identifications given in Table 
\ref{tab:ID-7}, the first term provides the up-type quark Yukawa 
coupling (\ref{eq:uYukawa}), and the fifth term the down-type quark 
and charged-lepton Yukawa couplings (\ref{eq:dYukawa}). 
The fourth term in the hermitian-conjugate representation is 
interpreted as the Dirac Yukawa coupling of neutrinos\footnote{
Note that right-handed neutrinos (in the identification A in Table 
\ref{tab:ID-7}) naturally arises in the spectrum, although 
we did not necessarily require their presence. Indeed, the existence of 
right-handed neutrinos does not necessarily favours SO(10) 
unification; the spinor representation of SO(10) certainly contains 
a right-handed neutrino, but it remains as a part of moduli 
(vector bundle moduli in Heterotic theory, and complex structure 
moduli in F-theory) in string theory terminology when SU(5) unified 
theory is considered instead of SO(10) theories. This is almost 
clear from the discussion in \cite{KcVf,6authors}. See also a recent 
article \cite{dona2}.} 
under the identification A, and the same term is interpreted 
as the trilinear coupling $W = S H({\bf 5}) \bar{H}(\bar{ \bf 5})$ 
of some extensions of the minimal supersymmetric standard model 
\cite{NMSSM,CPHW,LEP-limit} under the identification B.
\begin{table}[t]
\begin{center}
 \begin{tabular}{|c||c|c|c|c|c|}
\hline
 Bundles & $(\wedge^2{\bf 5},{\bf 2})^{(1,2)}$ & 
         $\overline{({\bf 5},{\bf 2})^{(1,-4)}}$ & 
         $\overline{(\wedge^4 {\bf 5},\wedge^2 {\bf 2})^{(2,4)}}$ & 
         $\overline{({\bf 5},{\bf 1})^{(0,6)}}$ & 
         $({\bf 1},{\bf 2})^{(3,0)}$ \\
\hline
 Particle ID A &  $(\bar{U},Q,\bar{E})$ & $\bar{\bf 5}=(\bar{D},L)$ &
         $H({\bf 5})$ & $\bar{H}(\bar{\bf 5})$ & $\bar{N}$ \\
 Particle ID B &  $(\bar{U},Q,\bar{E})$ & $\bar{H}(\bar{\bf 5})$ &
         $H({\bf 5})$ & $\bar{\bf 5}=(\bar{D},L)$ & $S$ \\
\hline
 \end{tabular}
\caption{\label{tab:ID-7}
Particle identification in $\mathfrak{e}_7/\mathfrak{su}(5)$: 
corresponding 
SU(5)$_{\rm GUT}\times$SU(2)$_1 \times$U(1)$_{q_6}\times$U(1)$_{q_7}$ 
irreducible representations in the $\mathfrak{e}_7$-{\bf adj.} 
is shown. There are 2 different possible identifications.} 
\end{center}
\end{table}

Note that there are no dimension-4 proton decay operators in this 
superpotential. It is true that (\ref{eq:16SUSY}) may not be the only 
interaction in the superpotential; there may be non-perturbative 
contribution as well. But it is inconceivable that the dimension-4
proton decay operator is generated, as we can see as follows. 
In either of the identification A or B, the operator (\ref{eq:dim4}) 
is an SU(2)$_2$ doublet, not a singlet. Since SU(2)$_2$ vector 
bundles in Heterotic terminology and corresponding objects 
in other descriptions of string theories can provide only spurions 
in SU(2)$_2$-{\bf adj.} representations, insertion of such spurions 
do not convert the doublet into a singlet. Even if we remain agnostic 
about the origin of corrections to the superpotential, as long as 
the corrections respect the underlying $E_7$ symmetry and its breaking 
is controlled by the spurions in 
SU(2)$_2\times$U(1)$_{q_6}\times$U(1)$_{q_7}$, the dimension-4 proton 
decay operators are forbidden by the $E_7$ symmetry. 
The absence of dimension-5 proton decay operators 
\begin{equation}
 W \ni {\bf 10}.{\bf 10}.{\bf 10}.\bar{\bf 5}
\label{eq:dim5}
\end{equation}
also follow from the same argument in the case of the identification B. 
In the case of identification A, these operators can be SU(2)$_2$
singlet, and only U(1) charges do not vanish. Since U(1) symmetry may
have mixed anomalies, Green--Schwarz fields may effectively supply the 
U(1) charges. Thus, it is not absolutely clear that the dimension-5 
proton decay operators are absent in the presence of non-perturbative 
corrections. But, on the other hand, it is worthwhile to note that 
a reasonable exponential suppression in non-perturbative effects is 
sufficient to make it consistent with the experimental limits, since 
the limits on the dimension-5 operators are not that stringent as 
those on the dimension-4 operators.

Thus, we have seen that the $E_7$ symmetry (or 2-cycles with 
$E_7$ intersection form) is minimal when we require 
all the Yukawa couplings from the Yang--Mills interaction, and 
different origins for $\bar{\bf 5}$ and $\bar{H}(\bar{\bf 5})$ in 
order to avoid the dimension-4 proton decay. 
It is not difficult to realise the $E_7$ symmetry and its breaking to 
SU(5)$_{\rm GUT}$ in any branches of string theories, 
just as in the case of $E_6$. For instance, the Figure 8 of \cite{GZ} 
can be regarded as $(p,q)$-string picture of the Dirac Yukawa 
coupling of neutrinos in the identification A, or 
the $W\ni S H \bar{H}$ interaction in the identification B.
Vector bundles (in Heterotic terminology) that break 
the $E_7$ symmetry down to SU(5)$_{\rm GUT}$ have to maintain 
the distinction between $({\bf 5},{\bf 2})^{(1,-4)}$ and 
$({\bf 5},{\bf 1})^{(0,6)}$. Thus, the bundle is supposed to have the 
structure group SU(2)$_2 \times$U(1)$_{q_6} \times$U(1)$_{q_7}$.

\subsection{Extension to the $E_8$ Symmetry}

The observation that the $E_7$ is the minimal underlying symmetry 
implies that the $E_8$ symmetry of the Heterotic theory is not 
strictly necessary in order to construct a ``realistic''\footnote{
Note, however, that we have not considered how the SU(5)$_{\rm GUT}$ 
symmetry is broken down to the standard model gauge group. 
See below.} model.
But all Heterotic, D=11 SUGRA and F-theory admit up to the 
$E_8$ symmetry, and it is also interesting as an alternative 
to think of identification in $\mathfrak{e}_8 / \mathfrak{su}(5)$.
This can be done simply by embedding the identification in the 
minimal $\mathfrak{e}_7/\mathfrak{su}(5)$ coset through 
\begin{equation}
 E_7 \times \SU(2)_1 \subset E_8, \qquad {\rm or} \qquad 
 \mathfrak{e}_7 + \mathfrak{su}(2)_1 \subset \mathfrak{e}_8.
\label{eq:7to8}
\end{equation}
\begin{equation}
 {\rm Res}^{E_8}_{E_7 \times \SU(2)_1} \left(\mathfrak{e}_8 
   \mbox{-}{\bf adj.} \right) \simeq 
  ({\bf adj.},{\bf 1}) \oplus ({\bf 1},{\bf adj.}) \oplus 
  ({\bf 56},{\bf 2}). 
\end{equation} 
Since we have already seen the irreducible decomposition of 
$\mathfrak{e}_7\mbox{-}{\bf adj.}$, we are left with 
$({\bf 56},{\bf 2})$, which decomposes into 
\begin{equation}
 {\rm Res}^{E_7\times \SU(2)_1}_{
   \SU(6)_1 \times\SU(2)_1 \times\SU(2)_2 \times \U(1)_{q_6}}
  ({\bf 56},{\bf 2}) \simeq 
  \left[
     ({\bf 6},{\bf 2},{\bf 2})^{-1}\oplus({\bf 6},{\bf 2},{\bf 1})^{2}
  \right] \oplus {\rm h.c.} 
  \oplus (\wedge^3 {\bf 6},{\bf 2},{\bf 1})^0. 
\end{equation}
After further decomposing into SU(5)$_{\rm GUT}$-irreducible pieces,
one can see that the $\mathfrak{e}_8$-{\bf adj.} representation 
provides three multiplets in SU(5)$_{\rm GUT}$-$\wedge^2{\bf 5}$ 
representation
\begin{equation}
(\wedge^2 {\bf 10},{\bf 1} \oplus {\bf 2}_1 \oplus {\bf 2}_2) = 
 \left( \begin{array}{r}
  (\wedge^2 {\bf 5},{\bf 1},{\bf 1})^{(-2,2)} \\
  (\wedge^2 {\bf 5},{\bf 2},{\bf 1})^{(0,-3)} \\
  {\bf 10}= (\wedge^2 {\bf 5},{\bf 1},{\bf 2})^{(1,2)} \\
	\end{array}\right),
\label{eq:10inE8}
\end{equation}
where $(R_5,R_1,R_2)^{(q_6,q_7)}$ on the right-hand side 
denotes the representation under the 
SU(5)$_{\rm GUT} \times \SU(2)_1 \times \SU(2)_2$ and 
the charge under U(1)$_{q_6}$ and U(1)$_{q_7}$. There are 5 multiplets 
in the SU(5)$_{\rm GUT}$-$\bar{\bf 5}$ representation 
\begin{equation}
 (\bar{\bf 5},\wedge^2 ({\bf 1} \oplus {\bf 2}_1 \oplus {\bf 2}_2)) 
\simeq
 \left(\begin{array}{rr|r}
\qquad & (\bar{\bf 5},{\bf 2},{\bf 1})^{(-2,-1)} 
& \bar{\bf 5} = (\bar{\bf 5},{\bf 1},{\bf 2})^{(-1,4)} \\
& \bar{H}(\bar{\bf 5}) = (\bar{\bf 5},{\bf 1},{\bf 1})^{(0,-6)}  
& (\bar{\bf 5},{\bf 2},{\bf 2})^{(1,-1)} \\
\hline 
& & H({\bf 5})^\dagger = (\bar{\bf 5},{\bf 1},{\bf 1})^{(2,4)} 
       \end{array}\right),
\label{eq:5inE8}
\end{equation}
and there are 3 SU(5)$_{\rm GUT}$-singlets
\begin{equation}
 ({\bf 1},{\bf adj.}({\bf 1} \oplus {\bf 2}_1 \oplus {\bf 2}_2)) 
\simeq
 \left(\begin{array}{cc|c}
   \qquad & ({\bf 1},{\bf 2},{\bf 1})^{(-2,5)} &  \\
   & &  \qquad \\ 
 \hline 
  \bar{N} = ({\bf 1},{\bf 1},{\bf 2})^{(3,0)}  &
  ({\bf 1},{\bf 2},{\bf 2})^{(1,5)}  &
        \end{array}\right).
\label{eq:1inE8}
\end{equation}
Corresponding low-energy multiplets are also noted above, following 
the identification A in Table \ref{tab:ID-7} 
in $\mathfrak{e}_7/\mathfrak{su}(5)$ and the embedding 
(\ref{eq:7to8}). 
The irreducible decompositions (\ref{eq:10inE8})--(\ref{eq:1inE8})
are a special cases of SU(5)-bundle compactification of the Heterotic 
$E_8\times E_8$ theory; the rank-5 SU(5) vector bundle $V_5$ is now 
split up into 
\begin{equation}
 V_5 \simeq {\bf 1} \oplus {\bf 2}_1 \oplus {\bf 2}_2 
\label{eq:221bdle}
\end{equation}
with the structure group reduced to 
SU(2)$_1\times$SU(2)$_2\times$U(1)$_{q_6}\times$U(1)$_{q_7}$. 
The singlets in (\ref{eq:1inE8}) are would-be vector bundle moduli if 
the structure group were SU(5). 

Since we only need to maintain the different origins for $\bar{\bf 5}$
and $\bar{H}({\bf 5})$ multiplets, 
the $\mathfrak{e}_8/\mathfrak{su}(5)$ coset does not have to be 
split up as much as in (\ref{eq:10inE8})--(\ref{eq:1inE8}).
In particular, the vector bundles $V_5 = {\bf 1} \oplus {\bf 2}_1 \oplus 
{\bf 2}_2$ can be a little more generic.
Let us  use the particle identification A in Table \ref{tab:ID-7} 
inside (\ref{eq:10inE8})--(\ref{eq:1inE8})
for concreteness\footnote{The following discussion does not 
change essentially when the ID B is adopted.} 
in the following discussion. 
Then, the distinction between the 2 multiplets in $\bar{\bf 5}$ 
representation is not lost when the SU(2)$_1 \times$U(1)$_{\tilde{q}_6}$ 
bundle\footnote{
The U(1) symmetries of $\tilde{q}_6$ and $\tilde{q}_7$ are different linear 
combinations of $q_6$ and $q_7$:
\begin{equation}
\left(\begin{array}{c} \tilde{q}_6 \\ \tilde{q}_7 \end{array} \right) = 
\frac{1}{3} \left(\begin{array}{cc} 2 & -1 \\ -5 & -2 \end{array} \right)
\left(\begin{array}{c} q_6 \\ q_7 \end{array}\right), \qquad 
\left(\begin{array}{c} q_6 \\ q_7 \end{array}\right) = 
\frac{1}{3} \left(\begin{array}{cc} 2 & -1 \\ -5 & -2 \end{array} \right)
\left(\begin{array}{c} \tilde{q}_6 \\ \tilde{q}_7 \end{array} \right).
\end{equation}
In the SU(5) structure group of $V_5$, the U(1) generators are
\begin{eqnarray}
{\bf q}_6 = \diag(-2,0,0,1,1), & \qquad & {\bf q}_7 = \diag(2,-3,-3,2,2), \\
{\bf q}_\chi = \diag(4,-1,-1,-1,-1), & \qquad & 
{\bf q}_\psi = \diag(0,-1,-1,1,1), \\
\tilde{\bf q}_6 = \diag(-2,1,1,0,0), & \qquad & 
\tilde{\bf q}_7 = \diag(2,2,2,-3,-3).
\end{eqnarray}
} ${\bf 1} \oplus {\bf 2}_1$ is replaced by a rank-3 SU(3) bundle 
${\bf 3}_2$, so that the rank-5 vector bundle is 
$V_5 = {\bf 3}_2 \oplus {\bf 2}_2$ with the structure group 
SU(3)$_2 \times$SU(2)$_2 \times$U(1)$_{\tilde{q}_7}$: 
\begin{eqnarray}
 ({\bf10},{\bf 3}_2 \oplus {\bf 2}_2) & \simeq &
  \left(\begin{array}{r}
	({\bf 10},{\bf 3}_2) 	\\ \hline 
        {\bf 10}= \, ({\bf 10},{\bf 2}_2)
	\end{array}
  \right), \\
 (\bar{\bf 5},\wedge^2 ({\bf 3}_2 \oplus {\bf 2}_2)) & \simeq &
  \left( \begin{array}{r|r}
   \bar{H}(\bar{\bf 5}) = (\bar{\bf 5},\wedge^2 {\bf 3}_2) & 
   \bar{\bf 5} = (\bar{\bf 5},{\bf 3}_2 \otimes {\bf 2}_2) \\ 
   \hline 
   & H({\bf 5})^\dagger = (\bar{\bf 5},\wedge^2 {\bf 2}_2)
	 \end{array}\right).
\end{eqnarray}

Likewise, the bundle ${\bf 2}_1 \oplus {\bf 2}_2$ can be replaced 
by a rank-4 vector bundle ${\bf 4}$ with the structure group 
SU(4)$\supset$SU(2)$_1 \times$SU(2)$_2 \times$U(1)$_{\psi}$, 
so that the rank-5 bundle becomes $V_5={\bf 1} \oplus {\bf 4}$ 
with the structure group U(1)$_{\chi}\times$SU(4). 
The reduced rank-5 bundles 
$V_5={\bf 3}_2 \oplus {\bf 2}_2$, and $V_5={\bf 1} \oplus {\bf 4}$ are much 
more generic than $V_5 = {\bf 1} \oplus {\bf 2}_1 \oplus {\bf 2}_2$, 
yet the original motivation is not lost---distinguishing $\bar{\bf 5}$ 
and $\bar{H}(\bar{\bf 5})$ in their origins.

The important difference between the two different choice of vector 
bundles is that $H({\bf 5})$ and $\bar{H}(\bar{\bf 5})$ are in 
a pair of hermitian-conjugate representation when 
$V_5={\bf 1} \oplus {\bf 4}$, but they are not when 
$V_5={\bf 3}_2 \oplus {\bf 2}_2$. Indeed, in $\mathfrak{e}_7/\mathfrak{su}(5)$,
they cannot be identified with a vector-like pair 
when all the Yukawa coupling are required to be generated.
But under the embedding $\mathfrak{e}_7/\mathfrak{su}(5)$ into 
$\mathfrak{e}_8/\mathfrak{su}(5)$ through (\ref{eq:7to8}), 
the SU(2)$_1$-singlet $\bar{H}(\bar{\bf 5})$ is regarded as  
$(\bar{\bf 5},\wedge^2 {\bf 2}_1,{\bf 1})$ representation of 
SU(5)$_{\rm GUT} \times$SU(2)$_1\times$SU(2)$_2$; 
since $H({\bf 5})$ is in $\overline{(\bar{\bf 5},{\bf 1},
\wedge^2 {\bf 2}_2)}$, both 
$H({\bf 5})$ and $\bar{H}(\bar{\bf 5})$ are from the same 
(up to conjugation) multiplet $(\bar{\bf 5},\wedge^2 {\bf 4})$ 
when the 2 rank-2 bundles ${\bf 2}_1$ and ${\bf 2}_2$ mix up 
into an irreducible SU(4) bundle ${\bf 4}$.

If the doublet--triplet splitting problem is to be solved 
by a Wilson line \cite{Witten23}, they have to be in a  
Hermitian-conjugate pair. A Wilson line does not make a difference 
in the Euler characteristics for triplets and doublets.
Thus, the Wilson line as the solution to the problem motivates 
to promote $\mathfrak{e}_7/\mathfrak{su}(5)_{\rm GUT}$ to 
$\mathfrak{e}_8/\mathfrak{su}(5)_{\rm GUT}$ for the origin 
of low energy particles.
Incidentally, $E_8$ is a maximal symmetry or singularity available 
in string theories. However, the missing-partner type mass matrix 
of the Higgs sector \cite{missing},\footnote{There is a variety 
in the D=4 field-theory realization of the missing partner mechanism, 
and the matter contents do not have to be the original one. There 
will be more variety in implementing the missing partner mechanism 
in string theories. Reference \cite{WY} is an attempt to embed a 
field-theory model \cite{IY} into Type IIB string theory, and is one 
of such variety.} another potential solution to the 
doublet--triplet splitting problem, does not necessarily require 
that they originate from a pair of conjugate representation.

In the next section, the idea of having reducible vector bundles 
(in Heterotic terminology) as a solution to the dimension-4 proton 
decay problem is implemented in Heterotic theory description. 
In later sections, M-theory and F-theory descriptions are given.

\section{Heterotic Vacua}
\label{sec:Het}


The Yukawa couplings have been discussed in the several previous 
approaches to Heterotic string compactification 
\cite{a2,dona1,Penn4,dona2}. The novelty in our story is that 
we consider simultaneously the Yukawa couplings and proton decay 
terms. In our consideration, we show that the presence of the 
former and absence of the latter appears naturally, thus providing 
a Heterotic picture of the discussion in the previous section. 
In the subsequent sections, we will provide M-theory and F-theory 
pictures for the same field theory constrains.

In compactification of Heterotic theory, in order to have D = 4 
effective field theories with unbroken SU(5)$_{\rm GUT}$ symmetry, 
one has only to turn on an SU(5) vector bundle in one of $E_8$ gauge 
group of $E_8\times E_8$.
The absence of dimension-4 proton decay 
operators in our vacuum suggests, however, that there will be more 
structure than that. Here we work on a possibility that the 
$\bar{\bf 5}=(\bar{D},L)$ and $\bar{H}(\bar{\bf 5})$ have different 
origins in the $\mathfrak{e}_8$-{\bf adj.} representation, so that 
the down-type quark Yukawa couplings exist whereas the dimension-4 
proton decay operators do not. This idea is implemented 
when the rank-5 SU(5) vector bundle on a Calabi--Yau 3-fold $Z$ 
becomes reducible and is split up into two irreducible vector 
bundles\footnote{We start our discussion with a 
direct sum splitting of the tank-5 vector bundle. Later we will consider the case when some neutrino fields acquire
expectation values and the rank-5 bundle becomes an extension of 
$U_4$ by L.} 
\begin{equation}
 V_5= L \oplus U_4, \quad {\rm with} \quad 
L \otimes {\rm det}U_4  \simeq {\cal O}_Z,
\label{eq:41bdle}
\end{equation}
where $U_4$ and $L$ are rank-4 and rank-1 vector bundles, respectively,
with the structure group U(4) and U(1). An alternative is 
\begin{equation}
 V_5 = U_3 \oplus U_2, \quad {\rm with} \quad 
{\rm det}U_3 \otimes {\rm det}U_2 \simeq {\cal O}_Z,
\label{eq:32bdle}
\end{equation}
where $U_3$ and $U_2$ are rank-3 and rank-2 vector bundles with 
structure group U(3) and U(2), respectively.
This can be regarded as the particular choice of vector-bundle 
moduli parameter of SU(5) vector bundles in \cite{Penn5}. 
The 2+2+1 reducible vector bundle (\ref{eq:221bdle}) corresponds 
to a particular choice of moduli parameter of (\ref{eq:41bdle}) 
and also of (\ref{eq:32bdle}).
The choice of the bundle (\ref{eq:41bdle}) reduces to \cite{Penn4} 
when the line bundle $L$ is a flat bundle. 
The reducible bundle (\ref{eq:41bdle}) was originally considered in 
\cite{Blumenhagen}.

Although the different bundles have different physical consequences, 
the string-theoretical study for each bundles is qualitatively the same. 
Thus, we discuss the case of (\ref{eq:41bdle}) extensively and 
only mention the differences in the cases of (\ref{eq:32bdle}) and 
(\ref{eq:221bdle}) bundles later.

\subsection{Compactification with 4+1 Vector Bundles}
\label{ssec:Het41}

\subsubsection{Spectrum and Yukawa Couplings}

The low-energy multiplets in SU(5)$_{\rm GUT}$-{\bf 10} representation 
arise from zero-modes (cohomology) of the vector bundle $V_5$, 
which is now split up into 2 irreducible pieces $U_4$ and $L$. 
Those in SU(5)$_{\rm GUT}$-$\bar{\bf 5}$ (or {\bf 5}) representation 
arise from $\wedge^2 V_5$, which now consists of 2 irreducible 
components $\wedge^2 U_4$ and $U_4 \otimes L$ (or bundles in the 
conjugate representations).
Finally, the SU(5) vector bundle moduli from {\bf adj.}$(V_5)$---which 
are SU(5)$_{\rm GUT}$-singlets---are split up into 2 pieces, namely 
the U(4) vector bundle moduli and SU(5)$_{\rm GUT}$-singlets 
from the bundle $U_4 \otimes L^{-1}$.
When the low-energy multiplets are from the bundles specified in Table
\ref{tab:ID-8A}, 
\begin{table}[tb]
\begin{center}
 \begin{tabular}{|c||c|c|c|c|c|c|c|}
  bundles & $U_4$ & $U_4 \otimes L$ & 
  $\overline{\wedge^2 U_4}$ & 
  $\wedge^2 U_4$ & $U_4 \otimes L^{-1}$ & 
  {\bf 1} & $L$ \\
\hline
  low-energy particles & $(Q,\bar{U},\bar{D})$ & $(\bar{D},L)$ 
     & $H$ & $\bar{H}$ & $\bar{N}$ & (({\bf adj.})) & ((${\bf 10}'$))
 \\
 \end{tabular}
\end{center}
\caption{Particle identification in $U_4 \oplus L$ bundle 
compactification of Heterotic $E_8 \times E_8$ theory. 
The last two columns mean that particles from the 2 bundles are not 
necessary in our low-energy effective theory. \label{tab:ID-8A}}
\end{table}
all the Yukawa coupling are generated from the $E_8$ Yang--Mills 
interaction, including the Dirac Yukawa coupling of neutrinos:
\begin{eqnarray}
 W & \ni & ({\bf 10},U_4) \otimes 
       ({\bf 10},U_4) \otimes 
       \overline{(\bar{\bf 5},\wedge^2 U_4)} 
       \longrightarrow {\bf 10}.{\bf 10}.H({\bf 5})
     \nonumber \\
  & &
    +  (\bar{\bf 5},U_4 \otimes L) \otimes 
       ({\bf 10},U_4) \otimes 
       (\bar{\bf 5},\wedge^2 U_4)  
 \longrightarrow \bar{\bf 5}\cdot {\bf 10} \cdot \bar{H}(\bar{\bf 5})
     \nonumber \\
  & & 
   +   (\bar{\bf 5},U_4 \otimes L) \otimes 
       ({\bf 1},U_4 \otimes L^{-1}) \otimes 
       \overline{(\bar{\bf 5},\wedge^2 U_4)} 
 \longrightarrow \bar{\bf 5} \cdot \bar{N} \cdot H({\bf 5}).        
\label{eq:Het-Yukawa}
\end{eqnarray}
On the other hand, the dimension 4 proton decay operator 
(\ref{eq:dim4}) is not generated, because 
\begin{equation}
 (U_4 \otimes L) \otimes U_4 \otimes (U_4 \otimes L)
\label{eq:Het-dim4}
\end{equation}
does not contain a trivial representation in its irreducible 
decomposition. The operator (\ref{eq:dim4}) is certainly invariant 
under SU(5)$_{\rm GUT}$, but not under full $E_8$. The symmetry 
U(1)$_\chi$ (or equivalently U(1)$_{\rm B-L}$) and the bundle 
structure group SU(4) guarantees the absence of the dangerous 
operators.

Note also that all of ${\bf 10}$, $\bar{\bf 5}$ and $\bar{N}$ 
would come from the same bundle $U_4$ if the line bundle $L$ 
were trivial (or flat). This was the situation discussed in SO(10) 
models of \cite{Penn4}. The line bundle $L \simeq \det U_4^{-1}$ is 
in the direction of U(1)$_\chi$. 

It is straightforward to obtain the formulae for the net chirality 
of low-energy multiplets. Let us take a Calabi--Yau 3-fold $Z$ 
and a vector bundle $V$ on it. The $\mathfrak{e}_8$-{\bf adj.} 
representation is split up into irreducible pieces under the 
the unbroken SU(5)$_{\rm GUT}$ and the structure group of the 
vector bundle. The number of chiral multiplets in the 
$(R_5,\rho(V))$ representation is given by $H^1(Z,\rho(V))$, and 
that of multiplets in the conjugate representation by 
$H^1(Z,\bar{\rho}(V))$. The net chirality at low energy is given 
by 
\begin{eqnarray}
 \chi(R_5) & \equiv & 
  \# (R_5,\rho(V)) - \# (\overline{R_5},\bar{\rho}(V)), \nonumber \\
 & = & h^1(Z;\rho(V)) - h^1(Z;\overline{\rho(V)}), \nonumber \\
 & = & - \chi(Z,\rho(V)) 
  = - \int_Z {\rm ch}(\rho(V)) {\rm Td}(TZ), 
\end{eqnarray}
where the last equality is the Hirzebruch--Riemann--Roch theorem 
\cite{Hirzebruch}.
In an application to the bundle of interest,\footnote{Integration 
over $Z$ is omitted, just for visual clarity.} 
\begin{eqnarray}
 \chi(Z,U_4) & = & 
   - \frac{1}{12}c_2(TZ)c_1(L)
   - \frac{1}{6} c_1(L)^3
   + \frac{1}{2} c_2(U_4)c_1(L)
   + \frac{1}{2} c_3(U_4),
\label{eq:chi-10A}\\
 \chi(Z,L) & = & 
   + \frac{1}{12}c_2(TZ)c_1(L)
   + \frac{1}{6} c_1(L)^3, 
\label{eq:chi-10B}\\
 \chi(Z,U_4 \otimes L) & = & 
   + \; \, \frac{1}{4}c_2(TZ)c_1(L)
   + \frac{1}{2}c_1(L)^3
   - \frac{1}{2}c_2(U_4)c_1(L)
   + \frac{1}{2}c_3(U_4),
\label{eq:chi-5DL}\\
 \chi(Z,\wedge^2 U_4) & = & 
   - \; \, \frac{1}{4}c_2(TZ)c_1(L)
   - \frac{1}{2}c_1(L)^3
   + \quad  c_2(U_4)c_1(L)   
\label{eq:chi-5HH}\\
 \chi(Z,U_4 \otimes L^{-1}) & = & 
   - \frac{5}{12}c_2(TZ)c_1(L)
   - \frac{11}{6}c_1(L)^3
   + \frac{3}{2}c_2(U_4)c_1(L)
   + \frac{1}{2}c_3(U_4).
\label{eq:chi-N} 
\end{eqnarray}
The Calabi--Yau condition $c_1(TZ)=0$ was also used in the above 
calculation. Note that the net chirality for the multiplets 
in SU(5)$_{\rm GUT}$-{\bf 10} representation is the same as that 
for those in SU(5)$_{\rm GUT}$-$\bar{\bf 5}$ representation:
\begin{equation}
 (\ref{eq:chi-10A}) + (\ref{eq:chi-10B}) = 
 (\ref{eq:chi-5DL}) + (\ref{eq:chi-5HH}) = 
\frac{1}{2} \int_Z c_3(U_4) + \frac{1}{2} \int_Z c_2(U_4)c_1(L) = 
\frac{1}{2} \int_Z c_3(V_5).
\end{equation}
This serves as a good check of the above calculation 
\cite{Blumenhagen}. 

Since we do not want a non-zero net chirality from the 
Higgs sector, our vacuum should be based on 
\begin{equation}
 \chi(Z,\wedge^2 U_4) = 0.
\label{eq:Higgsless}
\end{equation}
Since SU(5)$_{\rm GUT}$-${\bf 10}'$ multiplets from the bundle $L$ 
do not have appropriate Yukawa couplings, they cannot have mass 
terms. Since we have not seen exactly massless 
SU(5)$_{\rm GUT}$-charged particles, there should not be non-zero 
net chirality for this multiplet which tells us that:
\begin{equation}
 \chi({\bf 10}') = - \chi(Z,L) = 0.
\label{eq:10less}
\end{equation}
Combining these phenomenological inputs (\ref{eq:Higgsless})
and (\ref{eq:10less}) with general chirality formulae 
(\ref{eq:chi-10B}) and (\ref{eq:chi-5HH}), we see that 
our vacuum has to satisfy 
\begin{eqnarray}
\int_Z c_2(TZ) c_1(L) & = & - 2 \int_Z c_1(L)^3, \label{eq:ZL}\\
 \int_Z c_2(U_4) c_1(L) & = & 0. 
 \label{eq:4L}
\end{eqnarray}
Substituting these two relations back into (\ref{eq:chi-10A}), 
(\ref{eq:chi-5DL}) and (\ref{eq:chi-N}), the number of generations 
is given by 
\begin{eqnarray}
N_{gen}= \chi({\bf 10}) = \chi(\bar{\bf 5}) 
 & = & 
      - \int_Z \frac{1}{2}c_3(U_4), 
\label{eq:Ngen-QL} \\
 \chi(\bar{N}) & = & 
      - \int_Z \frac{1}{2}c_3(U_4) + \int_Z c_1(L)^3.
\label{eq:Ngen-N}
\end{eqnarray}
The ``chirality'' of right-handed neutrinos (\ref{eq:Ngen-N}) is 
no longer equal to that of other quarks and leptons, but 
there is nothing wrong phenomenologically. Since the SO(10) or
U(1)$_\chi$ gauge symmetry is already broken, there 
is no reason why exactly 3 copies of right-handed neutrinos exist.
Note also that (\ref{eq:Ngen-N}) represents the minimal number 
of right-handed neutrinos. 
 
The SU(5)$_{\rm GUT}$ symmetry may be broken by introducing 
a Wilson line. If there is a discrete symmetry group $\Gamma$ 
mapping $Z$ to itself without a fixed point, $Z/\Gamma$ can be 
used for compactification instead of $Z$. The Wilson line can be
introduced on the non-trivial $\pi_1(Z/\Gamma)$. In this case, 
the number of generation of quarks and leptons are given by 
\begin{equation}
 N_{gen}' = - \frac{1}{\# \Gamma}
\frac{1}{2} \int_Z c_3(U_4).
\label{eq:Ngen-quot}
\end{equation}
This is supposed to be 3 for our vacuum. 
$\# \Gamma$ should divide $- \int_Z c_1(L)^3 = \int_Z c_1(U_4)^3$.

\subsubsection{Anomalous U(1)$_\chi$ Symmetry and Instability}
\label{sssec:Het-anomalous}

As we saw in (\ref{eq:Ngen-N}), the ``chirality'' of right-handed 
neutrinos is generically different from that of quarks and leptons 
when $c_1(L) \neq 0$, indicating that the U(1)$_\chi$ symmetry 
is anomalous. All the necessary Green--Schwarz couplings are 
worked out in \cite{Blumenhagen} where a detailed account of 
the conventions can be found.
One can see there that various triangle anomalies involving this 
symmetry are cancelled by the shift of the imaginary part of 
dilaton and K\"{a}hler moduli chiral multiplets, $S$ and $T^k$, 
respectively. All of them come from the $B$ field.
Certain linear combination of the imaginary parts are eaten as the 
longitudinal mode of the anomalous U(1)$_\chi$ vector boson. The 
scalar--vector mixing interactions arise from the K\"{a}hler potential
\begin{equation}
 K = - M_G^2 \left(\ln 
    \left(\frac{1}{3!} \int \tilde{J} \wedge \tilde{J} \wedge \tilde{J}
    \right) + \ln \left( S + S^\dagger - Q^0 V \right) \right),
\end{equation}
where $M_G \simeq 2.4 \times 10^{18}$ GeV, $S$ is the dilaton chiral 
multiplet whose expectation value is $1/g_{\rm YM}^2$ of 
SU(5)$_{\rm GUT}$ in the absence of 1-loop corrections, and 
\begin{equation}
 \frac{1}{2\pi} \left(- J + iB \right) = l_s^2 T^k \omega_k, \qquad  
 \tilde{J} = - \pi l_s^2 (T^k + T^{k\dagger} - Q^k V) \omega_k,
\end{equation}
where $J$ is the K\"{a}hler form of the compactification manifold $Z$, 
$\omega_k$ ($k=1,\cdots,h^{1,1}(Z)$) are integral basis of the second 
cohomology of $Z$, $l_s = 2\pi \sqrt{\alpha'}$ and $T^k$ are chiral 
multiplets containing the $\omega_k$-component of the K\"{a}hler 
modulus and the $B$ field. Under the U(1) symmetry,the charged chiral
multiplets $\Psi_i$ such as $\bar{N}$, {\bf 10} and $\bar{\bf 5}$ 
and the moduli chiral multiplets transform as 
\begin{eqnarray}
  V & \rightarrow & V - i \Lambda + i \Lambda^\dagger, \\
 \Psi_i & \rightarrow & e^{iq_i \Lambda} \Psi_i, \\
 S & \rightarrow & S - i Q^0 \Lambda, \label{eq:S-shift}\\
 T^k & \rightarrow & T^k - i Q^k \Lambda, \label{eq:T-shift}
\end{eqnarray}
with\footnote{The charges adopted below differ from those in
\cite{Blumenhagen} by a factor of $(2\pi)$. Coefficients in this 
article are corrected in version 2, following \cite{Munich3}.} 
\begin{eqnarray}
 Q^0 & = & - \frac{2 \tr{}_f({\bf q}{\bf q}_L)}{32\pi^2}
   \int_Z c_1(L)\left(c_2(V_5)-\frac{1}{2}c_2(TZ)\right),  \\
 Q^k & = & - \frac{2 \tr{}_f({\bf q}{\bf q}_L)}{8\pi^2} q_L^k, \\
  c_1(L) & \equiv & q_L^k \omega_k. 
\end{eqnarray}
The trace is taken in the fundamental representation of the bundle 
group SU(5). In the case of the U(1)$_\chi$ anomalous symmetry, 
${\bf q}_L=\diag(1,-1/4,-1/4,-1/4,-1/4)$, and we can choose the
normalisation of the charges of U(1)$_\chi$ symmetry as 
${\bf q}={\bf q}_\chi \equiv \diag(4,-1,-1,-1,-1)$ for 
convenience.\footnote{The D-term scalar potential after completing 
square in (\ref{eq:D-term}) does not depend on the choice 
of the overall normalisation of ${\bf q}_\chi$ and $q_{\chi,i}$. 
This is just a matter of how charges and coupling constant are 
separated in U(1) gauge theories.} 
With this choice of ${\bf q}={\bf q}_\chi$, the {\bf 10}-representations 
from $U_4$ bundle have $-1$ unit charge, and 
$\bar{N}$ from $U_4 \otimes L^{-1}$ bundle have $-5$ units. 
If $h^1(Z;\overline{U}_4 \otimes L) \neq 0$, so that there is an 
anti-generation right-handed neutrino $\overline{\bar{N}}$ in the 
spectrum, its U(1)$_\chi$ charge is 5 units.

The Fayet--Iliopoulos parameter of such anomalous U(1) symmetries can 
be obtained by\footnote{$\xi= - (Q^k \partial K/\partial T^k + 
Q^0 \partial K / \partial S) = -(Q^k \partial K/\partial T^{k\dagger}
+Q^0 \partial K/\partial S^\dagger)$ is also the same. 
The derivative of the Killing potential 
\begin{equation} 
d \xi = 
- Q^k ( K_{T^k T^{l \dagger}} dT^{l\dagger} 
      + K_{T^l T^{k \dagger}} dT^l) 
- Q^0 (K_{SS^\dagger} dS^\dagger + K_{SS^\dagger} dS)
\end{equation} 
is consistent with the Killing vector 
$- iQ^k(\partial_{T^k}-\partial_{T^{k \dagger}}) 
 - iQ^0(\partial_S - \partial_{S^\dagger})$ corresponding to 
(\ref{eq:T-shift}) and (\ref{eq:S-shift}).}
${\cal L} \ni D \xi = V|_{\theta^2 \bar{\theta}^2} 
(\partial K/\partial V)|_{V=0}$, and in the case of U(1)$_\chi$, 
\begin{equation}
 \xi_\chi =  \frac{10 M_G^2}{32 \pi^2} \left[
   \frac{ 2\pi l_s^2}{{\rm vol}(Z)} \int c_1(L) \wedge J \wedge J 
   - \frac{g_{\rm YM}^2 e^{2\widetilde{\phi}_4}}{2} 
        \int c_1(L)\left(c_2(V_5) - \frac{1}{2}c_2(TZ)\right)
                                              \right].
\label{eq:FI41}
\end{equation}  
The first term is the tree level 
Fayet--Iliopoulos parameter, which comes from the derivative with 
respect to $T^k$, and depends on K\"{a}hler moduli $T^k$'s through the 
K\"{a}hler form $J$. The second term is the 1-loop correction, and 
depends on the fluctuation of the dilaton  $e^{\widetilde{\phi}_4}$.
This Fayet--Iliopoulos parameter is used as in 
\begin{equation}
 {\cal L} = \frac{1}{2g_{\rm YM}^2} 2 \tr{}_f({\bf q}_\chi^2) D_\chi^2
     + D_\chi \xi_\chi + D_\chi q_{\chi,i} \psi_i^\dagger \psi_i 
\rightarrow 
V = \frac{1}{2}\frac{g_{\rm YM}^2}{2\tr {}_f({\bf q}_\chi^2)}
 \left( \xi_\chi + q_{\chi,i}\psi_i^\dagger \psi_i \right)^2.
\label{eq:D-term}
\end{equation}

The equations of motions of Yang--Mills fields of the vector bundles 
$L$ and $U_4$ are given by \cite{Witten-rk-red}
\begin{equation}
2  g^{\bar{\beta}\alpha}\frac{F^{(L,U_4)}_{\alpha \bar{\beta}}}{2\pi} 
= \lambda^{(L,U_4)} {\bf id.}_{1\times 1, \; 4\times 4},
\label{eq:YM-eq}
\end{equation}
with some constants $\lambda^{(L)}$ and $\lambda^{(U_4)}$, which 
are equivalent to the Hermitian--Einstein condition for gauge fields
\begin{eqnarray}
 \frac{F^{(L)}}{2\pi} \wedge J \wedge J & = & \quad \;\;  \lambda^{(L)} \; 
 {\bf id.}_{1\times 1} \left(\frac{1}{3!} J \wedge J \wedge J \right),
\label{eq:EH-L} \\
 \frac{F^{(U_4)}}{2\pi} \wedge J \wedge J & = & - \frac{1}{4} 
   \lambda^{(L)} \;
{\bf id.}_{4\times 4} \left(\frac{1}{3!} J \wedge J \wedge J \right). 
\label{eq:EH-U4}  
\end{eqnarray}
The solutions to these equations minimize\footnote{Just like the
equation of motion (\ref{eq:YM-eq}) is implemented as the minimization 
of the D-term potential in D = 4 effective field theories, the 
condition for a bundle to be holomorphic (the absence of $(0,2)$ and 
$(2,0)$ components of the Yang--Mills field strength) is implemented 
as the F-term condition. Indeed, when the background Yang--Mills field
configuration has a non-vanishing $(0,2)$ component, $W \ni \int \Omega
\wedge (A \vev{dA})$ gives rise to a term linear in $A$ in the 
superpotential. The F-term potential requires to minimize the 
coefficient of the linear term, the (0,2) component of $\vev{dA}$. 
When the minimized (0,2) component of a U(1) bundle does not vanish, 
i.e., when the ``Fayet--Iliopoulos (or O'Raifeartaigh) F-term'' 
does not vanish, a phase transition of the spontaneous U(1) symmetry 
breaking is triggered through the superpotential 
$W \ni \int \Omega \wedge (A\vev{dA}-AAA)$. The story 
here is completely in parallel with that of the D-term.} 
the D-term potential from the tree-level Fayet--Iliopoulos parameter, 
leaving positive energy $(g_{\rm YM}^2/80)|\xi_\chi|^2=
(5/256)g_{\rm YM}^2(M_G^2 l_s^2\lambda)^2$. If one finds a holomorphic 
stable vector bundle $U_4$, Donaldson--Uhlenbeck--Yau's theorem 
guarantees that there exists a Yang--Mills field configuration 
satisfying (\ref{eq:EH-U4}).  There is no such  subtlety for the 
holomorphic line bundle $L$ because it is stable by definition.
Thus, a solution to the equation of motions exists for the reducible 
bundle (\ref{eq:41bdle}), as long as $U_4$ is chosen to be a 
holomorphic stable bundle. 

This solution, however, is not stable both in mathematical and 
physical sense. 
If $c_1(L)\wedge J \wedge J= -c_1(U_4) \wedge J \wedge J \propto 
\lambda$ is not zero, the rank-5 reducible holomorphic bundle $V_5$ 
(\ref{eq:41bdle}) is unstable in the mathematical definition.
In physics language, this ``instability'' has a clear meaning.
Non-zero $\lambda$ means a non-vanishing tree-level Fayet--Iliopoulos 
parameter (\ref{eq:FI41}), which requires 
(at least at the classical level, ignoring the 1-loop term in 
(\ref{eq:FI41})) that either one of 
U(1)$_\chi$-charged objects develop a non-zero expectation value to 
absorb the non-zero D-term, further minimizing the entire D-term,  
$\xi_\chi$ and $q_{\chi,i}\psi^\dagger_i \psi_i$ combined. 
The field responsible for this can be either $\bar{N}$ or 
$\overline{\bar{N}}$, the off-diagonal blocks of the bundle group 
SU(5). The Fayet--Iliopoulos parameter triggers the U(1)$_\chi$ 
symmetry breaking and the U(1)$_\chi$-symmetric vacuum is not stable. 

Another way to see the instability is to note that 
$F_{\alpha \bar{\beta}} g^{\bar{\beta}\alpha}$ in the U(1)$_\chi$
direction gives rise to the difference between the equation of motions 
of vector fields and gauge fermions. Fermions and vector fields have 
the same property under the SU(3) $\subset$ SO(6) of the Lorentz group
of the internal space, but not under the 
U(1) $\subset$ U(3) $\subset$ SO(6). When there is a zero mode 
of gauge fermion, its would-be supersymmetric partner has either 
positive or negative mass-squared in the equations of motions, 
depending on the U(1)$_\chi$ charge $q_\chi$. This explains why 
either one of $\bar{N}$ or $\overline{\bar{N}}$ have negative 
mass-squared when $\xi_\chi$ does not vanish. The cross terms 
of the D-term potential (\ref{eq:D-term}) lead to 
$m^2_i \propto \xi_\chi q_{\chi,i}$.

When $\int c_1(L)\wedge J \wedge J$ is negative, $\xi_\chi < 0$, and 
one of the fields with positive U(1)$_\chi$ charge, 
$\overline{\bar{N}}$, $\bar{\bf 5}$ and $H({\bf 5})$, develops 
an expectation value. Supersymmetry-breaking masses of those fields 
determine which field gets the expectation value.\footnote{In the simplest situation where all these 
chiral multiplets have equal gravity-mediated supersymmetry breaking 
mass-squared, potential minimisation shows that only 
$\overline{\bar{N}}$ develops a non-zero expectation value, 
none others. This is because $\overline{\bar{N}}$ has the largest 
positive the U(1)$_\chi$ charge among them.} 
Since the only possibility consistent with the real world is for 
$\overline{\bar{N}}$ to develop an expectation value, we assume 
this in the following.
This mathematically means that the rank-5 vector bundle $V_5$ is 
no longer a reducible bundle (\ref{eq:41bdle}), but is given 
by an extension \cite{FMW} of $U_4$ by $L$:
\begin{equation}
 0 \rightarrow L \rightarrow V_5 \rightarrow U_4 \rightarrow 0.
\label{eq:14ext}
\end{equation}
If $\int c_1(L) \wedge J \wedge J > 0$, then the rank-5 bundle is an 
extension of $L$ by $U_4$:
\begin{equation}
 0 \rightarrow U_4 \rightarrow V_5 \rightarrow L \rightarrow 0.
\label{eq:41ext}
\end{equation}
We prefer for phenomenological reasons that right-handed neutrinos 
$\bar{N}$ do not acquire large expectation values. Since their 
expectation values provide large mass terms between $H_u$ and lepton 
doublets $L$, either $H_u$ or $L$ would be left out of the low-energy 
spectrum. Dimension-4 proton decay operators are also generated 
in this case, as we shall see later. Thus,  
$\int c_1(L)\wedge J \wedge J$ should not be positive.\footnote{Here, 
we implicitly assume that (\ref{eq:Ngen-QL}) and (\ref{eq:Ngen-quot}) 
are negative. If they are positive, the definition of the particles 
and anti-generation particles are exchanged, and now 
$\int_Z c_1(L) \wedge J \wedge J$ should not be negative.}
If it is negative, then 
$h^1(Z,\overline{U}_4 \otimes L)$ should be non-zero as well, so that 
$\overline{\bar{N}}$ exists in the spectrum and absorbs the non-zero 
Fayet--Iliopoulos parameter. 
We shall see later that, in some cases, we can then relax the 
condition that the vector bundle be reducible without allowing 
a too rapid proton decay. 
 
Before going directly to the discussion of proton decay operators 
controlled under the spontaneously broken anomalous U(1)$_\chi$ 
symmetry, let us discuss the 1-loop 
term of the Fayet--Iliopoulos parameter (\ref{eq:FI41}). 
As one can see from \cite{Blumenhagen}, the 1-loop term in the 
Heterotic $E_8 \times E_8$ theory\footnote{In the compactification 
of Heterotic SO(32) theory with the spin connection embedded in 
the SU(3) subgroup of SO(6) $\subset$ SO(32), the bundle group 
leaves SO(26) $\times$ U(1)$_3$ gauge symmetry unbroken. 
The U(1)$_3$ factor is, however, anomalous; $Q_3^k=0$, but  
\begin{equation}
 Q_3^0 = \frac{c_3(V_3)}{8 \pi^2}. 
\end{equation} The corresponding Fayet--Iliopoulos parameter is given 
by $\xi = - M_G^2 g_{\rm YM}^2 \tr_{\rm LE.}(q_3)/192 \pi^2$, 
after using a relation $\tr_{\rm LE.} (q_3) = -(26-2)\chi(Z;V_3) = 
- 12c_3(V_3)$; here, tr$_{\rm LE.}(q_3)$ is the sum of U(1)$_3$ charge 
in the low-energy spectrum. This Fayet--Iliopoulos parameter is proportional 
to the gravitational anomaly $\tr_{\rm LE.}(q_3)$. In general, 
however, the 1-loop Fayet--Iliopoulos parameter is not expected to be 
proportional to gravitational anomaly; it is proportional only when 
the quadratic divergence of loops of all the charged particles are 
made finite exactly in the same way. 
In the SU(3)-bundle compactification of the Heterotic SO(32) theory, 
it is the case: all the charged particles originate either from 
rank-3 bundle $V_3$ or from its Hermitian conjugate. In the case of 
our interest, however, various particles originate from totally 
different vector bundles, and we cannot expect that all the loops of 
low-energy particles are rendered finite in the same way.
Perturbative Heterotic compactification can give rise to more than 
one anomalous U(1) gauge symmetries \cite{Blumenhagen}; it is also 
the case in the model in our section \ref{ssec:Het221}. 
Their Fayet--Iliopoulos parameters can have both tree and 1-loop level 
contributions, either (or both) of them can vanish for certain choice 
of moduli parameters or matter contents, just like in Type I or 
Type II string theories.} is proportional to the U(1)-SU(5)$_{\rm GUT}^2$ 
mixed anomaly,  
\begin{eqnarray}
 \sum_i q_{\chi,i} 2T_{R,i} & = & 
  3 \times (\ref{eq:chi-10A}) -12 \times (\ref{eq:chi-10B})
 - 3 \times (\ref{eq:chi-5DL}) + 2 \times (\ref{eq:chi-5HH}), 
 \nonumber \\
 & = & \tr {}_f ({\bf q}_\chi {\bf q}_L) 
   \int_Z c_1(L) \left(c_2(U_4) - c_1(L)^2 - \frac{1}{2}c_2(TZ)\right).
\label{eq:155anomaly}
\end{eqnarray}
Since we know that this anomaly vanishes for the U(1)$_\chi$ symmetry 
in the spectrum of the minimal supersymmetric standard model
(possibly with additional SU(5)$_{\rm GUT}$-singlets and vector-like
pairs), the net 1-loop contribution is absent; $Q_\chi^0=0$. 
One can also see this explicitly by using (\ref{eq:ZL}) and 
(\ref{eq:4L}) with $c_2(V_5) = c_2(U_4) - c_1(L)^2$. Therefore, 
no matter how moduli are stabilised (or even when moduli are not 
stabilised otherwise), the cancellation between the tree and 1-loop 
level contributions \cite{Blumenhagen} cannot happen in realistic 
vacua satisfying (\ref{eq:ZL}) and (\ref{eq:4L}).

Although we cannot expect cancellation between the two terms, the 
tree-level term itself may also vanish. The K\"{a}hler moduli can 
either be stabilised so that $\lambda$ in (\ref{eq:EH-L}) vanishes 
or some of it remains unstable and the D-term scalar 
potential of the U(1)$_\chi$ symmetry may take the moduli into some 
region of the K\"{a}hler cone so that $\lambda$ in (\ref{eq:EH-L}) 
vanishes. Either way, the Fayet--Iliopoulos parameter can vanish. 
In that case, some linear combination of the imaginary parts of 
the chiral multiplets $T^k$'s is absorbed as the longitudinal mode 
\cite{Witten-Stkbg,Witten-rk-red}. Since the kinetic term of the 
imaginary parts of $T^k$'s are 
\begin{eqnarray}
 {\cal L} & = & - K_{kl} \left(\partial (\Im T)^k - Q^k A \right)
\left(\partial (\Im T)^l - Q^l A \right), \\
 K_{kl} & = &  M_G^2 (2\pi l_s^2)^2 \frac{3}{2}
  \left[ - \frac{\int \omega_k \wedge \omega_l \wedge J}{\int J^3}
         + \frac{3}{2} \frac{\int \omega_k J^2 \int \omega_l J^2}
                            {(\int J^3)^2} \right], 
\end{eqnarray}
and the mass-squared of the gauge field is roughly of the order 
\begin{equation}
 m_\chi^2 \approx g_{\rm YM}^2 M_G^2 \left( \frac{l_s}{R}\right)^4,
\label{eq:Amass-S}
\end{equation}
where $R$ is a ``typical'' radius of compactification manifold, 
assuming that it is isotropic. The real-part scalar corresponding to 
the absorbed combination acquires a mass-squared from the D-term 
scalar potential. 

If all the K\"{a}hler moduli have already been stabilised by other 
means, such as non-perturbative potentials, then the tree-level 
Fayet--Iliopoulos parameter may not vanish. When $\xi_\chi$ is 
negative, a chiral multiplet $\overline{\bar{N}}$ of positive
U(1)$_\chi$ charge, $+5$, develops an expectation value, absorbing 
negative $\xi_\chi$. In this case, the gauge boson absorbs some linear 
combination of $T^k$'s and $\overline{\bar{N}}$ and removes just one 
flat direction. The non-zero expectation value of U(1)$_\chi$-charged 
particles, $\vev{\overline{\bar{N}}^\dagger \overline{\bar{N}}} 
\approx M_G^2 (l_s/R)^2 /(8\pi)$, gives rise to another mass term 
of the U(1)$_\chi$ gauge field:
\begin{equation}
 m_{\chi}^2 \approx \frac{5 g_{\rm YM}^2}{32\pi} M_G^2 
    \left(\frac{l_s}{R}\right)^2 = \frac{5\pi}{2} \frac{1}{R^2}.
\label{eq:Amass-H}
\end{equation}
For compactification manifold with $R$ sufficiently larger than $l_s$,
the latter Higgs-mechanism contribution dominates over 
(\ref{eq:Amass-S}). The mass scale of the U(1)$_\chi$ gauge boson 
is an important parameter in leptogenesis. From (\ref{eq:Amass-H}), 
we see that the mass is typically around the Kaluza--Klein scale.
Independent of whether $\xi=0$ or $\xi<0$, the anomalous U(1)$_\chi$ 
gauge boson tends to have a very large mass, and easily escape the 
mass bound $m_\chi \gtrsim 600$ GeV \cite{PDG} from low-energy 
experiments.

\subsubsection{Effective Superpotential}
\label{sssec:Het-super}

Let us now turn our attention to the effective superpotential, with 
or without the expectation value of the scalar field in 
$\overline{\bar{N}}$. If both $\overline{\bar{N}}$ or $\bar{N}$ 
have vanishing expectation value, the Dirac Yukawa couplings in 
(\ref{eq:Het-Yukawa}) are the only sources of the masses of neutrinos. 
The U(1)$_\chi$ symmetry remains unbroken as a global symmetry 
at the perturbative level, and forbids Majorana masses of right-handed 
and left-handed neutrinos. The dimension-4 proton decay operators 
are also forbidden. Although the U(1)$_\chi$ symmetry
is broken at non-perturbative level, the SU(4) structure group of 
the bundle $U_4$ still controls the effective operators. Any effective 
operators should be written by low-energy multiplets and spurions 
in the $\mathfrak{su}(4)$-{\bf adj.} representation. The operators 
have to be symmetric under SU(4) and SU(5)$_{\rm GUT}$ and SU(4)  
symmetry is broken only by the Yang--Mills field configuration, 
whose effects are taken into account by the spurions. 

Under this constraint on the effective operators, one cannot find 
an operator for Majorana neutrino masses or dimension-4 proton decay 
operators. This is a blessing in a sense that proton does not decay
too rapidly. On the other hand, the absence of the Majorana masses 
of neutrinos require\footnote{The origin of baryon asymmetry of our 
universe is another problem in the pure Dirac scenario of the neutrino 
masses. Standard scenario of leptogenesis \cite{FY} does not work. 
Some ideas have been proposed for the leptogenesis in the scenario 
with Dirac neutrino masses \cite{DiracLG}.} that 
the Dirac Yukawa couplings of neutrinos be as small as 
$10^{-11}$ or even less, in order to account for the very tiny 
neutrino masses observed in neutrino oscillation experiments.
The Dirac Yukawa couplings in (\ref{eq:Het-Yukawa}) originate from 
$E_8$ Yang--Mills interactions. It is a challenging problem to find 
out how they can be so small.
 
Let us now consider a situation where either the anti-generation 
right-handed neutrinos $\overline{\bar{N}}$ or right-handed neutrinos
$\bar{N}$ obtain non-zero expectation values. These situations 
correspond to $\int_Z c_1(L) \wedge J \wedge J < 0$ and 
$\int_Z c_1(L) \wedge J \wedge J > 0$, respectively.

Large expectation value in $\bar{N}$ are not welcomed 
phenomenologically, partly because we do not want large mass terms 
of the form $W \ni \vev{\bar{N}} \bar{\bf 5} . H({\bf 5})$. Not 
all the moduli of the rank-5 vector bundle $V_5$ turn on masses to 
all vector-like pair of SU(5)$_{\rm GUT}$-{\bf 5}+$\bar{\bf 5}$ 
representations \cite{Penn5}. Thus, there may be a way out of 
this problem. 

Another serious problem is the dimension-4 
proton decay. In the rank-5 vector bundle $V_5$ given as an extension 
of $L$ by $U_4$ as in (\ref{eq:41ext}), $U_4$ remains a sub-bundle, 
and $L$ ceases to be a sub-bundle of $V_5$. This means that the 
zero-mode wave functions of the SU(5)$_{\rm GUT}$-{\bf 10} multiplets 
from the bundle $U_4$ can still be confined in $U_4$, but those of 
the SU(5)$_{\rm GUT}$-${\bf 10}'$ multiplets (if exist) are no longer 
confined in ``a sub-bundle $L$.'' Rather they take values in the entire
bundle $V_5$. The same is true for the wave functions of 
the SU(5)$_{\rm GUT}$-$\bar{\bf 5}$ multiplets as they are no longer 
confined in ``a sub-bundle $U_4 \otimes L$'' of $\wedge^2 V_5$, but 
take values in $U_4 \otimes V_5$. In particular, the $\bar{\bf 5}$ 
multiplets have wave functions partially in ``$\wedge^2 U_4$'' part 
and acquire properties of the $\bar{H}(\bar{\bf 5})$ multiplet. 
Thus, the second term of (\ref{eq:Het-super}) generates the 
dimension-4 proton decay operators by picking up the wave function 
of one of $\bar{\bf 5}$ multiplets from the ``$\wedge^2 U_4$'' part, 
just like the down-type quark and charged-lepton Yukawa couplings
are generated. 

The effective-field theory language can also capture how the dimension-4 
operators are generated. The expectation values in the vector bundle 
moduli $\bar{N}$ force the wave functions of zero-modes to be 
modified, and for $\bar{\bf 5}$ multiplets, for instance, their new 
wave functions are something like 
\begin{equation}
 \bar{\bf 5}_{\rm new} \approx \bar{\bf 5}_{\rm red} 
+ \vev{\bar{N}}\bar{\bf 5}_{\rm red},
\end{equation} 
treating the deviation from that in the reducible limit as a small  
perturbation. The second term takes its value in the ``$\wedge^2 U_4$''
part, like the $\bar{H}(\bar{\bf 5})$ multiplet. Thus, 
effective operators 
\begin{equation}
 W \ni \bar{\bf 5}_{\rm red}.{\bf 10}_{\rm red}.
 (\vev{\bar{N}}\cdot \bar{\bf 5}_{\rm red})
\end{equation}
that are invariant under the SU(4)$\times$U(1)$_\chi$ structure group 
at the reducible limit give rise to dimension-4 proton decay operators.

On the contrary, expectation values of $\overline{\bar{N}}$ do not 
create this problem. The sub-bundle $L$ remains to be well-defined, 
and the wave functions of the $\bar{\bf 5}$ multiplets are confined 
in $V_5 \otimes L$, which does not contain a ``$\wedge^2 U_4$'' part 
that have properties of the $\bar{H}(\bar{\bf 5})$ multiplet. The
effective-field theory analysis using $\vev{\overline{\bar{N}}}$ 
insertions leads to the same answer.\footnote{As long as 
$\chi({\bf 10}')$ is negative, such chiral matters can form a 
mass term $W\ni {\bf 10} \cdot (\vev{\overline{\bar{N}}} 
\overline{{\bf 10}'})$ and do not necessarily appear in the low-energy 
spectrum. Thus, (\ref{eq:10less}) does not have to be imposed 
when the Fayet--Iliopoulos parameter is non-zero.} In summary, 
in order to avoid dimension-4 proton decay, a rank-5 holomorphic 
stable vector bundle $V_5$ should have 
\begin{center}
\begin{tabular}{lll}
 a rank-1 sub-bundle $L$ & with ($\int_Z c_1(L)\wedge J \wedge J < 0$) 
 & when $\int_Z c_3(V_5) < 0$ \\
 a rank-4 sub-bundle $U_4$ & with ($\int_Z c_1(\det \; U_4) \wedge 
 J \wedge J < 0$) & when $\int_Z c_3(V_5) > 0$. 
\end{tabular} 
\end{center}

It is interesting to see how the dimension-4 operators are forbidden 
in the scenario with non-vanishing $\overline{\bar{N}}$ expectation 
values. Two major ideas for getting rid of the dimension-4 operators 
have been the $R$-parity and $\Z_2$-$\chi$ (or $B{\rm -}L$) symmetry. 
But, the anti-generation right-handed neutrinos $\overline{\bar{N}}$ 
have the same $R$-parity and $\Z_2$-$\chi$ charges as the right-handed 
neutrinos, and are odd. Thus, both symmetries are broken by the 
expectation value of $\overline{\bar{N}}$, and the insertion of 
the expectation value can supply any $R$-parity or $\Z_2$-$\chi$ 
charges in effective operators. It is rather the continuous 
U(1)$_\chi$ symmetry that is broken only by a positively charged field 
that essentially protects protons from decaying rapidly.
Since the symmetry breaking of U(1)$_\chi$ (and $B{\rm -}L$) 
is triggered not by an F-term, but by the D-term, fields with opposite 
charges do not have to develop expectation values even when the 
symmetry is spontaneously broken.

Such terms in effective superpotential 
\begin{equation}
 W \ni \kappa \; \overline{\bar{N}}\overline{\bar{N}} \bar{N}\bar{N}
\label{eq:4N}
\end{equation}
are perfectly consistent with the underlying $E_8$ symmetry, and 
it is not surprising if such interactions are generated in some 
way ({\it e.g.} \cite{DSWW}). Once such terms are generated, 
the non-vanishing expectation value of $\overline{\bar{N}}$ gives 
rise to the Majorana mass term of right-handed neutrinos \cite{DIN}. 
Since the natural scale of the expectation value of 
$\overline{\bar{N}}$ is very high, the Majorana mass is also very 
large unless the coefficient of the operator (\ref{eq:4N}) 
is extremely suppressed. Such heavy right-handed Majorana neutrinos 
naturally explain tiny left-handed neutrino masses observed in neutrino
oscillation experiments through the see-saw mechanism \cite{see-saw}. 
A flat direction $\vev{|\overline{\bar{N}}|^2} - \vev{|\bar{N}|^2} 
= - \xi_\chi / 5$ is also removed by the operator (\ref{eq:4N}), 
if it exists, and $\vev{\bar{N}}=0$ and 
$\vev{|\overline{\bar{N}}|^2} = - \xi_\chi/5$ becomes the only 
solution. Therefore, if the operator (\ref{eq:4N}) exists, 
it not only provides the Majorana masses of right-handed neutrinos and also
explains tiny left-handed neutrino masses through the see-saw 
mechanism, but also prevents the dimension-4 proton decay operators 
from being generated. 

We have seen that the Yang--Mills interaction 
$W \ni \int \Omega \wedge (A\wedge A \wedge A)$ does not generate 
dimension-4 proton decay operators when $\vev{\bar{N}}=0$ and 
$\vev{\overline{\bar{N}}} \neq 0$, or when the rank-5 bundle $V_5$ 
is given by the extension (\ref{eq:14ext}). The effective-field 
theory arguments are more powerful and say that no  
perturbative (tree level in practise) processes can give rise to 
the dangerous dimension-4 operators. Non-perturbative processes, 
however, may generate them. For example, an operator 
\begin{equation}
 W \ni \epsilon \;  
  \bar{\bf 5}. {\bf 10}. \bar{\bf 5}. \vev{\overline{\bar{N}}}^3
\label{eq:dim-7}
\end{equation}
is invariant under SU(4) structure group and SU(5)$_{\rm GUT}$ 
symmetry, and is of U(1)$_\chi$ charge 20. Since the chiral multiplets 
of K\"{a}hler moduli $T^k$ transform as in (\ref{eq:T-shift}), 
if the coefficient $\epsilon$ contains a non-perturbative factor 
such as some linear combinations of $T^k$'s on the exponent, 
U(1)$_\chi$ charges may be supplied and the operator above is 
consistent with the underlying 
SU(4)$\times$U(1)$_\chi \times$SU(5)$_{\rm GUT} \subset E_8$ symmetry.
World-sheet instanton \cite{DSWW} amplitudes have such properties, 
and in particular, an amplitude from a world-sheet wrapping on a curve 
$C$ has a prefactor 
\begin{equation}
 \epsilon \propto \exp \left[ \frac{1}{2\pi\alpha'}\int_C (-J + iB) 
  \right]
 = \exp \left[ (2\pi)^2 \int_C \omega_k T^k \right].
\end{equation}
Such a prefactor transforms linearly under the U(1)$_\chi$ gauge 
transformation with charge 
\begin{equation}
q_{\chi,C}=  -(2\pi)^2 Q^k \int_C \omega_k = 5 \int_C c_1(L).
\end{equation} 
Thus, for instance, a dimension-7 operator like (\ref{eq:dim-7}) 
may be generated from a curve $C$ with $\int_C c_1(L)= - 4$. 
Such operators tend to be numerically suppressed because the prefactors
are exponentially small $\exp [- {\rm vol}(C)/(2\pi \alpha')]$
and it is not immediately clear whether such operators are excluded 
by the experimental limits by the data available so far.

\subsection{Compactification with 3+2 Vector Bundles}
\label{ssec:Het32}

\subsubsection{Spectrum}

If the reducible rank-5 vector bundle (\ref{eq:32bdle}) with the 
structure group SU(3)$_2 \times$SU(2)$_2 \times$U(1)$_{\tilde{q}_7}$ 
is introduced instead of SU(4)$\times$U(1)$_\chi$ bundle 
(\ref{eq:41bdle}), the origin of the low-energy particles are 
identified with the irreducible bundles shown in Table \ref{tab:ID-8B}.
Discussion in section \ref{sec:BottomUp} guarantees that the trilinear 
Yukawa couplings of quarks and leptons arise from the $E_8$ 
Yang--Mills interactions, while the dimension-4 operators for proton 
decay remain absent.

\begin{table}[t]
\begin{center} 
\begin{tabular}{|c|c|c|c|c|c|c|}
\hline
 Bundles &
  $U_2$ & $U_3$ & $U_3 \otimes U_2$ & $\overline{\wedge^2 U_2}$ & 
          $\wedge^2 U_3$ & $U_2 \otimes \overline{U_3}$ \\
\hline 
 Particles ID A & 
  $(\bar{U},Q,\bar{E})$ & ((${\bf 10}'$)) & $\bar{\bf 5}=(\bar{D},L)$ &
  $H({\bf 5})$ & $\bar{H}(\bar{\bf 5})$ & $\bar{N}$ \\
 Particles ID B & 
  $(\bar{U},Q,\bar{E})$ & ((${\bf 10}'$)) &  $\bar{H}(\bar{\bf 5})$ &
  $H({\bf 5})$ & $\bar{\bf 5}=(\bar{D},L)$ & $S$ \\
\hline
 \end{tabular}
\end{center} 
\caption{\label{tab:ID-8B}Particle identification in 
SU(3)$_2\times$SU(2)$_2\times$ U(1)$_{\tilde{q}_7}$ bundle 
compactification. There are 2 phenomenologically viable 
identifications. There are no particles in the 
SU(5)$_{\rm GUT}$-${\bf 10}'$ representation coming from the bundle 
$U_3$ in the low-energy spectrum of our world.}
\end{table}

Chirality formulae are given by 
\begin{eqnarray}
\chi(U_2) & = & 
  - \frac{1}{12}c_2(TZ)c_1(U_3)
  - \frac{1}{6}c_1(U_3)^3
  + \frac{1}{2}c_2(U_2)c_1(U_3), \\
\chi(U_3) & = & 
  + \frac{1}{12}c_2(TZ)c_1(U_3)
  + \frac{1}{6}c_1(U_3)^3
  - \frac{1}{2}c_2(U_3)c_1(U_3)
  + \frac{1}{2}c_3(U_3), \\
\chi(\overline{\wedge^2 U_2}) & = &
  + \frac{1}{12}c_2(TZ)c_1(U_3)
  + \frac{1}{6}c_1(U_3)^3, \\
\chi(U_3 \otimes U_2) & = &
   - \frac{1}{12}c_2(TZ)c_1(U_3)
   - \frac{1}{6}c_1(U_3)^3
   + \frac{1}{2}c_2(U_2)c_1(U_3)
   + c_3(U_3), \\
\chi(\wedge^2 U_3) & = &
   + \frac{2}{12}c_2(TZ)c_1(U_3)
   + \frac{2}{6}c_1(U_3)^3
   - \frac{1}{2}c_2(U_3)c_1(U_3)
   - \frac{1}{2}c_3(U_3), \\
 \chi(U_2 \otimes \bar{U}_3) & = & 
   - \frac{5}{12}c_2(TZ)c_1(U_3)
   - \frac{11}{6}c_1(U_3)^3
   + \frac{5}{2}c_2(U_3)c_1(U_3)
   - c_3(U_3)   \nonumber \\
& & \qquad \qquad \qquad \qquad \qquad \qquad \quad 
   + \frac{4}{2}c_2(U_2)c_1(U_3).
\end{eqnarray}
Here, $c_1(U_2) = - c_1(U_3)$ is used.

The same argument leading to (\ref{eq:10less}) requires 
\begin{equation}
\chi(U_3) = 0,  
\end{equation}
and the same net chirality for ${\bf 10}$ and $\bar{\bf 5}$ 
(or equivalently for $H({\bf 5})$ and $\bar{H}(\bar{\bf 5})$) requires 
\begin{equation}
 \chi(U_2) = \chi(U_3 \otimes U_2), 
\quad {\rm or~equivalently} \quad 
 \chi({\rm det} \; U_2^{-1})= \chi(\wedge^2 U_3)
\label{eq:hh-510}
\end{equation}
under the particle identification A in Table \ref{tab:ID-8B}.
These phenomenological information constrains the possible choice 
of topology for our vacuum as 
\begin{eqnarray}
c_2(TZ)c_1(U_3) & = & \left[6 c_2(U_3) - 2 c_1(U_3)^2 \right]c_1(U_3),
\\
c_3(U_3) & = & 0,
\end{eqnarray}
where we used $c_1(U_2) = - c_1(U_3)$. Under these conditions we have 
\begin{eqnarray}
 \chi({\bf 10})=\chi(\bar{\bf 5}) & = & 
  \frac{1}{2} \left[ c_2(U_2) - c_2(U_3) \right] c_1(U_3), \\
 \chi(H({\bf 5}))=\chi(\bar{H}(\bar{\bf 5})) & = & 
  \frac{1}{2} c_2(U_3) c_1(U_3), \\
 \chi(\bar{N}) & = & - c_1(U_3)^3 -\frac{1}{2}c_2(U_3)c_1(U_3) 
    + \frac{5}{2}c_2(U_2)c_1(U_3).
\end{eqnarray}
Similar analysis can be done for the case of the identification B in 
Table~\ref{tab:ID-8B}.

\subsubsection{Anomalous U(1)$_{\tilde{q}_7}$ Symmetry and Effective 
Superpotential}

The U(1)$_{\tilde{q}_7}$ gauge field has Green--Schwarz coupling 
because of non-trivial $c_1(U_3)=-c_1(U_2)$, just like the gauge 
field of the U(1)$_\chi$ symmetry does in section 
\ref{sssec:Het-anomalous}. Essentially the same analysis can be done 
for the U(1)$_{\tilde{q}_7}$ symmetry. The gauge boson acquires 
a large mass either from the Green-Schwarz interactions or 
from the Higgs mechanism, and can be absent from low-energy spectrum. 
In this section, we just only refer to the difference from 
what was discussed in \ref{sssec:Het-anomalous}.

The Fayet--Iliopoulos parameter given in (\ref{eq:FI41}) is valid for 
the U(1)$_{\tilde{q}_7}$ symmetry after replacing $c_1(L)$ by 
$c_1(U_3)$; the prefactor $10$ happens to be the same. 
The 1-loop contribution does not necessarily vanish for the 
U(1)$_{\tilde{q}_7}$ symmetry, and the Fayet--Iliopoulos parameter 
vanishes when the tree-level and 1-loop contributions cancel 
\cite{Blumenhagen}. 
In that case, the reducible vector bundle (\ref{eq:32bdle}) is 
semi-stable at the 1-loop level.

When the bundle (\ref{eq:32bdle}) is at the reducible limit, not only 
the dimension-4 but also dimension-5 proton decay operators are 
forbidden by the underlying symmetry. Indeed, neither
\begin{equation}
 U_2 \otimes U_2 \otimes U_2 \otimes (U_3 \otimes U_2) 
 \quad [\rm ID~A], \quad {\rm nor} \quad 
 U_2 \otimes U_2 \otimes U_2 \otimes (\wedge^2 U_3)  
 \quad [\rm ID~B]
\label{eq:Het-dim5-3+2}
\end{equation}
contains a trivial bundle. Since these bundles are not even singlet 
under SU(3)$_2$, these operators are forbidden even at the 
non-perturbative level.

The particle identification B in Table~\ref{tab:ID-8B} may yield  
chiral multiplets $S$. Since we have not made any specific choices of 
vector bundles, we do not know how many chiral multiplets 
are available (or if there is a vacuum without such a multiplet) 
around the electroweak scale. But, if there are, then they play 
some roles in the Higgs-related physics, since they has a coupling 
$W \ni S H_u H_d$. Thus, it is interesting to know more about the 
interactions of the multiplet $S$. If there are only multiplets $S$,
without their Hermitian-conjugate $\bar{S}$, then the underlying 
SU(3)$_2\times$SU(2)$_2\times$U(1)$_{\tilde{q}_7}$ symmetry kills 
any kinds of terms in the superpotential only in $S$. At the
non-perturbative level, where we ignore the U(1)$_{\tilde{q}_7}$-charge 
conservation, some terms may be generated:
\begin{equation}
 W \ni S H_u H_d + \epsilon \; S^6 + \epsilon' \; S^{12}+\cdots.
\label{eq:appPQ}
\end{equation}
The dimension-7 term $W \ni S^6$ is the leading correction, and 
the cubic term $W \ni S^3$ of the next-to-minimal supersymmetric 
standard model \cite{NMSSM} is not allowed even at the 
non-perturbative level.\footnote{Here, we assume that the underlying 
$E_8$ symmetry is broken only by the vector bundle, and not directly 
by any non-perturbative processes.} Phenomenology of an effective 
theory without $S^3$ term has been studied in \cite{CPHW,LEP-limit}. 

If the values of K\"{a}hler moduli and dilaton are such that the 
Fayet--Iliopoulos parameter does not vanish, then the 
U(1)$_{\tilde{q}_7}$ symmetry is spontaneously broken. 
The absence of dimension-4 proton decay operators suggests the presence 
of the Fayet--Iliopoulos parameter. Under the identification A
in Table~\ref{tab:ID-8B}, $\vev{\overline{\bar{N}}}=0$, 
$\vev{\bar{N}} \neq 0$ and the rank-5 bundle is an extension
\begin{equation}
 0 \rightarrow U_2 \rightarrow V_5 \rightarrow U_3 \rightarrow 0.
\label{eq:23ext}
\end{equation}
The bundle $U_2$ has to remain well-defined as a sub-bundle, so that 
the $\bar{\bf 5}$-representation is still distinguished from 
$\bar{H}(\bar{\bf 5})$.
Under the identification B, $\vev{S}=0$, $\vev{\bar{S}}\neq 0$ and 
the rank-5 bundle is an extension 
\begin{equation}
 0 \rightarrow U_3 \rightarrow V_5 \rightarrow U_2 \rightarrow 0. 
\label{eq:32ext}
\end{equation}
The bundle $U_3$ has to be a well-defined sub-bundle for the same 
reason as in the case of the identification A.

If the Fayet--Iliopoulos parameter is of order $M_G^2 (l_s/R)^2$, as 
indicated from the tree-level contribution, it is not small as 
long as $1/R$ is around the GUT scale. Thus, the expectation values 
of $\bar{N}$ and $\bar{S}$ are much larger than the electroweak scale. 
In the former case, $\bar{\bf 5}$ and $H({\bf 5})$ multiplets decouple 
from the low-energy physics in pair, by acquiring mass terms from the 
Dirac Yukawa coupling with the expectation value of $\vev{N}$.
The expectation value of a vector bundle moduli $\bar{N}$, however, 
does not necessarily add masses to all the vector-like pair in the 
SU(5)$_{\rm GUT}$-{\bf 5}+$\bar{\bf 5}$ representations \cite{Penn5}. 
Thus, some of those multiplets may remain in the low-energy spectrum, 
and a large expectation value in one of $\bar{N}$ may not be 
inconsistent with the reality. 
The analysis based on (\ref{eq:hh-510}) does not take account of this,
and the constraints on topology obtained there have to be modified 
appropriately.

Neutrinos may have masses through 
\begin{equation}
 W \ni \frac{1}{2} \left(L,\bar{N}',\overline{\bar{N}}\right)
   \left( \begin{array}{ccc}
    0 & y \; \vev{H_u} &  0 \\
	 y \; \vev{H_u} & 0 & M \\
	 0 & M & \kappa \vev{\bar{N}^2}
	  \end{array}\right)\left(\begin{array}{c}
	  L \\ \bar{N}' \\ \overline{\bar{N}}
				  \end{array}\right)
\label{eq:double}
\end{equation}
under the identification A, where a vector-like mass term $W \ni M
\bar{N}' \overline{\bar{N}}$ and an operator like (\ref{eq:4N}) 
exist. $\bar{N}'$ collectively stand for chiral multiplets from 
the bundle $U_2 \otimes \overline{U_3}$ except the one that develops 
an expectation value to absorb the Fayet--Iliopoulos parameter.
This type of mass matrix explains the tiny neutrino masses through 
so-called the double-see-saw mechanism.
On the other hand, in the identification B, the origin of neutrino 
masses is not understood.

Effective field theory analysis can be carried out for 
the dimension-5 proton decay operators by allowing the expectation 
values of $\bar{N}$ or $\bar{S}$ to be inserted in effective operators.
It turns out that the underlying symmetry 
SU(5)$_{\rm GUT}\times$SU(3)$_2 \times$SU(2)$_2 \times$
U(1)$_{\tilde{q _7}}$ in the identification A does not allow such 
an operator. Hence the dimension-5 proton decay is absent at 
the perturbative level.
On the other hand, in the identification B, effective operators of 
the form 
\begin{equation}
 W \ni {\bf 10}.{\bf 10}.({\bf 10}\vev{\bar{S}}).\bar{\bf 5}
\label{eq:32B-dim5}
\end{equation}
are consistent with the symmetry SU(5)$_{\rm GUT}\times$SU(3)$_2 
\times$SU(2)$_2\times$U(1)$_{\tilde{q}_7}$, and hence may be generated 
at the perturbative level. Since the experimental limit on the 
dimension-5 proton decay is not very severe, such operators may be 
consistent with the reality if the Fayet--Iliopoulos parameter is 
not very large.

\subsection{Compactification with 2+2+1 Vector Bundles}
\label{ssec:Het221}

In 2+2+1 bundle compactifications, 
$H({\bf 5})$ and $\bar{H}(\bar{\bf 5})$ do not arise 
from a Hermitian-conjugate pair of irreducible representations, 
just like in 3+2 bundle compactifications. Thus, the Wilson line 
cannot work as a solution to the doublet--triplet splitting problem. 
Not only dimension-4 proton decay operators but also dimension-5 
operators are forbidden by the underlying symmetry, just like 
in the 3+2 bundle compactifications. 
Since the 2+2+1 vector bundles are regarded as particular choice of 
vector-bundle moduli of 3+2 bundles and 4+1 bundles, they inherit 
stronger properties of 3+2 and 4+1 bundles.  

When the vector-bundle moduli are at the reducible limit, there are 
two linearly independent anomalous U(1) symmetries: $q_6$ and $q_7$, 
$\chi$ and $\psi$, or $\tilde{q}_6$ and $\tilde{q}_7$. 
When the Fayet--Iliopoulos parameters of those U(1) symmetries are 
non-zero, some of fields in the off-diagonal blocks in 
(\ref{eq:1inE8}) develop non-zero expectation values so that 
the D-term potential energy vanishes. Depending on which field does, 
different types of effective operators are potentially generated. 
Which field develops an expectation value depends also on 
supersymmetry breaking masses, and we are not going into that 
discussion here. 

\section{M-theory Vacua}
\label{sec:M}

In this section, we will provide M-theory implementations of the 
general idea for the origin of low-energy particles and Yukawa 
couplings developed in terms of field theory in section 
\ref{sec:BottomUp}. The $E_7$ or $E_8$ vector multiplet does 
not exist in the eleven-dimensional supergravity, as opposed to 
the case in Heterotic theory where $E_8$ vector multiplet exists 
from the beginning. 
In M theory vacua, the $E_7$ or $E_8$ symmetry has to arise from 
singularity of $G_2$ holonomy manifold\footnote{We abuse the word
``manifold'' even when a ``manifold'' has singularity.} used for 
the compactification yielding a vacuum with ${\cal N} = 1$ 
supersymmetry.
The geometry of the compactification manifold has to be properly 
engineered to obtain the right particle spectrum and the right form 
of interactions at low energies.
In this section we present the geometric origin of the low-energy particles 
and of the absence of the very rapid proton decay.

It is known that the compactification of Heterotic String on $T^3$ is dual to 
M-theory compactification on $K3$ \cite{wittendual}. This duality is further
exploited by fibring $T^3$ and $K3$ on a common base manifold $Q$. 
When $Q$ is a 3 dimensional, Heterotic compactification on $T^3$-fibred 
Calabi--Yau 3-fold and M-theory compactification on $K3$-fibred $G_2$
holonomy manifold become relevant to the description of the real world 
(if low-energy supersymmetry is confirmed in a near future experiment). 
Neither all the Calabi--Yau 3-fold have $T^3$-fibred structure, nor 
all the $G_2$ holonomy manifold have $K3$-fibred structure. 
Thus, some vacua have descriptions both in terms of Heterotic theory 
and M-theory, some may have only one in Heterotic theory, others 
may be described only in terms of M-theory compactification. 
Such vacua in different parts of the moduli space of the entire string 
theory may have different phenomenological aspects, especially when 
it comes to the pattern of Yukawa matrices and 
this is why we consider that it is important not only to develop 
string phenomenology in Heterotic theory but also in M-theory 
compactification in order to cover more vacua and more 
variety in phenomenology. 

\subsection{Local Model of Up-type Quark Yukawa Coupling}
\label{ssec:M6}

By comparing with complex geometries as Calabi--Yau 3-folds or 4-folds, 
little is known about the classification of $G_2$ holonomy manifold.
Thus, it does not seem promising to search for realistic M-theory 
vacua in a top down approach. Although M-theory vacua with Heterotic 
dual may be studied by using the terminology of complex geometry 
in the Heterotic side, this approach does not suit for the purpose 
of exploring M-theory vacua that may not have Heterotic dual 
descriptions. 
What is more, we do not necessarily have to start in M-theory 
with an $E_8$ symmetry, as we saw in section \ref{sec:BottomUp} where $E_7$ was
the minimal choice for the visible particle physics sector. 
Inflation and the supersymmetry breaking sector does not have to be 
constructed out of $E_8$ symmetric theory. 

We adopt the philosophy of a local construction approach (or bottom-up
construction approach) \cite{local}, instead. In M-theory and F-theory 
vacua, the SU(5)$_{\rm GUT}$ gauge fields propagate on a subvariety 
of the compactification geometry and the geometry (and D-brane 
configuration in Type IIA and Type IIB string theories) that is not in 
direct contact with the cycles of the visible sector has little effect
on the visible sector. Since virtually no constraint comes from 
observational or experimental tests for the invisible sectors, 
the geometry of that part is almost arbitrary. In order to study 
various aspects of particle physics, it is only sufficient to 
construct the part of geometries that relates to experiments. 
One can neglect at this stage the mechanism of supersymmetry breaking and 
concentrate on the particle physics of quarks and leptons.
By doing so, one can capture some generic features of phenomenology, while 
avoiding the limitation to a small subclass of geometries and of 
phenomenology models.\footnote{Needless to say, constructing explicit 
geometry is also important, to confirm that one has concrete examples 
in order to gain a firm understanding of what is really going on. } 
This is the basic philosophy of the local construction. \footnote{Other approach to 
using local geometries to get phenomenology models appeared in \cite{verlinde}.}
One can start 
with a configuration with an appropriate gauge symmetry, and add the
matter content one by one. Matters are added not by 
hand, but by constructing geometry, and the anomaly cancellation
condition is replaced by consistency conditions of geometry. They are
the difference from conventional model building in effective field
theories. References \cite{WittenG2} are notable examples of this 
approach in the context of M/$G_2$ compactification.

The subject of relating compactifications of M-theory and particle physics has 
been treated extensively in the last years, see \cite{ag} for a review.
The geometric building blocks are used in this section to
implement our ideas of section \ref{sec:BottomUp} in M-theory 
compactification. The goal is
 
i) to choose the right underlying symmetry and its breaking pattern,  

ii) to maintain different origin for  $\bar{\bf 5}$ and $\bar{H}(\bar{\bf 5})$ in 
order to avoid a too fast dimension-4 proton decay, 

iii) to obtain all the necessary Yukawa couplings generated from Yang--Mills interactions. 

\begin{figure}[t]
\begin{center}
\includegraphics[width=.5\linewidth]{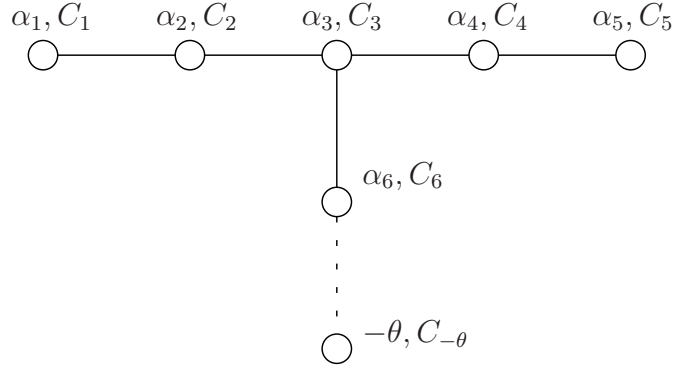}  
\begin{picture}(0,0)
 \put(-245,130){$\alpha_1, C_1$}
 \put(-190,130){$\alpha_2, C_2$}
 \put(-135,130){$\alpha_3, C_3$}
 \put(-80,130){$\alpha_4, C_4$}
 \put(-25,130){$\alpha_5, C_5$}
 \put(-112,70){$\alpha_6, C_6$}
 \put(-112,10){$-\theta, C_{-\theta}$}
\end{picture}
\caption{\label{fig:e6dynkin}
The (extended) Dynkin diagram of $E_6$, which not only describes the 
Lie algebra of $E_6$ symmetry but also the intersection form of 
$E_6$-type singularity. Each node is assigned a simple root $\alpha_i$ 
and a 2-cycle $C_i$. The $\mathfrak{su}(2)_2$ and 
$\mathfrak{su}(5)_{\rm GUT}$ subalgebra of 
$\mathfrak{su}(2)_2+\mathfrak{u}(1)+\mathfrak{su}(5)_{\rm GUT} \subset 
\mathfrak{e}_6$ in section \ref{sec:BottomUp} are generated by
 $\alpha_1$ and $\alpha_{3,4,5,6}$, respectively. 
The highest root $\theta$ is not linearly independent from other 
six simple roots:
$\theta=\alpha_1+2\alpha_2+3\alpha_3+2\alpha_4+\alpha_5+2\alpha_6$. 
}
\end{center}
\end{figure}

Let us first construct the local model of the up-type quark Yukawa 
coupling in M-theory, as we did in section \ref{sec:BottomUp}.
In M-theory description, the $E_6$ symmetry is realised by compactifying
the M-theory on an ALE space containing six 2-cycles with the 
$E_6$-type intersection form. M2-branes wrapped on the 2-cycles 
correspond to gauge fields in the simple roots assigned to the 
nodes of the Dynkin diagram (see Fig.~\ref{fig:e6dynkin}). 
The D = 4 effective theories with ${\cal N} = 1$ supersymmetry are 
obtained by compactifying the M-theory on a $G_2$ holonomy manifold 
$X$. We are interested in a manifold $X$ with an associative  
3-cycle $Q$ such that  $X$ is locally ALE fibration over $Q$ with the 
six 2-cycles in the fibre. Four of them---$C_3, C_4, C_5$ and 
$C_6$---are not resolved anywhere on $Q$. The locus of the $A_4$ 
singularity is identified with $Q$ itself. The SU(5)$_{\rm GUT}$ 
vector multiplet of ${\cal N}$ = 1 supersymmetry is localised 
along this locus \cite{Acharya}. 

In order to describe the geometry for matter fields explicitly, we use 
the following description of ALE spaces \cite{Kronheimer} 
(for a brief summary, see appendix \ref{sec:ALE}). 
$E_6$ type ALE space and its metric are encoded by data 
$\vec{\zeta}^i=(\zeta^i_1,\zeta^i_2,\zeta^i_3) \in \R^3$ for 
$i=0,\cdots,6$, which are under a constraint 
\begin{equation}
   \vec{\zeta}^0 + \vec{\zeta}^1 + 2 \vec{\zeta}^2 + 3 \vec{\zeta}^3 
 + 2 \vec{\zeta}^4 + \vec{\zeta}^5 + 2 \vec{\zeta}^6 = 0.
\label{eq:HKdata-6}
\end{equation}
Each of $\vec{\zeta}^i$ describes the ``size'' of the corresponding 
2-cycle $C_i$, and this is also true for $i=0$, if we consider  
that $\alpha_0 \equiv -\theta$ and $C_0 \equiv C_{-\theta} \equiv 
-(C_1+2C_2+3C_3+2C_4+C_5+2C_6)$, where $-\theta$ is the negative 
of the highest root. The ALE space has an $A_4$ singularity 
when $\vec{\zeta}^3 = \vec{\zeta}^4 = \vec{\zeta}^5 = \vec{\zeta}^6 
= \vec{0}$, and the 2-cycles $C_3$, $C_4$, $C_5$ and $C_6$ are 
of zero size.

Let $\vec{y}=(y_1,y_2,y_3)$ be a set of local coordinates of the 
associative 3-cycle $Q$. The ALE fibre is allowed to vary 
over the base manifold $Q$, while keeping the $E_6$ intersection 
form and $A_4$ singularity for the SU(5)$_{\rm GUT}$ gauge field.
Thus, the local geometry of $G_2$ holonomy of our interest is 
described by specifying $\vec{\zeta}^1(\vec{y})$ and 
$\vec{\zeta}^2(\vec{y})$ as functions on $Q$. $\vec{\zeta}^0(\vec{y})$ 
is determined by (\ref{eq:HKdata-6}). The base manifold $Q$ is 
identified with the locus of $A_4$ singularity.

At some points on the base 3-fold $Q$, $\vec{\zeta}^2(\vec{y})$ or 
$\vec{\zeta}^{1}(\vec{y})+\vec{\zeta}^2(\vec{y})$ may become 
$\vec{0}$. There, either $C_2$ or $C_1+C_2$ shrinks. 
Since the intersection form of $C_3$, $C_4$, $C_5$ and $C_6$ along 
with an extra vanishing cycle $C_2$ or $C_1+C_2$ is the negative 
of the $D_5$ Cartan matrix, the $A_4$ singularity on $Q$ is enhanced 
to $D_5$ at such points.
The local geometry around the point of enhanced gauge symmetry has 
a Type IIA interpretation, after reduction along the $S^1$ fibre; 
five D6-branes are on top of one another, and intersect with 
an O6-plane \cite{Sen-M}. One massless chiral multiplet either 
in the SU(5)$_{\rm GUT}$-$\overline{\bf 10}$ representation or in the 
SU(5)$_{\rm GUT}$-{\bf 10} is localised at the locus of 
enhanced gauge symmetry (singularity) \cite{BDL}.
The chirality depends on the sign of Jacobian  
$|\partial \vec{\zeta}^2(\vec{y}) / \partial \vec{y}|$ or 
$|\partial (\vec{\zeta}^1(\vec{y})+\vec{\zeta}^2(\vec{y})) /
\partial \vec{y}|$ at the loci of enhanced singularity. 

The local description of the ALE fibration with local coordinates 
$(y_1,y_2,y_3)$ and data $\vec{\zeta}^1(\vec{y})$ and 
$\vec{\zeta}^2(\vec{y})$ is glued together between adjacent patches 
on $Q$ to give the global description of the ALE fibration over 
the entire $Q$. When two adjacent descriptions are glued, the 2-cycles
on one patch should be matched with those of the other so that the 
intersection form is preserved, and further more, $C_3$, $C_4$, $C_5$ 
and $C_6$ should be matched with those of the other, since they 
have zero size. Furthermore, over the entire $Q$ we need to make sure 
that $C_5$ and $C_6$ remain the same everywhere on $Q$. 
In principle, one can trace $C_6$ moving over $Q$ and when 
one comes back to the original point on $Q$, $C_6$ may have become 
$C_5$. The $A_4$ intersection form is preserved through the entire 
process. But, in such geometry, not all of the SU(5)$_{\rm GUT}$ gauge 
symmetry is maintained, as in the non-split case discussed in 
\cite{6authors}. This is not what we want for the description 
of our world.

For the cycles $C_2$ and $C_1$, there is a little more degree of 
freedom in how to glue them between two adjacent patches. 
The Weyl reflection associated with the root $\alpha_1$ sends 
$\alpha_2$ to $\alpha_1+\alpha_2$, and $\alpha_1$ to $-\alpha_1$, 
keeping all other simple roots intact. This corresponds to a different 
choice of a Weyl chamber, and we could have chosen $-\alpha_1$ instead 
of $\alpha_1$, and $\alpha_1+\alpha_2$ instead of $\alpha_2$ as 
the simple roots. The same is true for the choice of independent 
homology basis of the ALE space. The intersection form is preserved 
under the ``Weyl reflection'' among the 2-cycles. The data may be 
matched on a common subset of a patch $\alpha$ and $\beta$ as 
\begin{equation}
\left(\begin{array}{c}
 \vec{\zeta}^2(\vec{y})\\ \vec{\zeta}^1(\vec{y})+\vec{\zeta}^2(\vec{y})
      \end{array}\right)_\alpha
= \left(\begin{array}{cc}
   1 & \\ & 1 
	\end{array}\right)
\left(\begin{array}{c}
 \vec{\zeta}^2(\vec{y})\\ \vec{\zeta}^1(\vec{y})+\vec{\zeta}^2(\vec{y})
      \end{array}\right)_\beta
\end{equation}
or 
\begin{equation}
\left(\begin{array}{c}
 \vec{\zeta}^2(\vec{y})\\ \vec{\zeta}^1(\vec{y})+\vec{\zeta}^2(\vec{y})
      \end{array}\right)_\alpha
= \left(\begin{array}{cc}
    & 1 \\ 1 &  
	\end{array}\right)
\left(\begin{array}{c}
 \vec{\zeta}^2(\vec{y})\\ \vec{\zeta}^1(\vec{y})+\vec{\zeta}^2(\vec{y})
      \end{array}\right)_\beta
\end{equation}
Thus, the 2-cycle $C_2$ on one patch may be $C_1+C_2$ on another, 
and vice versa. Thus, in the case the distinction between $C_2$ and 
$C_1+C_2$ is lost globally on $Q$ and there is no distinction at all 
between the SU(5)$_{\rm GUT}$-{\bf 10} representations arising from 
$C_2$-collapsed singularities and those from $(C_1+C_2)$-collapsed 
singularities. The doublet of data 
($\vec{\zeta}^2(\vec{y})$, 
$\vec{\zeta}^1(\vec{y})+\vec{\zeta}^2(\vec{y})$) becomes a 2-fold 
cover over $Q$, on which $\mathfrak{S}^{(2)}_2$, the Weyl group of 
SU(2)$_2$ acts. The fact that the {\bf 10} representations arise 
from either $C_2$ or $(C_1+C_2)$-collapsed singularities reflects 
that these representation comes from the $({\bf 10},{\bf 2})$ 
irreducible component of $\mathfrak{e}_6$-{\bf adj.} under 
the SU(2)$_2\times$U(1)$_6\times$SU(5)$_{\rm GUT}$ subgroup.
The 2-fold cover describes the degree of freedom of the geometry for 
the M-theory compactification, just like did the spectral cover for the 
description of vector bundles on elliptic fibred manifolds 
in Heterotic theory. It is not surprising since 
the Heterotic--M-theory duality holds between Heterotic 
compactifications on $T^3$-fibred Calabi--Yau 3-folds and M-theory 
compactifications on $K3$-fibred $G_2$ holonomy manifolds.
But the description in terms of the 2-fold cover is general in 
any M-theory compactifications on $G_2$ holonomy manifolds, not only 
for those with the $K3$-fibration structure.

Now let us move on to another matter multiplet, $H({\bf 5})$. The 
2-cycle $C_{-\theta}$ shrinks at points where $\vec{\zeta}^0(\vec{y})
= -(\vec{\zeta}^1(\vec{y})+2\vec{\zeta}^2(\vec{y}))$ becomes $\vec{0}$.
We can also say that the 2-cycle $(C_1+2C_2)$ shrinks there.
The intersection form of the four 2-cycles $C_3, C_4, C_5$ and $C_6$, 
along with another vanishing 2-cycle $C_{-\theta}$ is the negative of 
$A_5$ Cartan matrix.
The singularity is enhanced from $A_4$ to $A_5$. 
The Type IIA interpretation locally exists: one D6-brane intersects 
at this point with a stack of five D6-branes, and one massless chiral 
multiplet either in the SU(5)$_{\rm GUT}$-{\bf 5} or -$\bar{\bf 5}$ 
representation is localised there. The chirality depends on the sign 
of Jacobian 
$|\partial (\vec{\zeta}^1(\vec{y})+2\vec{\zeta}^2(\vec{y}))/
\partial \vec{y}|$ at the intersection point and, 
if the chirality comes out right, we have a candidate for $H({\bf 5})$.
The irreducible decomposition of the $\mathfrak{e}_6$-{\bf adj.} 
algebra along the the particle identification in section 
\ref{sec:BottomUp} also says that the $H({\bf 5})$ multiplet should 
come from the irreducible piece with roots $\alpha_1+2\alpha_2$ in 
addition to some linear combinations of $\alpha_{3,\cdots,7}$, 
confirming that this is the right geometric origin of $H({\bf 5})$ 
in the M-theory description.

Since the existence of massless chiral multiplets and their chirality 
depend only on the local geometry, the D6-brane and O6-plane 
interpretation that exists only locally can determine the multiplicity 
and chirality of massless multiplets, despite the absence of global 
Type IIA interpretation in compactification involving $E_d$-type 
intersection form. 

We have seen that there are candidates for {\bf 10} and $H({\bf 5})$ 
multiplets in the M-theory compactified on an $E_6$-type ALE fibration 
on a 3-cycle. The up-type quark Yukawa coupling (\ref{eq:uYukawaE6}) 
exists 
because the sum of 2-cycles $-C_2$, $-(C_1+C_2)$ and $(C_1+2C_2)$ 
is topologically trivial; M2-branes wrapped on these 2-cycles can 
merge together to disappear. Thus, the candidates really have the 
right properties to be ${\bf 10}$'s and $H({\bf 5})$.

When the $S^1$ fibre of the ALE space is negligibly small, this is 
qualitatively similar to the Type IIA description of the origin of 
trilinear Yukawa couplings: a worldsheet spanning 3 points of D6--D6 
intersections. The Yukawa coupling is exponentially small, if 
the area spanned by the worldsheet is large. When the $S^1$ fibre 
is not necessarily small, the area is replaced by 3-volume swept 
by a M2-brane. The Yukawa couplings have non-zero complex phases  
when the integration of the Ramond--Ramond 3-form over the 3-volumes 
$\int C^{(3)}$ are non-zero. Since the relevant 3-volumes are 
different for different entries of up-type quark Yukawa matrix, 
different entries may have different complex phases. 
Thus, the transformation matrix that brings 3 families of chiral 
multiplets $Q$ into the mass eigenstates can have complex phases, 
leading to the source of CP violation in the 
Cabbibo--Kobayashi--Maskawa matrix.
The mechanism of generating Yukawa coupling in Type IIA string / M 
theory \cite{CS} is now generalized above so that it is valid even 
when $E_n$-type intersection of 2-cycles are involved in the 
$G_2$-holonomy manifold.

One thing that one immediately notices is that the diagonal entries 
of the up-type quark Yukawa matrix may be suppressed, rather than 
unsuppressed.
This is because the sum of 2-cycles $-C_2 -C_2 + (C_1+2C_2)$ is not 
topologically trivial. Diagonal entries can be generated only by M2 
branes that once sweep a broad region over the base manifold $Q$ 
so that discrete holonomy between $\vec{\zeta}^2(\vec{y})$ and 
$\vec{\zeta}^1(\vec{y})+\vec{\zeta}^2(\vec{y})$ of the 2-fold cover 
can convert $C_2$ to $C_1+C_2$ or vice versa. Thus, when diagonal 
entries are generated, they are suppressed exponentially by the 
volume M2 branes swept. Here, we assume that the locus of enhanced 
gauge symmetry of $(C_1+2C_2)$ is different from all of those of 
$C_2$ or $C_1+C_2$.
Whether this feature can be consistent with the hierarchical and 
mixing pattern of quarks of the real world is left to a future 
investigation.

One can also attempt to connect our results to the ones of \cite{bb} where the 
$E_6$ was broken geometrically to $\SU(3)^2 \times \SU(2)$. 
This will consist of further breaking of the GUT group $SU(5)$. 
We leave this for a further work. 

\subsection{Minimal $E_7$ Model for All the Yukawa Couplings}
\label{ssec:M7}

Let us now proceed to the $E_7$ model, where all the low-energy
particles ${\bf 10}$, $\bar{\bf 5}$, $H({\bf 5})$ and 
$\bar{H}(\bar{\bf 5})$ are obtained along with all the necessary 
Yukawa couplings. We consider a $G_2$ holonomy manifold $X$ as 
an ALE fibration over an associative 3-cycle $Q$ such that 
each ALE fibre space contains seven 2-cycles with $E_7$ intersection form. 
Four of the cycles, namely $C_3$, $C_4$, $C_5$, and $C_6$, remain intact 
over the entire $Q$, forming $A_4$ singularity and 
yield the SU(5)$_{\rm GUT}$ gauge fields.
There should be no non-trivial holonomy flipping both ends of 
the $A_4$ Dynkin diagram of $\alpha_{3,\cdots,6}$, as we discussed 
in section \ref{ssec:M6}.
\begin{figure}[t]
\begin{center}
\includegraphics[width=.5\linewidth]{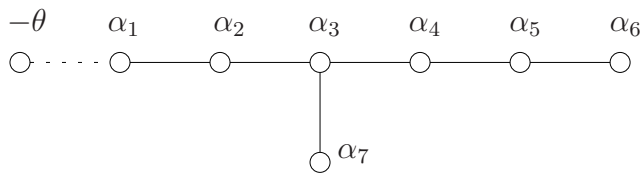}  
\begin{picture}(0,0)
 \put(-240,55){$-\theta$}
 \put(-202,55){$\alpha_1$}
 \put(-162,55){$\alpha_2$}
 \put(-126,55){$\alpha_3$}
 \put(-88,55){$\alpha_4$}
 \put(-50,55){$\alpha_5$}
 \put(-12,55){$\alpha_6$}
 \put(-115,7){$\alpha_7$}
\end{picture}
\caption{\label{fig:e7dynkinE}
The (extended) Dynkin diagram of $E_7$. 
The $\mathfrak{su}(2)_2$ is generated by $\alpha_1$, and 
$\mathfrak{su}(6)_1$ by $\alpha_{3,4,5,6,7}$ 
($\mathfrak{su}(5)_{\rm GUT}$ without $\alpha_7$). The highest root 
$\theta$ satisfies 
$-\theta + 2\alpha_1+3\alpha_2+4\alpha_3+3\alpha_4+2\alpha_5+\alpha_6
+2\alpha_7=0$.}
\end{center}
\end{figure}

Over each point on $Q$, the ALE fibre is described by  
$\vec{\zeta}^{0,1,2,7}(\vec{y})$, which describe the ``size'' of  
the 2-cycles $C_{0,1,2,7}$.
They obey the constraint 
\begin{equation}
 \vec{\zeta}^0 + 2\vec{\zeta}^1 + 3\vec{\zeta}^2 + 2\vec{\zeta}^7 = 0.
\end{equation}
The two 2-cycles $C_2$ and $C_1+C_2$ are a doublet of the Weyl group 
$\mathfrak{S}^{(2)}_2$ of $\mathfrak{su}(2)_2$, generated by 
$\alpha_1$ in the sense stated for the $E_6$ case.
Chiral matters arise at points where some linear combinations of 
those 2-cycles shrink, and singularity is enhanced. For different 
types of enhanced singularity and symmetry at different points on $Q$, 
we have different chiral multiplets. 

The correspondence between the extra vanishing cycles and chiral 
matters can be understood by carefully following the symmetry breaking 
of $E_7$ down to SU(5)$_{\rm GUT}\times$SU(2)$_2\times$U(1)$_{q_6}
 \times $ U(1)$_{q_7}$, and the decomposition of the representations. 
This is carried out in 
the appendix \ref{sec:note}, and the results are shown in 
Table~\ref{tab:e7cycle-matter}. As long as the loci of enhanced 
singularity (and gauge symmetry) are mutually isolated, the enhanced 
gauge symmetry is either $D_5$ or $A_5$, and the Type IIA 
interpretation that holds locally around each locus determines 
the multiplicity (just one) and the chirality depending on the sign 
of the Jacobian $|\partial \vec{\zeta}(\vec{y}) / \partial \vec{y}|$ 
around the intersection point of the relevant D6-branes 
(and O6-planes). 
\begin{table}[t]
\begin{center} 
\begin{tabular}{|c|c|c|c|c|c|}
particles & {\bf 10} & $\bar{\bf 5}$ & $H({\bf 5})$ & 
   $\bar{H}(\bar{\bf 5})$ & $\bar{N}$ \\
\hline
2-cycles & $-(C_2+C_3+C_7)$, & $C_2$, & $C_1+2C_2+2C_7$ & $C_7$ &  
   $C_{-\theta}$, \\ 
 & $-(C_1+C_2+C_3+C_7)$ & $(C_1+C_2)$ & $+3C_3+2C_4+C_5$ & & 
   $C_{-\theta}+C_1$ \\
\hline
singularity & $D_5$ & $A_5$ & $A_5$ & $A_5$ & $A_1+A_4$
 \end{tabular}
\caption{\label{tab:e7cycle-matter}Correspondence between the low-energy
particles and the 2-cycles on which M2-brane are wrapped. The relevant 
2-cycles are determined only up to adding some linear combinations of 
$C_{3}$, $C_4$, $C_5$, and $C_6$. For Higgs multiplets that are not 
doublets of SU(2)$_2$, only one 2-cycle is shown, but there are four 
others, which can be read out from the appendix \ref{sec:note}. For
 other multiplets, which are SU(2)$_2$ doublets, two 2-cycles are shown.
The last row shows the enhanced singularity when the relevant 2-cycle 
shrinks. Note that the particle identification in this table is based on 
the particle identification  pattern A in Table \ref{tab:ID-7}.}
\end{center} 
\end{table} 

The trilinear Yukawa couplings appear when three M2 branes merge and 
disappear, in other words when the sum of the 2-cycles on which M2
branes are wrapped become topologically trivial. We can see that all the 
necessary Yukawa couplings are generated in that way:
\begin{eqnarray}
 u{\rm \mbox{-}Yukawa}: & {\bf 10}.{\bf 10}.H({\bf 5}) & 
   -(C_1+C_2+C_7) - (C_2+C_7) + (C_1+2C_2+C_7) \equiv 0, 
\label{eq:M7-YukawaU} \\
 d,e{\rm \mbox{-}Yukawa}: & \bar{\bf 5}.{\bf 10}.\bar{H}(\bar{\bf 5}) & 
   C_2[+C_1] - (C_2+C_7[+C_1]) + C_7 \equiv 0, \\
 \nu{\rm \mbox{-}Yukawa}: & \bar{N}.\bar{\bf 5}.H({\bf 5}) & 
   C_{-\theta} + (C_1+C_2) + (C_1+2C_2+2C_7) \equiv 0,
\label{eq:M7-YukawaN}
\end{eqnarray}
where the equalities hold mod $+C_{3,4,5,6}$; the last line is for the 
Dirac mass terms for neutrinos. On the other hand, the dangerous operators 
leading to proton decay are not generated by merging M2-branes, since 
the sums of relevant 2-cycles are not topologically trivial:
\begin{eqnarray}
 {\rm dimension\mbox{-}4}: & \bar{\bf 5}.{\bf 10}.\bar{\bf 5} & 
   -(C_2+C_7[+C_1])+2(C_2[+C_1]) \nequiv 0,  \label{eq:M7-dim4} \\
 {\rm dimension\mbox{-}5}: & {\bf 10}.{\bf 10}.{\bf 10}.\bar{\bf 5} &
   -3(C_2+C_7[+C_1])+(C_2[+C_1]) \nequiv 0.
\label{eq:M7-dim5}
\end{eqnarray}
In Heterotic theory, the existence of Yukawa couplings and absence of 
too rapid proton decay was controlled by the existences or not the 
trivial bundles in the tensor products of vector bundles for relevant particles. 
In M-theory compactification, the topological triviality of 2-cycles 
replaces the bundle terminology. No matter which language is used,
physical results should be the same. Particle physics does not depend so much on whether 
the vacuum is described by Heterotic theory or M-theory but by what the underlying symmetry is 
and how it is broken---the idea described in section \ref{sec:BottomUp}.
As we saw in section \ref{sec:BottomUp}, $E_7$ is the minimal symmetry 
to accommodate all the low-energy particles, and there, both dimension-4 
and dimension-5 proton decay operators are forbidden, while all the 
necessary Yukawa couplings are allowed.
In the M-theory language, the SU(2)$_2$ part is taken care by the 
$C_2$ and $C_1+C_2$ components and the U(1) part by the $C_7$ component, 
in the argument lead to the absence of proton decay. 

Statements on the hierarchy and CP phases written for the up-type 
quark Yukawa couplings also hold for all the Yukawa matrices in 
(\ref{eq:M7-YukawaU})--(\ref{eq:M7-YukawaN}).
For vacua with M-theory (or Type IIA) description, chiral matters are 
localised at some points on a real 3-dimensional cycle in the M-theory 
(Type IIA) picture, and the Yukawa couplings are generated by spanning 
M2 branes (world sheets) between three points on the 3-cycle. This
picture is certainly more intuitively tractable\footnote{This does not
mean that the vacua with M-theory description are phenomenologically 
better than those without one.} than calculating 
zero-mode wavefunctions on a Calabi--Yau complex 3-fold with unknown
metric. Since the minimal model presented here has known source of 
Yukawa matrices and is guaranteed to be free of dangerous proton decay, 
it deserves further investigation.\footnote{It should be kept in mind
that the Wilson line on a quotient geometry obtained by freely acting 
symmetry cannot immediately solve the doublet--triplet splitting
problem in this model, as noted in section \ref{sec:BottomUp}.}

\subsection{Local Model with $E_8$ Intersection Form}
\label{ssec:M8}

The ALE geometry in the fibre allows chains of 2-cycles with $E_8$
intersection form so we can also consider M-theory
compactification with the $E_8$ symmetry.
The geometry of $E_8$ type ALE space is described by a set of 
data $\vec{\zeta}^{-\theta,1,2,7,8}(\vec{y})$, satisfying 
\begin{equation}
 \vec{\zeta}^{-\theta} + 2 \vec{\zeta}^1 + 4 \vec{\zeta}^2 
+ 2 \vec{\zeta}^7 + 3 \vec{\zeta}^8 = \vec{0}.
\label{eq:HKdata-8}
\end{equation}
The SU(5)$_{\rm GUT}$ gauge field comes from vanishing 2-cycles 
$C_3$, $C_4$, $C_5$ and $C_6$. Chiral matters arise where some 
linear combinations of $C_1$, $C_2$, $C_7$, $C_8$ (and $C_{-\theta}$)
shrinks and singularity (and gauge symmetry) is enhanced. It is 
a straightforward procedure to determine the vanishing 2-cycles 
corresponding to the low-energy multiplets. One can just read out the 
roots in (\ref{eq:root1inE8})--(\ref{eq:root5inE8}) for the 
irreducible pieces identified with the low-energy particles 
in (\ref{eq:10inE8})--(\ref{eq:1inE8}). 
We do not explicitly check whether the necessary Yukawa 
couplings are generated because our argument in section 
\ref{sec:BottomUp} guarantees it. The existence or absence 
of Yukawa couplings is just determined by $E_8$ algebra. 
The same is true for the dimension-4 proton decay operators.

Let us take a closer look at the geometry. 
There are three different choices of vector bundles in the Heterotic 
compactifications, namely (\ref{eq:41bdle}), (\ref{eq:32bdle}) and 
(\ref{eq:221bdle}) and the first two bundles are extensions of 
the last one. In the rest of this subsection, we clarify how the 
relations of these ``bundles'' are described in M-theory geometric 
terminology.

\begin{figure}[t]
\begin{center}
\includegraphics[width=.5\linewidth]{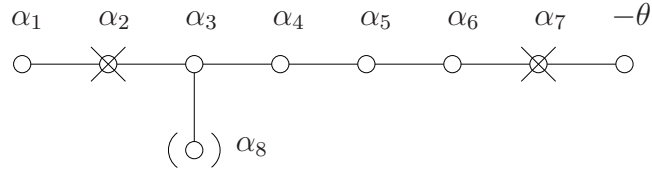}  
\begin{picture}(0,0)
 \put(-240,55){$\alpha_1$}
 \put(-207,55){$\alpha_2$}
 \put(-174,55){$\alpha_3$}
 \put(-141,55){$\alpha_4$}
 \put(-108,55){$\alpha_5$}
 \put(-75,55){$\alpha_6$}
 \put(-42,55){$\alpha_7$}
 \put(-13,55){$-\theta$}
 \put(-155,7){$\alpha_8$}
\end{picture}
\caption{\label{fig:e8dynkin221}
The extended Dynkin diagram of $E_8$ (I). 
The subalgebra $\mathfrak{su}(2)_2+\mathfrak{su}(2)_1+\mathfrak{su}(6)_1$
is generated by the simple roots $\alpha_1$, $-\theta$, and 
$\alpha_{3,4,5,6,7,8}$, respectively.}
\end{center}
\end{figure}

The SU(2)$_2\times$SU(2)$_1\times$U(1)$\times$U(1)-bundle 
compactification of Heterotic theory corresponds to the symmetry 
breaking pattern of removing the nodes $\alpha_2$, $\alpha_7$ and 
$\alpha_8$ from the extended Dynkin diagram 
in Fig.~\ref{fig:e8dynkin221}. The SU(2)$_2$ and SU(2)$_1$ are 
generated by $\alpha_1$ and $-\theta$, respectively. 
Their SU(2) bundles in Heterotic compactification corresponds to 
2-fold covers of the data $(\vec{\zeta}^2(\vec{y}),
\vec{\zeta}^1(\vec{y})+\vec{\zeta}^2(\vec{y}))$---a doublet of 
the Weyl group $\mathfrak{S}^{(2)}_2$ of SU(2)$_2$---and 
($\vec{\zeta}^7(\vec{y})$, 
$\vec{\zeta}^7(\vec{y})+\vec{\zeta}^0(\vec{y})$)---a doublet of the 
Weyl group $\mathfrak{S}^{(1)}_2$ of SU(2)$_1$. 
The remaining $\vec{\zeta}^8$ is determined from the four 
others through (\ref{eq:HKdata-8}) and the cycle $C_8$ is 
topologically equivalent to a certain linear combination 
of $C_1$, $C_2$, $C_7$ and $C_{-\theta}$. We reinterpret the 
geometry by thinking of another $\mathfrak{S}^{(1),(2)}_2$-singlet 
$C_8 + (2C_7 + C_{-\theta})=-(2C_8+2C_1+4C_2)$ instead of $C_8$ 
because this choice of the 2-cycle will shortly turn out to be 
more convenient. The data are glued between two adjacent patches 
through 
\begin{equation}
 \left(\begin{array}{c}
  -(2\vec{\zeta}^8+2\vec{\zeta}^1+4\vec{\zeta}^2) \\ 
 \hline \vec{\zeta}^7 \\  \vec{\zeta}^7+\vec{\zeta}^0 \\
 \hline  -(\vec{\zeta}^1+\vec{\zeta}^2) \\  -\vec{\zeta}^2
       \end{array}\right)_\alpha = 
 \left( \begin{array}{c|c|c}
  1 & & \\ \hline & \mathfrak{S}^{(1)}_2 & \\ 
  \hline
   & & \mathfrak{S}^{(2)}_2
	\end{array}\right)
 \left(\begin{array}{c}
  -(2\vec{\zeta}^8+2\vec{\zeta}^1+4\vec{\zeta}^2) \\ 
 \hline \vec{\zeta}^7 \\  \vec{\zeta}^7+\vec{\zeta}^0 \\
 \hline  -(\vec{\zeta}^1+\vec{\zeta}^2) \\  -\vec{\zeta}^2
       \end{array}\right)_\beta. 
\label{eq:221glue}
\end{equation}

\begin{figure}[t]
\begin{center}
\includegraphics[width=.5\linewidth]{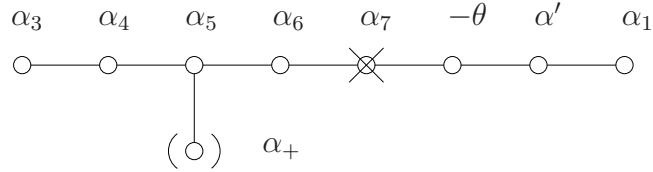}  
\begin{picture}(0,0)
 \put(-240,55){$\alpha_3$}
 \put(-207,55){$\alpha_4$}
 \put(-174,55){$\alpha_5$}
 \put(-141,55){$\alpha_6$}
 \put(-108,55){$\alpha_7$}
 \put(-75,55){$-\theta$}
 \put(-42,55){$\alpha'$}
 \put(-9,55){$\alpha_1$}
 \put(-145,7){$\alpha_+$}
\end{picture}
\caption{\label{fig:e8dynkin41}
The extended Dynkin diagram of $E_8$ (II), and its maximal subalgebra 
$\mathfrak{so}(10)+\mathfrak{su}(4)$. 
$\alpha_+ \equiv L_0-(L_3+L_4+L_5)$.}
\end{center}
\end{figure}

The SU(4)$\times$U(1)$_\chi$ bundle compactification of Heterotic theory 
corresponds to the symmetry breaking pattern of removing $\alpha_7$
in the extended Dynkin diagram Fig.~\ref{fig:e8dynkin41}. The SU(4) 
structure group of the vector bundle is generated by the simple roots 
$-\theta$, $\alpha'$ and $\alpha_1$, where a root $\alpha'$ is defined 
in the appendix \ref{sec:note}. As the two rank-2 vector bundles 
${\bf 2}_1$ and ${\bf 2}_2$ are extended to a rank-4 bundle $U_4$ 
in Heterotic description, the two 2-fold cover data, or two doublets 
of 2-cycles $(C_7,C_7+C_{-\theta})$ and $(C_2,C_1+C_2)$ should be 
paired up into a quartet in M-theory description; indeed, the 
quartet should be $(C_7,C_7+C_{-\theta},C_7+C_{-\theta}+C_{\alpha'},
C_7+C_{-\theta}+C_{\alpha'}+C_1)$ and since 
\begin{equation}
 \alpha_7 + (-\theta) +\alpha' \equiv -(\alpha_1+\alpha_2) 
\end{equation}
mod $\alpha_{3,4,5,6}$, it is the quartet 
$(C_7,C_7+C_{-\theta},-(C_1+C_2),-C_2)$ that mix together over the 
entire base manifold $Q$. The data $(\vec{\zeta}^7,
\vec{\zeta}^7+\vec{\zeta}^{-\theta},-(\vec{\zeta}^1+\vec{\zeta}^2),
-\vec{\zeta}^2)$ is a 4-fold cover of $Q$. Note that the other 2-cycle 
$-(2C_8+2C_1+4C_2)$ is $C_7 - C_{\alpha^{''}}$, where the root 
$\alpha^{''}$ is defined in the appendix \ref{sec:note}.
The transition function in (\ref{eq:221glue}) is enlarged into 
${\bf 1} \times \mathfrak{S}_4$.

\begin{figure}[t]
\begin{center}
\includegraphics[width=.5\linewidth]{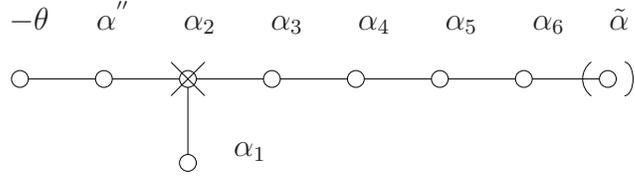}  
\begin{picture}(0,0)
 \put(-240,55){$-\theta$}
 \put(-207,55){$\alpha^{''}$}
 \put(-174,55){$\alpha_2$}
 \put(-141,55){$\alpha_3$}
 \put(-108,55){$\alpha_4$}
 \put(-75,55){$\alpha_5$}
 \put(-42,55){$\alpha_6$}
 \put(-13,55){$\tilde{\alpha}$}
 \put(-155,7){$\alpha_1$}
\end{picture}
\caption{\label{fig:e8dynkin32}
Extended Dynkin diagram of $E_8$ (III), and its maximal subalgebra 
$\mathfrak{su}(3)_2 + \mathfrak{su}(2)_2+\mathfrak{su}(6)_2$. 
$\tilde{\alpha} \equiv 2L_0 - (L_1 +\cdots + L_6)$. }
\end{center}
\end{figure}

The SU(3)$_2 \times$SU(2)$_2 \times$U(1)$_{\tilde{q}_7}$ bundle 
compactification of Heterotic theory corresponds to the symmetry 
breaking pattern of removing the node $\alpha_2$ in the extended 
Dynkin diagram in Fig.~\ref{fig:e8dynkin32}. $\mathfrak{su}(3)_2$ is 
generated by the roots $\alpha^{''}$ and $-\theta$, and 
$\mathfrak{su}(2)_2$ by $\alpha_1$ (see the appendix \ref{sec:note}). 
The replacing the bundle ${\bf 1}\oplus {\bf 2}_1$ by $U_3$ 
[(\ref{eq:221bdle}) by (\ref{eq:32bdle})] in the Heterotic side 
corresponds to mixing up the 2-cycle $-(2C_8+2C_1+4C_2)\equiv 
C_7-C_{\alpha^{''}}$ with the $\mathfrak{S}^{(1)}_2$ doublet 
$(C_7,C_7+C_{-\theta})$ globally on $Q$ to be a triplet of the Weyl 
group $\mathfrak{S}_3$ generated by $\alpha^{''}$ and $-\theta$. 
The data 
$(-(2\vec{\zeta}^8+2\vec{\zeta}^1+4\vec{\zeta}^2),\vec{\zeta}^7,
\vec{\zeta}^7+\vec{\zeta}^{-\theta})$ gives a 3-fold cover of $Q$.
The $\mathfrak{S}^{(2)}_2$-doublet $(-(C_1+C_2),-C_2)$ remains intact 
in the process of lifting the ``2+2+1 bundles'' to ``3+2 bundles.''
The transition function in (\ref{eq:221glue}) is enlarged into 
$\mathfrak{S}_3 \times \mathfrak{S}_2^{(2)}$.

The quintet data $(-(2\vec{\zeta}^8+2\vec{\zeta}^1+4\vec{\zeta}^2),
\vec{\zeta}^7,\vec{\zeta}^7+\vec{\zeta}^{-\theta},
-(\vec{\zeta}^1+\vec{\zeta}^2),-\vec{\zeta}^2)$ are mutually separated 
by
\begin{eqnarray}
 \vec{\zeta}^{\alpha^{''}} & \equiv & 
   2\vec{\zeta}^1 + 4\vec{\zeta}^2 + \vec{\zeta}^7 +2\vec{\zeta}^8, \\ 
 &  \vec{\zeta}^{-\theta},  &  \\
 \vec{\zeta}^{\alpha^{'}} & \equiv & 
    \vec{\zeta}^1 + 3\vec{\zeta}^2 + \vec{\zeta}^7 +3\vec{\zeta}^8, \\ 
 &  \vec{\zeta}^{1}, &  \\
 \vec{\zeta}^{\alpha^{'''}} & \equiv & 
   -( 2\vec{\zeta}^1 + 3\vec{\zeta}^2 +2\vec{\zeta}^8). 
\end{eqnarray}
Since these separations are defined in a cyclic order, they satisfy
a constraint 
\begin{equation}
 \vec{\zeta}^{\alpha^{''}}+\vec{\zeta}^{-\theta}+\vec{\zeta}^{\alpha'}
 +\vec{\zeta}^1 + \vec{\zeta}^{\alpha^{'''}} = 0,
\end{equation}
which is equivalent to the constraint (\ref{eq:HKdata-8}), and is also 
in the form of the traceless condition for the $A_4$ type ALE space 
\cite{DM}, corresponding to the SU(5) structure group of the Heterotic 
description. 
Irreducible vector bundles with full SU(5) structure group correspond 
to irreducible 5-fold cover of $Q$ given by the quintet data 
in the M-theory description. Following the idea in section 
\ref{sec:BottomUp}, we suggest that the absence of dimension 4 proton 
decay is an indication that the 5-fold cover is not irreducible 
at some level of approximation; it is split up into irreducible 
components of either (2-fold, 2-fold, single) covers, (4-fold, single) 
covers or (3-fold, 2-fold) covers.

We have seen at the end of section \ref{ssec:M6} that the diagonal 
entries of up-type quark Yukawa matrix tend to be suppressed. The 
argument there does not depend specifically on the fact that 
$(\vec{\zeta}^2,\vec{\zeta}^1+\vec{\zeta}^2)$ is a \underline{2-fold} 
cover; rather the argument is more general, and is valid also in the 
case of 4-fold cover. Thus, the suppressed diagonal entries can be 
regarded as generic predictions of M-theory vacua with SU(5)$_{\rm GUT}$ 
unification.

\section{F-theory Vacua}
\label{sec:F}

The Heterotic theory compactified on $T^2$ has the same moduli space
as that of the F-theory on elliptic K3 \cite{evidence}. 
Likewise, the Heterotic theory on elliptic Calabi--Yau 3-fold 
\begin{equation}
 \pi_Z: Z \rightarrow B 
\label{eq:Het-T2}
\end{equation}
has the same moduli space\footnote{The overlap between Heterotic vacua 
and F-theory vacua is more than this. See, for example, 
\cite{MV2,BM,Rajesh}.} 
as that of the F-theory on elliptic Calabi--Yau 4-fold 
\begin{equation}
 \pi_X: X \rightarrow B
\label{eq:F-T2}' 
\end{equation}
whose base manifold $B'$ is a $\P^1$-fibration on the common $B$, 
\begin{equation}
\pi^{''}: B' \rightarrow B, 
\label{eq:F-P1}
\end{equation}
so that 
\begin{equation}
 \pi':X \rightarrow B
\label{eq:F-K3}
\end{equation}
is a $K3$-fibration \cite{MV1,MV2,FMW,Het-F-4D}. 
 
For vacua that fall into the category of Heterotic--F-theory dual, 
some region of the moduli space is described better by the Heterotic 
theory but other regions are better described by F-theory.
Furthermore, there will be vacua that do not fall into the shared 
moduli space, and in particular, some of F-theory compactification 
on a non-K3-fibred elliptic Calabi--Yau 4-fold may not have Heterotic 
descriptions. Therefore, in this section, we take a step toward 
translating what has been discussed in earlier sections into F-theory 
language, so that we get prepared for exploring vacua that are 
well-described by the F-theory.

\subsection{Trilinear Yukawa Couplings}
\label{ssec:F-Yukawa}

One way to see the Heterotic--F-theory duality is through 
the del Pezzo fibration. Vector bundles on an elliptic Calabi--Yau 
3-fold $\pi: Z \rightarrow B$ with the structure group in $E_8$ are
described\footnote{To be more precise, not all the information is
encoded in the del Pezzo fibration. Discrete topological choices that 
changes the chirality are one example.} by $dP_8$-fibration 
$W \rightarrow B$. One point in $dP_8$ fibre is blown up to be 
$dP_9$, and the $dP_9$ fibres for the visible and hidden sectors 
add up to form the elliptic $K3$ fibration $\pi': X \rightarrow B$ 
of the F-theory \cite{FMW}. 

When we discuss Yukawa couplings in the visible sector, we do not 
pay attention to the del Pezzo fibration for the hidden sector.
The commutation relation of $E_8$ that determines Yang--Mills 
interaction of the Heterotic theory is now encoded in the intersection 
form of the 2-cycles of the $dP_8$ (or $dP_9$). Those surfaces contain 
the chains of 2-cycles that we discussed in section \ref{sec:M} and 
the identifications of particles and interactions there can also be 
valid for the F-theory. Thus, the $E_8$ version of the discussion 
around (\ref{eq:M7-YukawaU})--(\ref{eq:M7-dim5}) is also valid in 
the F-theory vacua. As is clear from the discussion in section 
\ref{sec:M}, the existence of trilinear Yukawa couplings and absence
of dangerous proton decay operators can be discussed in that way, 
as long as the local geometry in $X$ around the holomorphic 4-cycle 
of the visible sector gauge group has the fibration of eight 
(or seven) 2-cycles with the right intersection forms. The validity 
of this way of thinking is not limited to F-theory vacua that have 
Heterotic dual.

\subsection{F-theory Dual of Heterotic 
$E_8 / \SU(5)_{\rm GUT} \times \vev{\SU(4) \times \U(1)_\chi}$ Model}
\label{ssec:F-41}

If one is only interested in knowing whether Yukawa couplings exist 
or proton decay operators are absent, there is no essential 
difference between the Heterotic, M and F-theory. Such aspects were 
determined by the underlying symmetry such as $E_7$ or $E_8$ and 
how it is broken. The above argument is fine for F-theory vacua.
But when one wants to extract more physical consequences of F-theory 
vacua, we need more detailed descriptions. For that purpose, we begin 
with translating the Heterotic models discussed in section \ref{sec:Het} 
into F-theory language through the Heterotic--F-theory duality.
Section \ref{ssec:F-41} is devoted to the translation of 
$E_8/ \SU(5)_{\rm GUT} \times \vev{\SU(4) \times \U(1)_\chi}$ model, and 
section \ref{ssec:F-32} to $E_8 / \SU(5)_{\rm GUT} \times \vev{\SU(3)_2 
\times \SU(2)_2 \times \U(1)_{\tilde{q}_7} }$ and 
$E_8 / \SU(5)_{\rm GUT} \times \vev{\SU(2)_1 \times \SU(2)_2 \times 
\U(1) \times \U(1)}$ models. Later in section \ref{ssec:F-gen}, 
we discussed how various descriptions obtained through translation 
can be generalized to F-theory vacua that may not have Heterotic dual.

\subsubsection{Matter Curves at the Reducible Limit}

An intriguing aspect of F-theory vacua is that charged matters are 
localized in complex curves in the internal space $B'$ \cite{FMW,6authors}.
When one wants to have better understanding of the pattern of Yukawa 
matrices, for instance, it is crucial to know the wavefunctions of 
the chiral zero modes. Thus the determination of the matter curve 
is an important issue. Although the localisation pattern of matters 
in SU(5)$_{\rm GUT}$ unbroken F-theory vacua is well-known in the 
literature, we repeat this separately for the case where the structure 
group of the bundle is reduced to SU(4)$\times$U(1)$_\chi$ subgroup 
of SU(5). Although the unbroken gauge symmetry is the same 
SU(5)$_{\rm GUT}$, we see the matter curves are not of the generic 
SU(5) unified theories.

The rank-$n$ vector bundles on an elliptic Calabi--Yau 3-fold 
$\pi: Z \rightarrow B$ for Heterotic compactification are constructed 
out of a set of data $(C,{\cal N})$, where $C \subset Z$ is a $n$-fold 
spectral cover 
$\pi_C: C\rightarrow B$ and ${\cal N}$ is a line bundle on $C$.
A bundle for $(C,{\cal N})$ is given by 
\begin{equation}
V=p_{2*} (p^*_1({\cal N})\otimes {\cal P}_B),
\label{eq:FM-transf} 
\end{equation}
where $p_{1,2}$ are maps associated with a fibre product 
\begin{equation}
 \begin{array}{ccc}
 & C \times_B Z &  \\
 p_1 \swarrow & & \searrow p_2  \\
 C & & Z  \\
 \pi_C \searrow & & \swarrow \pi  \\ 
 & B & 
\label{eq:fib-prod}
 \end{array},
\end{equation}
$q \equiv \pi_C \; \circ \; p_1 = \pi \; \circ \; p_2$, and 
${\cal P}_B$ is the Poincare line bundle 
${\cal O}_{C \times_B Z}(\Delta - \sigma_1 -  \sigma_2 + q^* K_B)$
 \cite{FMW}. 
The first Chern class of the vector bundle $V$ is given by 
\begin{equation}
 c_1(V)  =  \pi_{C*}\left( c_1({\cal N}) -  \frac{1}{2} r \right)
\label{eq:c1}
\end{equation}
where $r \equiv \omega_{C/B} \equiv K_C - \pi_C^* K_B$ is the 
ramification divisor on $C$ of $\pi_C: C \rightarrow B$ 
\cite{Hartshorne}, and $c_1(V)$ is a 2-form on the base 2-fold $B$. 
When one thinks of a bundle with the structure group U($n$), rather 
than SU($n$), (\ref{eq:c1}) does not have to be zero \cite{U(n)}.
The second Chern character is 
\begin{equation}
 {\rm ch}_2(V) = - \sigma \cdot \eta + \pi^* \omega,
\end{equation}
where $\sigma$ is a section $\sigma: B \rightarrow Z$ of the elliptic
fibration, $\eta$ is a divisor on $B$ that determines the $n$-fold 
spectral cover $C \sim n \sigma + \eta$, and 
$\omega$ is some 2-form on $B$.

When the line bundle $L$ of (\ref{eq:41bdle}) on an elliptic 
Calabi--Yau 3-fold is given by spectral data $(C_1,{\cal N}_1)$,
the spectral cover $C_1$ is just $\sigma$. Indeed, if it had a 
component of a divisor $\eta_1$ on $B$, then ch$_2(L)$ of the line 
bundle $L$ would have a component $\sigma \cdot \eta_1$ that is not 
strictly on the base $B$ despite $c_1(L)$ strictly on the base \cite{U(n)}. 
Thus, $C_1 \sim \sigma$. The line bundle ${\cal N}_1$ on 
$\sigma \sim B$ is identified with the line bundle $L$ itself.
We abuse the notation a little bit hereafter, and $L$ stands for 
both the line bundle on $B$ and $Z$.

Let the bundle $U_4$ be given by the data $(C_4,{\cal N}_4)$, and 
$C_4 \sim 4\sigma + \eta_4$. 
Then the data for the bundles in various representations are given by 
\begin{eqnarray}
 U_4 & {\rm for~}{\bf 10} & (4\sigma + \eta_4,{\cal N}_4), \\
 L   & {\bf 10}' {\rm ~absent~in~LE.} & (\sigma,L), \\
 \wedge^2 U_4 & {\rm for~}\bar{H}(\bar{\bf 5}) {\rm ~and~} 
   H({\bf 5})^\dagger & (6\sigma+2\eta_4,{\cal N}_{\wedge^2 U_4}), 
    \label{eq:H-cover} \\
 U_4 \otimes L & {\rm for~}\bar{\bf 5} & 
   (4\sigma+\eta_4,{\cal N}_4 \otimes L), \\
 U_4 \otimes L^{-1} & {\rm for~}\bar{N} & 
   (4\sigma+\eta_4,{\cal N}_4 \otimes L^{-1}).
\end{eqnarray}
Equation (\ref{eq:H-cover}) is due to 
\begin{equation}
C_{\wedge^2 V_r} \sim \frac{r(r-1)}{2} \sigma + (r-2) \eta_r
\label{eq:asym-cover} 
\end{equation}
for the anti-symmetric representation of a rank-$r$ bundle $V_r$ 
with a spectral surface $C_r \sim r \sigma + \eta_r$ \cite{Penn5}. 
Note that the sum of the spectral cover of $U_4$ and $L$ is 
$C_{V_5} \sim 5\sigma+\eta_4$, and that of $\wedge^2 U_4$ and 
$U_4 \otimes L$ is $C_{\wedge^2 V_5} \sim 10\sigma + 3\eta_4$. 
The reducible limit of the vector bundle (\ref{eq:41bdle}) corresponds 
to the case where the spectral surfaces of the SU(5) bundle become 
reducible, and $C_5 \sim 5K_B +\eta_4$ splits into irreducible pieces 
$C_4\sim 4K_B + \eta_4$ and $C_1 \sim \sigma$ \cite{ovrutpark}.

In Heterotic theory, the cohomology of the vector bundles, and hence 
massless modes, can be calculated by the spectral sequence, 
evaluating the cohomology in the elliptic fibre direction first, and 
in the base manifold later.
For a bundle $\rho(V)$, the cohomology vanishes only over a curve 
$\bar{c}_{\rho(V)} \equiv C_{\rho(V)} 
\cdot \sigma$ on the base $B$ \cite{FMW,DI}.\footnote{This 
localization is is not an artifact in calculation, but is indeed physical,
when the volume of the $T^2$ fibre in the Heterotic compactification 
is not large. Such region of the Heterotic moduli space is better 
described by the F-theory, where matters are expected to localised.} 
Thus, the massless matters are localised on curves on $B$ given by  
\begin{eqnarray}
 \bar{c}_{\bf 10} = \bar{c}_{\bar{\bf 5}} = \bar{c}_{\bar{N}} & = & 
   4 K_B + \eta_4 \equiv \bar{c}_{\bf 16},\label{eq:curve-SO(10)-16}\\
 \bar{c}_{H,\bar{H}} & = & 6 K_B + 2\eta_4 + \bar{c}_{\bf vec.},
                                          \label{eq:curve-SO(10)-10} \\
 \left(\bar{c}_{{\bf 10}'} \right. & = & 
 \left. K_B, \qquad {\rm spread~out~in~}B \right), 
\label{eq:curve-SO(10)-root}
\end{eqnarray}
where $\sigma \cdot \sigma \sim \sigma \cdot K_B$ \cite{FMW} was used.

In equation (\ref{eq:curve-SO(10)-root}) we see that the fields $\bar{5}$ and 
$H$ have different localisation. Therefore  the argument of the existence of the 
Yukawa couplings and absence of the proton decay holds well to F-theory picture.
 
The matter curves $\bar{c}_{\bf 10} = 5K_B + \eta_4$ for 
SU(5)$_{\rm GUT}$-{\bf 10} representation and 
$\bar{c}_{\bar{\bf 5}}=10K_B+3\eta_4$ for 
SU(5)$_{\rm GUT}$-$\bar{\bf 5}$ representation split into irreducible 
pieces. Although $(Q,\bar{U},\bar{E})$, $(\bar{D},L)$ 
and $\bar{N}$ originate from different irreducible bundles, their 
matter curves are the same, as in the case of SU(4)-bundle 
compactification with unbroken SO(10) symmetry. 
This is because $L$ and det$U_4$ are line bundles only on the base 
manifold $B$ and the only source of the distinction among {\bf 10}, 
$\bar{\bf 5}$ and $\bar{N}$ representations is the twist of those two 
bundles. Their effects come in only on the matter curves 
$\bar{c}_{\bf 16}$ and not at the level of changing the spectral surfaces 
or matter curves themselves.

In the F-theory dual of these Heterotic vacua obtained in  
(\ref{eq:F-T2})--(\ref{eq:F-K3}), and the $\P^1$ fibration 
(\ref{eq:F-P1}) is such that the $z=0$ section of (\ref{eq:F-P1}) 
$\Sigma_0 \subset B'$ ($\Sigma_0$ is isomorphic to $B$) satisfies 
\cite{MV1,MV2,FMW,Het-F-4D,Rajesh}:
\begin{equation}
 \Sigma_0 \cdot \Sigma_0 \sim (6K_B + \eta_4). 
\label{eq:old-def}
\end{equation}
The divisor $\Sigma_0$ of $B'$ is where the unbroken SU(5)$_{\rm GUT}$ 
gauge field propagates in F-theory. Over $\Sigma_0$ there are loci of 
enhanced singularity and enhanced gauge symmetry, and this is where 
matters are localised, and they are the same as the matter curves 
obtained in the Heterotic argument above \cite{FMW,6authors}. 
The reducible limit of vector bundles in Heterotic theory correspond 
to taking some coefficients of the polynomial defining (\ref{eq:F-T2}) 
to zero. The coefficients for SU(5) bundles are given by global 
holomorphic sections $g \in H^0(B,{\cal O}(\eta))$, 
$f \in H^0(B,{\cal O}(2K_B + \eta))$, 
$q \in H^0(B,{\cal O}(3K_B + \eta))$, 
$H \in H^0(B,{\cal O}(4K_B + \eta))$ and 
$h \in H^0(B,{\cal O}(5K_B + \eta))$ \cite{6authors}, and 
the dual of SU(4)$\times$U(1)$_\chi$ bundle with 
$m =0$ in (\ref{eq:L-fibre}) corresponds to taking $h=0$.
The loci of enhanced singularity $\bar{c}_{\bf 10}$ and
$\bar{c}_{\bar{\bf 5}}$ rearrange themselves in this limit, 
so that they are grouped as $\bar{c}_{\bf 16}$, $\bar{c}_{\bf vec.}$ 
while the irreducible piece (\ref{eq:curve-SO(10)-root}) supplements 
the roots in $\mathfrak{so}(10)/\mathfrak{su}(5)$, so that the locus 
$\Sigma_0$ support SO(10) gauge group.

The rearrangement argument of the matter curves still misses one piece:
where has the SU(5)$_{\rm GUT}$-singlet part of the SO(10)-{\bf 16} 
representation come from? This piece is for the right-handed 
neutrinos, and hence it should be related to the vector bundle moduli 
of SU(5) bundle that are set to zero in the reducible limit---in
Heterotic language. Thus, it corresponds to 
$h \in H^0(B,{\cal O}(5K_B+\eta_4))$ in F-theory language. Those 
degrees of freedom are essentially localised on a curve defined 
by the intersection $C_4 \cdot C_1 \sim C_4 \cdot \sigma$, which 
is also the definition of the $\bar{c}_{\bf 16}$. This is because 
$h$ describes the reconnection of two irreducible pieces $C_4$ and 
$C_1$. The other way to see the localisation on $\bar{c}_{\bf 16}$ 
is to use the exact sequence 
\begin{equation}
 0 \rightarrow {\cal O}_B(C_5-C_4) \rightarrow {\cal O}_B(C_5) 
   \rightarrow {\cal O}_{\sigma \cdot C_4}(C_5) \rightarrow 0.
\end{equation}
Note that $C_5-C_4 \sim \sigma \sim K_B$ on $B$. 
Since $h^0(B,{\cal O}(K_B))=h^2(B,{\cal O}_B)$ and 
$h^1(B,{\cal O}(K_B))=h^1(B,{\cal O}_B)$ are zero in elliptic 
fibration (\ref{eq:Het-T2}) that leaves only ${\cal N}=1$ 
supersymmetry \cite{Sethi} we see that  
\begin{equation}
 H^0(B,{\cal O}(5K_B + \eta_4)) \simeq 
 H^0(\bar{c}_{\bf 16},{\cal O}(5K_B + \eta_4)).
\end{equation}
Thus, $h \in H^0(B,{\cal O}(5K_B + \eta_4))$ is associated with 
a global holomorphic section of a sheaf on $\bar{\bf c}_{\bf 16}$. 

U(1)$_\chi$ bundle $L$ is turned on on the discriminant locus 
$\Sigma_0 \simeq B$, and the SO(10) gauge symmetry is broken down 
to SU(5)$_{\rm GUT}$. The Fayet--Iliopoulos parameter due to U(1) 
flux itself is calculated in \cite{Louis} in Type IIB string theory 
orientifold compactification, and although F-theory is not exactly 
the same as the Type IIB theory, it may be roughly the same. In Type 
IIB calculation \cite{Louis} it is 
\begin{equation}
\xi \approx l_s^{-4} \int c_1(L) \wedge J
\label{eq:FI-IIB}
\end{equation}
after restoring proper dimensionality; numerical factors being ignored. 
This Fayet--Iliopoulos parameter may vanish when $c_1(L)$ is orthogonal 
to the K\"{a}hler form $J$ on the discriminant locus $\Sigma_0 \simeq B$. 
In this case, the global U(1)$_\chi$ symmetry is left unbroken, and 
the U(1)$_\chi$ gauge boson is massive, due to the generalized 
Green-Schwarz interactions.
If this Fayet--Iliopoulos parameter is non-zero, the U(1)$_\chi$-symmetry 
breaking phase transition is triggered and this symmetry is broken.
Because of the same reason as in section \ref{sec:Het}, either one of 
the sign of $\int_{\Sigma_0} c_1(L) \wedge J$ is phenomenologically 
acceptable, because otherwise $\bar{\bf 5}$ multiplets are no longer 
distinguished from the $\bar{H}(\bar{\bf 5})$ multiplet, and the 
dimension-4 proton decay operators are generated. 

\subsubsection{Chiral Matter}
\label{sssec:chiral}

The massless modes $H^1(Z;\rho(V))$ from vector bundles $\rho(V)$ 
of Heterotic theory can be obtained by cohomology on matter curves 
$\bar{c}_{\rho(V)}$, when we consider elliptic fibred Calabi--Yau 
3-fold $Z \rightarrow B$ \cite{FMW,DI}:
\begin{equation}
 H^1(Z;\rho(V)) \simeq H^0(\bar{c}_{\rho(V)};{\cal F}_{\rho(V)}), 
\end{equation}
where ${\cal F}$ is a sheaf on the matter curve $\bar{c}_{\rho(V)}$ 
such that $R^1\pi_* \rho(V) = i_{\bar{c}_{\rho(V)} *} {\cal F}$ for 
$i_{\bar{c}_{\rho(V)}} : \bar{c}_{\rho(V)} \rightarrow B$; 
in particular, it is given in terms of spectral cover data by
\begin{equation}
 {\cal F} = j^* {\cal N}_{\rho(V)} \otimes i^* K_B,
\label{eq:sheaf-F}
\end{equation}
where $j : \bar{c}_{\rho(V)} \rightarrow C_{\rho(V)}$. The net 
chirality (\ref{eq:chi-10A})--(\ref{eq:chi-N}) can be 
expressed as Euler characteristic on the matter curves as well 
\cite{DI}. 
\begin{eqnarray}
 \chi(R_5) = - \chi(Z,\rho(V))
 & = &  h^1(Z,\rho(V)) - h^2(Z,\rho(V)),             \nonumber \\
 & = &  h^0(\bar{c}_{\rho(V)},{\cal F}) 
       - h^1(\bar{c}_{\rho(V)},{\cal F}) 
       =  \chi(\bar{c}_{\rho(V)},{\cal F}). 
\label{eq:chirality-curve}
\end{eqnarray}
All above is valid as long as $c_1(\rho(V))$ is trivial in the fibre 
direction, and hence applicable to all the irreducible bundles
considered in section \ref{ssec:Het41}.
The net chirality of matters can be calculated 
on the matter curves. Since they are localised on the matter curves 
in F-theory vacua, there should be such expressions formulated on the 
matter curves; the net chirality does not depend on the global geometry 
of the elliptic Calabi--Yau 4-fold $X$ but only on geometric 
information along the matter curves.

The net chirality of quarks, leptons and right-handed neutrinos are 
given by
\begin{eqnarray}
 \# \bar{N} - \# \overline{\bar{N}} & = & 
     h^0(\bar{c}_{\bf 16};
         {\cal O}(K_{\bar{c}_{\bf 16}})^{1/2} \otimes 
         {\cal L}_\gamma \otimes L^{-1})
   - h^0(\bar{c}_{\bf 16};
         {\cal O}(K_{\bar{c}_{\bf 16}})^{1/2} \otimes 
         {\cal L}_\gamma^{-1} \otimes L) \nonumber \\
  & = & \int_{\bar{c}_{\bf 16}} j^* \gamma - c_1(L), 
      \label{eq:chi-F-N} \\
 \# {\bf 10} - \# \overline{\bf 10} & = & 
     h^0(\bar{c}_{\bf 16};
         {\cal O}(K_{\bar{c}_{\bf 16}})^{1/2} \otimes 
         {\cal L}_\gamma )
   - h^0(\bar{c}_{\bf 16};
         {\cal O}(K_{\bar{c}_{\bf 16}})^{1/2} \otimes 
         {\cal L}_\gamma^{-1}) \nonumber \\
  & = & \int_{\bar{c}_{\bf 16}} j^* \gamma, 
      \label{eq:chi-F-10} \\
 \# \bar{\bf 5} - \# \overline{\bar{\bf 5}} & = & 
     h^0(\bar{c}_{\bf 16};
         {\cal O}(K_{\bar{c}_{\bf 16}})^{1/2} \otimes 
         {\cal L}_\gamma \otimes L)
   - h^0(\bar{c}_{\bf 16};
         {\cal O}(K_{\bar{c}_{\bf 16}})^{1/2} \otimes 
         {\cal L}_\gamma^{-1} \otimes L^{-1}) \nonumber \\
  & = & \int_{\bar{c}_{\bf 16}} j^* \gamma + c_1(L),  
       \label{eq:chi-F-5}
\end{eqnarray}
where a 2-form $\gamma$ on $C_4$ determines $c_1({\cal N}_4)$ by 
\begin{equation}
 c_1({\cal N}_4) = \frac{1}{2}r + \gamma,
\end{equation}
and ${\cal L}_\gamma$ is a line bundle on $\bar{c}_{\bf 16}$  
whose first Chern class is $j^* \gamma$. In derivation of 
(\ref{eq:chi-F-N})--(\ref{eq:chi-F-5}), we have used 
Hirzebruch--Riemann-Roch theorem \cite{Hirzebruch}\footnote{
One has to pay more attention to whether the theorem is applicable 
in each explicit example. Here, we only deal with situations that 
do not require special treatment.} and 
\begin{equation}
 \frac{1}{2} j^* \left(K_{C_{\rho(V)}} - \pi^* K_B \right) + i^* K_B 
= \frac{1}{2}
 K_{\bar{c}_{\rho(V)}} = -  {\rm td}_1(T\bar{c}_{\rho(V)}).
\label{eq:adj-2}
\end{equation}
One could also separate 
the det$U_4$ piece and set $j^*\gamma = j^* \gamma' -c_1(L)/4$; then 
\begin{equation}
 \chi = \int_{\bar{c}_{\bf 16}}\left(j^* \gamma' 
 + \frac{q_\chi}{4}c_1(L)\right).
\end{equation} 
$\gamma$ is translated into the primitive $G^{(2,2)}$ flux in F-theory
language \cite{CD}. 
The phenomenological constraint (\ref{eq:ZL}) reduces to\footnote{
Note that $c_2(TZ) = \sigma \cdot 12 c_1(TB) + \cdots$, where ellipses 
stand for the fibre class. $c_1(L)^3=0$ because $c_1(L)$ is purely 
on $B$. Note also that $c_2(U_4) = \sigma \cdot \eta_4 + \cdots$.}
$c_1(TB) \cdot c_1(L) = - K_B \cdot c_1(L) = 0$, and 
(\ref{eq:4L}) to $\eta_4 \cdot c_1(L) = 0$ in elliptic Calabi--Yau 
compactification with $m=0$ in (\ref{eq:L-fibre}). Thus, 
$c_1(L)$ does not contribute to the net chirality of any of quarks, 
leptons and right-handed neutrinos.

\subsubsection{Approach from Parabolic (Extension) Construction}
\label{sssec:parabora}

The spectral cover construction can only describe vector bundles 
with the first Chern class purely on the base manifold. This is 
reasonable because the $n$-point fibre of an $n$-fold spectral cover 
describes $n$ Wilson lines in the fibre direction, not a non-trivial 
field strength. 

Purely from the viewpoint of Heterotic compactification, however, 
there seems to be no problem in considering a line bundle $L$ whose 
first Chern class has a fibre component:
\begin{equation}
 c_1(L) = m \sigma + \pi^* \omega_1,
\label{eq:L-fibre}
\end{equation}
where $\omega_1$ is a divisor on $B$ and $m$ an integer, and indeed, 
such compactification was discussed in \cite{Blumenhagen,Blumenhagen2}.
Our discussion in section \ref{sec:Het} does not exclude this 
possibility, either. However, when we consider a small fibre limit, 
the tree-level Fayet--Iliopoulos parameter is approximated by 
\begin{equation}
 \xi_\chi \approx M_G^2 l_s^2 \frac{m \int_B J\wedge J}
{\int_Z J\wedge J \wedge J} \approx m M_G^2 
\left(\frac{l_s^2}{{\rm vol.} (T^2)} \right),
\end{equation}
and cannot vanish when $m \neq 0$. Either $\bar{N}$ or 
$\overline{\bar{N}}$ develops an expectation value, as we saw 
in section \ref{sssec:Het-anomalous}, and the vector bundle 
cannot remain reducible. 
Thus, in the small fibre limit, we cannot consider the reducible limit 
in Heterotic compactification if $m\neq 0$. In the construction of 
the moduli space of vector bundles in \cite{FMW}, such a reducible limit 
with $m \neq 0$ was cut out from the beginning. The moduli of spectral 
cover construction provides the description of vector bundles obtained 
after the U(1)$_\chi$ breaking phase transition \cite{FMW}.

The absence of dimension-4 proton decay requires that 
$\int_Z c_1(L) \wedge J \wedge J < 0$, or $\xi_\chi < 0$, 
assuming $N_{gen} > 0$. On elliptic-fibred Calabi--Yau 3-folds, 
this condition simplifies; $m < 0$ (or $m=0$ and $\xi < 0$ in 
(\ref{eq:FI-IIB})). 

Let us consider the simplest case, $m = -1$. In particular, 
when the rank-4 bundle $U_4$ is ${\cal W}_4 \otimes {\cal M}$ with 
${\cal W}_4$ constructed in \cite{FMW,FMW2} and a line bundle 
${\cal M}$ on $B$, and $L \simeq {\cal O}(-\sigma + 6K_B) \otimes {\cal
M}^{-4}$ the U(1)$_\chi$ breaking phase transition is just 
the extension of $U_4 \simeq {\cal W}_4 \otimes {\cal M}$ by 
${\cal W}_1 \otimes {\cal O}(6K_B) \otimes{\cal M}^{-4}$.
In the fibration structure of the vector bundle moduli \cite{FMW}
\begin{equation}
0 \rightarrow H^1(B;R^0\pi_* {\rm adj.}(V_5)) \rightarrow 
 H^1(Z;{\rm adj.}(V_5)) \rightarrow H^0(B;R^1 \pi_* {\rm adj.}(V_5)) 
\rightarrow 0,
\label{eq:bdle-moduli}
\end{equation}
we have 
\begin{eqnarray}
 H^0(B;R^1\pi_* {\rm adj.}(V_5)) & \simeq & 
 H^1(Z;\overline{U}_4 \otimes L), \label{eq:red-nbbar}\\
 H^1(B;R^0\pi_* {\rm adj.}(V_5)) & \simeq & 
 H^1(Z;U_4 \otimes L^{-1}). \label{eq:red-nbar}
 \end{eqnarray}
Since $U_4$ or ${\cal W}_4$ does not have a vector bundle moduli, by 
construction, all the moduli of spectral cover 
$H^0(B;R^1\pi_* {\rm adj.}(V_5))$ comes from the off-diagonal block, 
and they are the degree of freedom called anti-generation right-handed 
neutrinos $\overline{\bar{N}}$ in section \ref{ssec:Het41}. They are 
a part of complex structure moduli in F-theory description \cite{FMW}. 
Fibering upon this moduli are $H^1(Z;U_4 \otimes L^{-1})$, which are 
now right-handed neutrinos $\bar{N}$. They are identified with the 
value of Ramond--Ramond 3-form field integrated over (2,1) cycles of 
an elliptic Calabi--Yau 4-fold of F-theory compactification. 
Note that the sign of $m$ and hence the constraint from the 
dimension-4 proton decay determines which is the base and which 
is the fibre. Such an identification of the origin of right-handed 
neutrinos in F-theory terminology will help us understand the nature 
of these particles better. 

\subsection{F-theory Dual of $E_8 / \SU(5)_{\rm GUT}\times 
\vev{\SU(3)_2 \times \SU(2)_2 \times \U(1)_{\tilde{q}_7}}$ and 
$E_8 / \SU(5)_{\rm GUT}\times \vev{\SU(2)_1 \times \SU(2)_2 \times 
\U(1) \times \U(1)}$ Models}
\label{ssec:F-32}

Let us now turn our attention to F-theory dual description of Heterotic 
compactification with a 3+2 bundle (\ref{eq:32bdle}) or a 2+2+1 bundle 
(\ref{eq:221bdle}). The Heterotic configuration at the reducible limit 
can be easily mapped to the F-theory description, just as in the case 
of 4+1 bundle compactification. The reducible limit can be realized 
only when the first Chern classes of U(1)$_{\tilde{q}_7}$ and 
U(1)$_{\tilde{q}_6}$ do not have a fibre component, just like  
like in the 4+1 bundle compactification with $m=0$ in (\ref{eq:L-fibre}).

The discriminant locus $\Sigma_0$ for the unified gauge group is of 
$A_5$ type since SU(6)$_2$ is the maximal group that commutes 
with the non-Abelian part of the structure group SU(3)$_2\times$SU(2)$_2$. 
Likewise, the F-theory dual of 2+2+1 bundles (\ref{eq:221bdle}) has 
$D_6$ singular locus, since SO(12) $\subset E_8$ is the maximal group 
that commutes with SU(2)$_1 \times$SU(2)$_2$, the non-Abelian part 
of the structure group.
 
One can also see this by following the irreducible decomposition of 
spectral surfaces and matter curves. Let us begin with the 3+2 bundle 
compactification. Let the spectral data of vector bundles $U_3$ and 
$U_2$ be denoted by $(C_3,{\cal N}_3)$ and $(C_2,{\cal N}_2)$. 
The spectral surfaces 
\begin{equation}
C_3 \sim 3\sigma + \pi^* \eta_3, \qquad C_2 \sim 2 \sigma + \pi^* \eta_2,
\end{equation}
are irreducible components of $C_5 = C_3+C_2$. The spectral surface for the 
vector bundle $\wedge^2 V_5$, 
$C_{\wedge^2 V_5} \sim 10 \sigma + 3 \pi^*(\eta_3+\eta_2)$ is split into 
irreducible components 
\begin{equation}
C_{\wedge^2 U_3} \sim 3\sigma + \pi^* \eta_3, \quad 
C_{U_3 \otimes U_2} \sim 6 \sigma + 2 \pi^* \eta_3 + 3\pi^* \eta_2, \quad 
C_{\wedge^2 U_2} \sim \sigma.
\end{equation}
Here, the first and last relations are due to (\ref{eq:asym-cover}), and 
one in the middle is due to\footnote{Note that 
${\rm ch}(V_i) = r_i - \sigma \cdot \pi^* \eta_i + \cdots$, 
and ${\rm ch}(V_1 \otimes V_2) = {\rm ch}(V_1){\rm ch}(V_2) = 
r_1r_2 - \sigma \cdot (r_1 \eta_2 + r_2 \eta_1) + \cdots$.
, where 
Hence (\ref{eq:tensor}). An alternative intuitive way to understand 
(\ref{eq:tensor}) is as follows: when a spectral cover 
$C_{V_i}$ define $r_i$ points $\{p_{n_i}\}$ ($n_i=1,\cdots,r_i$) on 
a fibre $E_b \equiv \pi^{-1}(b)$ for $b \in B$, 
the spectral cover of the bundle $V_1 \otimes V_2$ is given by 
$\{ p_{n_1} \boxplus p_{n_2}\}$ ($n_1 = 1,\cdots,r_1$ and 
$n_2=1,\cdots,r_2$), where $\boxplus$ means addition under the group 
law of the elliptic curve $E_b$. The surface given by the $r_1r_2$ points 
in each fibre define a spectral surface given by (\ref{eq:tensor}).}  
\begin{equation}
 C_{V_1 \otimes V_2} \sim r_1 r_2 \sigma + r_1 \eta_2 + r_2 \eta_1
\label{eq:tensor}
\end{equation}
for vector bundles $V_i$ with spectral covers 
$C_{V_i} \sim r_i \sigma + \pi^* \eta_i$.

Chiral matters in the low-energy spectrum should arise from 
matter curves (assuming the identification A in Table~\ref{tab:ID-8B})
\begin{eqnarray}
 \bar{c}_{\bf 10} & = & 2K_B + \eta_2 = \bar{c}_{\wedge^3 {\bf 6}}, 
     \label{eq:curve-SU(6)-20} \\
 \bar{c}_{\bar{\bf 5}} = \bar{c}_{\bar{N}} & = & 6K_B + 2\eta_3 + 3 \eta_2 
   = \bar{c}_{\bf 6}, \label{eq:curve-SU(6)-6} \\
 \bar{c}_{\bar{H}(\bar{\bf 5})} (= \bar{c}_{{\bf 10}'}) & = & 
   3K_B + \eta_3 = \bar{c}_{\wedge^2 {\bf 6}}, \label{curve-SU(6)-15} \\  
 \bar{c}_{H({\bf 5})} & = & K_B \qquad ({\rm spread~out~in~} B).
      \label{eq:curve-SU(6)-root}
\end{eqnarray}
Various irreducible pieces of the matter curves are grouped into 
the matter curves of SU(6)$_2$ multiplets in \cite{6authors}.
The up-type Higgs multiplet $H({\bf 5})$ is a part of SU(6)$_2$-{\bf adj.} 
representation and propagates over the entire base manifold $B$.
The chiral ${\bf 10}$ representations come from SU(6)$_2$-$\wedge^3 {\bf 6}$
representation, $\bar{\bf 5}$ and $\bar{N}$ from SU(6)$_2$-$\bar{\bf 6}$ 
(and {\bf 6}) and the down-type Higgs multiplet $\bar{H}(\bar{\bf 5})$ from 
SU(6)$_2$-$\overline{\wedge^2 {\bf 6}}$. The Yukawa couplings arise 
from interactions\footnote{Note that the SU(6)$_2$ symmetry here is a
different subgroup from the SU(6)$_1$ in (\ref{eq:poten}).}
\begin{eqnarray}
W & \ni & \wedge^3 {\bf 6} \cdot {\bf adj.} \cdot \wedge^3 {\bf 6}
   + \bar{\bf 6} \cdot \wedge^3 {\bf 6}
      \cdot \overline{\wedge^2 {\bf 6}}
   + \bar{\bf 6} \cdot {\bf adj.} \cdot {\bf 6}, \\
\left[ \right. & \rightarrow & \left. \quad {\bf 10} \cdot H \cdot {\bf 10} 
\quad + \bar{\bf 5} \cdot {\bf 10} \cdot \bar{H}
\quad + \bar{\bf 5} \cdot H \cdot \bar{N}, \right] \\
\left[ \right. & \rightarrow & \left. \quad {\bf 10} \cdot H \cdot {\bf 10} 
\quad + \bar{H} \cdot {\bf 10} \cdot \bar{\bf 5}
\quad + \bar{H} \cdot H \cdot S. \right]
 \label{eq:F32-Yukawa}
\end{eqnarray}
Since the {\bf 10} representations and $\bar{\bf 5}$ (and $\bar{N}$) 
are localised on separate matter curves, there is no reason to expect 
that the family structures of those representations are the same \cite{WY}. 
The up-type Higgs multiplet propagates in the bulk of the discriminant 
locus $\Sigma_0$.
The second term, the Yukawa couplings for down-type quarks and charged 
leptons, involves three different matter curves, but in general, three 
curves on a complex 2-fold $\Sigma_0$ do not necessarily meet at 
one point. Thus, these Yukawa couplings are generated by a membrane 
spanning non-locally, and may be a little small. The masses 
of bottom quark and tau lepton are way below the electroweak scale, 
and their Yukawa couplings may be small, indeed.

We have already seen that the right-handed neutrinos in the 
identification A in Table \ref{tab:ID-8B} (and singlet chiral 
multiplets in the identification B) are localised on a matter curve 
$6K_B + 2\eta_3 + 3 \eta_2$. The other way to see this is to remind 
ourselves that these fields describe the deformation of the vector 
bundle from the reducible limit (\ref{eq:32bdle}).  
Using an exact sequence 
\begin{equation}
 0 \rightarrow {\cal O}_{X_1}(-X_2) \rightarrow {\cal O}_X 
   \rightarrow {\cal O}_{X_2} \rightarrow 0 
\end{equation}
for $X= X_1 \cup X_2$ with irreducible components $X_1$ and $X_2$, 
substituting $X_1 \rightarrow C_2$, $X_2 \rightarrow C_3$, and 
tensoring ${\cal O}(C_5)$, we see that 
\begin{equation}
 0 \rightarrow {\cal O}_{C_2}(C_2) \rightarrow {\cal O}_{C_5}(C_5)
   \rightarrow {\cal O}_{C_3}(C_5) \rightarrow 0
\end{equation}
is also exact. Thus, the deformation of the spectral surface $C_5$, 
$H^0(C_5;{\cal O}(C_5))$, is related to the deformation of $C_2$ by
\begin{equation}
 0 \rightarrow H^0(C_2;{\cal O}(C_2)) 
   \rightarrow H^0(C_5;{\cal O}(C_5)) 
   \rightarrow H^0(C_3;{\cal O}(C_5)).
\end{equation}
$H^0(C_3;{\cal O}(C_5))$ is further related to the deformation of the 
spectral surface $C_3$ by another exact sequence 
\begin{equation}
0 \rightarrow {\cal O}_{C_3}(C_5-C_2)
  \rightarrow {\cal O}_{C_3}(C_5)
  \rightarrow {\cal O}_{C_3 \cdot C_2}(C_5).
\end{equation}
It follows that 
\begin{equation}
0 \rightarrow H^0(C_3;{\cal O}(C_3)) 
  \rightarrow H^0(C_3;{\cal O}(C_5))
  \rightarrow H^0(C_3 \cdot C_2;{\cal O}(C_5)) 
\end{equation}
is also exact. Thus, the deformation of the spectral surface $C_5$ 
is decomposed into three pieces. Two of them are the deformation of 
the spectral surfaces of the bundle $U_2$ and $U_3$, 
$H^0(C_2;N_{C_2|Z})$ and $H^0(C_3;N_{C_3|Z})$. The other one 
$H^0(C_2 \cdot C_3;{\cal O}(C_5))$ describes the deviation from 
the reducible limit and is localised at the intersection of 
$C_2 \cdot C_3$. Its projection on $B$ is $6K_B + 2\eta_3 + 3\eta_2$, 
which is the same as $\bar{c}_{\bar{N}}$ determined above.

In the 2+2+1 bundle compactification of the Heterotic theory, the spectral 
surface $C_3 \sim 3\sigma + \pi^* \eta_3$ further splits into irreducible 
pieces $\sigma$ and $2\sigma+ \pi^* \eta_3$. The matter curves rearrange 
themselves accordingly, and become 
\begin{eqnarray}
 \bar{c}_{\bf 10} = \bar{c}_{\bar{\bf 5}} = \bar{c}_{\bar{N}} & = & 
   2K_B+ \eta_2 = \bar{c}_{{\bf 32}'}, \label{eq:curve-SO(12)-32p} \\
 ({\rm absent~in~LE.}) & & 2 (2K_B + \eta_3 + \eta_2) = \bar{c}_{\bf vec.}, 
  \label{eq:curve-SO(12)-12} \\
 ({\rm absent~in~LE.}) & & 2K_B + \eta_3 = \bar{c}_{\bf 32}, 
  \label{eq:curve-SO(12)-32} \\
 \bar{c}_{H({\bf 5})} = \bar{c}_{\bar{H}(\bar{\bf 5})} & = & 
  K_B \qquad \qquad ({\rm spread~out~in~} B). \label{eq:curve-SO(12)-root}
\end{eqnarray}
Here, both the $H({\bf 5})$ and $\bar{H}(\bar{\bf 5})$ multiplets 
arise within the SO(12) gauge group, and their wavefunctions are not 
confined on a matter curve.

The SU(6) gauge symmetry in the F-theory dual of the 3+2 bundle 
compactification or SO(12) gauge symmetry dual to the 2+2+1 bundle 
compactification are broken by the U(1) bundles on the discriminant 
locus $\Sigma_0$. If the Fayet--Iliopoulos parameters of those 
U(1) symmetries are non-zero, some fields charged under the symmetry 
develop expectation values, and this correspond to the extension 
structure of vector bundles in (\ref{eq:23ext}) and (\ref{eq:32ext}) 
in Heterotic theory vacua.
The same constraint on the sign of the Fayet--Iliopoulos parameters 
as in section \ref{sec:Het} has to be imposed here.

A story similar to section \ref{sssec:parabora} can be told about the 
F-theory dual of 3+2 bundle and 2+2+1 bundle compactification of Heterotic 
theory, but we just do not do it here.

\subsection{Toward a Description Intrinsic to F-theory}
\label{ssec:F-gen}

\subsubsection{Characterising Matter Curves within F-theory}

We have so far provided some examples of F-theory vacua that do not 
have dimension-4 proton decay operators. All of them are dual to 
certain Heterotic vacua, since they were obtained through the 
Heterotic--F-theory duality. But not all the F-theory vacua are 
dual to Heterotic vacua. If we had only handles on vacua that 
are dual to Heterotic theory, we would lose a chance to study phenomenology 
of F-theory vacua that do not have Heterotic dual.
Thus, at the end of this section, we take a little step to try to 
phrase some aspects of F-theory vacua in a way intrinsic to F-theory.
That will help further explore F-theory vacua that are phenomenologically 
viable.

Matter curves, or loci of enhanced singularity, have been discussed, but 
they were described by divisors $\eta$ on the base manifold $B$ and its 
definition was associated with the spectral cover (or the second Chern 
class) of vector bundles in Heterotic compactification. It is rather 
unsatisfactory that the matter curves, which are generic features of 
F-theory vacua, are described in language intrinsic to Heterotic theory.
It would be nice if the matter curves are characterized without referring 
to such objects in Heterotic theory as vector bundles. 

In the case of 4+1 model at the reducible limit, the line bundles $L$ 
and det$U_4$ add twists only on the matter curves and they do not change 
the matter curves themselves. The discriminant locus $\Sigma_0 \simeq B$
has $D_5$ singularity, supporting SO(10) gauge symmetry if seen locally.
The matter curves are those of theories with unbroken SO(10) symmetry.
We do not question if this result found in the Heterotic--F-theory dual 
region holds true in generic F-theory vacua without dimension-4 proton decay.
But we would like to ask the following question: how we define $\eta$ 
to begin with, and whether there is an extra freedom in choosing matter curves, 
when an elliptic fibration $\pi_X: X \rightarrow B'$ of an F-theory vacuum 
is not of a $K3$-fibration. The following content is closely related to 
what was discussed in \cite{Sadov}.

Let us consider an F-theory compactification on an elliptic fibred 
CY$_4$ $\pi:X \rightarrow B'$ given by 
\begin{equation}
 y^2 = x^3 + f(z_1,z_2,z_3) x + g(z_1,z_2,z_3),
\end{equation}
where $(z_1,z_2,z_3)$ are local coordinates of the base manifold $B'$,
$(y,x)$ are those of the elliptic fibre, and $(x,y,f,g)$ are sections 
of ${\cal O}(- 2 K_{B'})$, ${\cal O}(- 3 K_{B'})$, 
${\cal O}(- 4 K_{B'})$, and ${\cal O}(- 6 K_{B'})$, respectively. 
The determinant of this elliptic curve $\Delta \equiv 4 f^3 + 27 g^2$ 
is a section of ${\cal O}(- 12 K_{B'})$, whose zero locus determines 
the discriminant locus on the base manifold $B'$. 
Thus, the following consistency condition has to be satisfied \cite{MV1}
\begin{equation}
 {\rm div} \Delta = - 12 K_{B'}.
\label{eq:RRgen}
\end{equation}
This condition reduces to the Ramond--Ramond charge cancellation 
condition of 7-branes in the Type IIB limit.

Now let us assume that one of irreducible components of 
the discriminant locus is of $D_5$ type singularity, supporting SO(10) 
unbroken gauge symmetry. Let us denote the irreducible component 
by $\Sigma$, i.e., 
\begin{equation}
 {\rm div}\Delta  = 7 \Sigma + \cdots.
\end{equation}
At the loci where other irreducible components intersect $\Sigma$, 
singularity and gauge symmetry are enhanced, and matter is localised.
The consistency condition (\ref{eq:RRgen}) has to be satisfied in 
any F-theory compactification, and in particular, when evaluated 
on any divisors of the base manifold $B'$. Assuming that there 
are only matters\footnote{We restrict ourselves to this limited matter 
contents because we are interested in F-theory geometry relevant 
to the real world in this article; there is no strong phenomenology 
motivation (except in an application to some models of 
Dimopoulos--Wilczek mechanism) 
to think of geometry that yields multiplets in the symmetric tensor 
representation.} 
in the SO(10)-{\bf 16} and -{\bf vec.}, 
the condition (\ref{eq:RRgen}) becomes 
\begin{equation}
 7 \Sigma \cdot \Sigma
 +3 \bar{c}_{\bf 16} + \bar{c}_{\bf vec.}= 
 -12 K_{B'} \cdot \Sigma 
\label{eq:tot-discr-SO(10)}
\end{equation}
when evaluated on $\Sigma$. The integer coefficients in front of matter
curves are read out from the behaviour of the determinant around 
the matter curve found in \cite{6authors}.
Let us now define a divisor $\eta$ on the subvariety $\Sigma$.
Since $\Sigma \subset B'$ defines a divisor on $\Sigma$ itself through 
the self-intersection, let us define a divisor $\eta$ on $\Sigma$ 
through \cite{Rajesh} 
\begin{equation}
 \eta \equiv  \Sigma \cdot \Sigma - 6K_{\Sigma}.
\label{eq:new-def}
\end{equation}
This is a generalization of (\ref{eq:old-def}).
Using the adjunction formula 
\begin{equation}
 K_{\Sigma} = K_{B'}|_\Sigma + \Sigma,
\end{equation}
we see that the general solution to (\ref{eq:tot-discr-SO(10)}) is  
of the form 
\begin{equation}
 \bar{c}_{\bf 16} = \eta + 4K_\Sigma + \eta',  \quad {\rm and} \quad 
 \bar{c}_{\bf vec.} = 2\eta + 6 K_\Sigma -3\eta',
\end{equation}
where $\eta'$ is an arbitrary divisor on $\Sigma$. 

Another theoretical constraint is the box anomaly cancellation. 
On an arbitrary curve in $\Sigma$, the budget of the inflow of irreducible 
box anomalies should balance. 
It is equivalent to the Ramond--Ramond charge cancellation in 
Type IIB string theory, but it seems it is not equivalent to 
(\ref{eq:RRgen}) in F-theory in general. Since the [SO(10)]$^4$ 
irreducible anomaly comes from $-2$ units from hypermultiplets 
in {\bf 16} representation and $+1$ units from half-hypermultiplets 
in {\bf vec} representation, the arbitrariness $\eta'$ in 
the above general solutions changes the total amount of the 
irreducible box anomaly. Thus, the anomaly cancellation condition 
eliminates the arbitrariness, and since we know that $\eta'=0$ is 
a consistent solution, that should be the only solution.
This implies that the contribution to the irreducible box anomaly 
from SO(10)-{\bf adj.} representation, including both 
the contribution from the vector multiplet and hypermultiplets, 
is given by $K_\Sigma$ so that
\begin{equation}
  2  \bar{c}_{\bf adj.} - 2  \bar{c}_{\bf 16}
  +  \bar{c}_{\bf vec.} 
=  2 K_{\Sigma} 
 -2 \left(\eta + 4 K_\Sigma \right)
 + \left(2 \eta + 6 K_\Sigma  \right) = 0.
\end{equation}
Therefore, the matter curves given in (\ref{eq:curve-SO(10)-16}) and 
(\ref{eq:curve-SO(10)-10}) are of general form in F-theory vacua, 
valid even when there is not Heterotic dual.  
The divisor $\eta$ is now defined only from local information of the 
geometry of F-theory through (\ref{eq:new-def}), and the matter 
curves are determined by this divisor. 

Exactly the same argument can be developed for the matter curves on 
a discriminant locus $\Sigma$ with SU(6) or SO(12) gauge symmetry on 
it. This is a situation relevant to the reducible limit of the 3+2 and 
2+2+1 bundle compactification of Heterotic theory. 
The divisor $\eta$ on $\Sigma$ is defined by (\ref{eq:new-def}). 
There are three different matter curves to think about, namely 
$\bar{c}_{\wedge^3 {\bf 6}}$, $\bar{c}_{\bf 6}$ and 
$\bar{c}_{\wedge^2 {\bf 6}}$ for SU(6), and $\bar{c}_{{\bf 32}'}$, 
$\bar{c}_{\bf vec}$ and $\bar{c}_{\bf 32}$ for SO(12). 
On the other hand, there are two independent constraints, 
(\ref{eq:RRgen}) and the irreducible box anomaly cancellation:
\begin{eqnarray}
6 \Sigma \cdot \Sigma + 3 \bar{c}_{\wedge^3 {\bf 6}} + 
\bar{c}_{\bf 6} + 4 \bar{c}_{\wedge^2 {\bf 6}} & = & 
 -12 K_{B'} \cdot \Sigma, \\
6 K_\Sigma -3 \bar{c}_{\wedge^3 {\bf 6}} 
+ \bar{c}_{\bf 6} -2 \bar{c}_{\wedge^2 {\bf 6}} & = & 0,
\end{eqnarray} 
for SU(6) and 
\begin{eqnarray}
8 \Sigma \cdot \Sigma + 2 \bar{c}_{{\bf 32}'} + \bar{c}_{\bf vec.} 
+ 2 \bar{c}_{\bf 32} & = & -12 K_{B'} \cdot \Sigma, \\
4 K_\Sigma  -2 \bar{c}_{{\bf 32}'} + 
\bar{c}_{\bf vec.} -2 \bar{c}_{\bf 32} & = & 0,
\end{eqnarray} 
for SO(12). Thus, general solutions have one degree of freedom left 
in choosing divisors. They are given by 
(\ref{eq:curve-SU(6)-20})--(\ref{eq:curve-SU(6)-root}) and 
(\ref{eq:curve-SO(12)-32p})--(\ref{eq:curve-SO(12)-root}), respectively, 
with the freedom of how to split the divisor $\eta$ defined 
in (\ref{eq:new-def}) into $\eta_2$ and $\eta_3$.
Thus, even in generic F-theory vacua that may not have Heterotic dual, 
the matter curves are constrained just as in vacua with Heterotic dual. 


\subsubsection{Chirality Formula}

There is another expression that allows straightforward generalization 
to F-theory vacua. Let us look at the expression for the chirality 
for the {\bf 10}$'$ representation in (\ref{eq:curve-SO(10)-root}),  
in a way presented in section \ref{sssec:chiral}.
Although it arises from a part of the SO(10) gauge field of the 
discriminant locus $\Sigma_0 \sim B$, and is not localised on a 
specific matter curve, $\sigma \cdot C_1 \sim \sigma \cdot K_B$ 
can be formally used as its matter curve. 
The line bundle ${\cal N}_1$ is $L$ itself.
Applying (\ref{eq:chirality-curve}) and (\ref{eq:sheaf-F}) formally, 
along with (\ref{eq:adj-2}), we obtain\footnote{The phenomenological 
constraint (\ref{eq:ZL}) becomes $K_B \cdot c_1(L) = 0$ in elliptic 
Calabi--Yau compactification of the Heterotic theory. Thus 
(\ref{eq:chi-self-F}) becomes zero. This fact itself is
nothing more than a straightforward consequence of the 
phenomenological requirement (\ref{eq:10less}).} 
\begin{equation}
 \# {\bf 10}' - \# \overline{\bf 10}' = \int_{K_{\Sigma_0}} c_1(L).
\label{eq:chi-self-F}
\end{equation}
This formula is a generalisation of the corresponding expression 
in the Type IIB string theory. Indeed, let us think of a Type IIB 
configuration where five D7-branes are wrapped 
on a holomorphic 4-cycle $\Sigma$ along with an O7-plane and five D7-branes 
as the orientifold mirror image of the original D7-branes. Those 
branes support SO(10) gauge group. When a U(1) flux $F$ is turned on 
on the five D7-branes (and on their images) so that SU(5) gauge group 
is left unbroken, we have 
\begin{equation}
 \# {\bf 10}' - \# \overline{\bf 10}' = 
   2 \int_{\Sigma \cdot \Sigma = K_\Sigma} 
      \left(\frac{F}{2\pi} - \frac{B}{(2\pi)^2\alpha'} \right) 
 = - \int_\Sigma c_1(T\Sigma) 
   2 \left(\frac{F}{2\pi} - \frac{B}{(2\pi)^2\alpha'} \right).
\label{eq:chi-self-IIB} 
\end{equation}
The difference in the factor 2 is due to the normalisation of U(1) 
charges: the {\bf 10}$'$ representations in (\ref{eq:chi-self-F}) 
have $q_L=1$ unit of charge under the line bundle $L$ on the 4-cycle 
$\Sigma_0 \sim B$, whereas the normalisation of $F$ in 
(\ref{eq:chi-self-IIB}) is set so that the {\bf 10}$'$-representations 
have 2 units of charge---one unit from each end of open strings
connecting D7-branes on the both sides of an O7-plane.
We have ignored the $B$-field.
Thus, (\ref{eq:chi-self-F}) suggests that the expression in Type IIB 
string theory (\ref{eq:chi-self-IIB}) is generalized straightforwardly 
in the F-theory as in (\ref{eq:chi-self-F}). It should be kept 
in mind that we should use $K_\Sigma$, rather than 
$\Sigma \cdot \Sigma$, in the
F-theory version. They are not the same in the F-theory, since 
the half of Calabi--Yau 3-folds for the Type IIB orientifold 
compactifications are projected out to be $B'$, which  
is not a Calabi--Yau manifold.


The chirality formula (\ref{eq:chi-self-F}) has an immediate 
application. One can ask a following question: certainly it is 
a generic feature of F-theory vacua that matter multiplets are 
localised on codimension-1 subspace, matter curves, in a discriminant 
locus; but is it possible to have a vacuum where all chiral matters 
are not localised but propagate over the entire ``bulk'' $\Sigma$? 
In other words, is it possible to think of a vacuum where all matters 
arise from a stack of parallel 7-branes with $E_7$ or $E_8$ gauge 
group on it, rather than from a system of intersecting 7-branes?
Reference \cite{CY4} contains examples of F-theory geometry of 
parallel 7-branes supporting $E_7$ and $E_8$. When a vector bundle 
is turned on as in the discussion in section \ref{sec:BottomUp} and 
\ref{sec:Het}, various low-energy particles are obtained with 
the right Yukawa couplings. Thus, one can think of such a vacuum 
theoretically. But we see by using the chirality formula above, 
that parallel 7-branes with $E_8$ gauge symmetry cannot yield a net 
chirality in the low-energy spectrum.

It is important in (\ref{eq:chi-self-F}) that a) only the first Chern
class matters and b) only the first Chern classes on one curve 
$K_\Sigma$ in $S$ are relevant. In all the three symmetry breaking 
patterns in section \ref{sec:BottomUp} with the underlying $E_8$ 
symmetry, multiplets in the SU(5)$_{\rm GUT}$-{\bf 10} representation 
arise from a rank-5 bundle with the structure group in SU(5). 
Thus, the overall net chirality is given by the sum of 
$\int_{K_\Sigma} c_1$ of each irreducible bundle, which vanishes 
because $\det V_5 \simeq {\cal O}_\Sigma$. Thus, we cannot think 
of a vacuum where all the matter curves of SU(5)$_{\rm GUT}$-{\bf 10} 
representation 
are $K_\Sigma$ in order to have a net chirality in the low-energy 
spectrum. Some multiplets in the SU(5)$_{\rm GUT}$-{\bf 10}
representation have to be localized on $\Sigma$. This also means 
that the isolated discriminant locus with $E_8$ gauge symmetry  
does not lead to a realistic vacuum. 

The matter curves and the gauge symmetries on the discriminant locus 
$\Sigma$ are relevant to particle physics, primarily through the 
determination of Yukawa couplings. Although they are important, 
we have little chance to see them directly, as long as the 
compactification radius remains very small compared with the length 
scale of the electroweak scale. But there is a situation where 
the gauge symmetry in the bulk is of more importance: 
the gauge mediation of supersymmetry breaking. 
The gravitino problem of supersymmetric theories suggests 
that either the gravitino mass is very heavy or very light 
($m_{3/2} \lsim 10 \; \EV$). 
Supersymmetry breaking should be mediated through gauge interactions 
in the latter scenario, and there is a phenomenology model that 
naturally realizes gauge mediation with very low gravitino mass 
\cite{Nomura}. 
It may be realized in Type IIB string / F-theory with a resolved 
conifold geometry with warped factor \cite{PzT} where D7-branes 
for the SU(5)$_{\rm GUT}$ gauge group are wrapped on $S^3$. 
In such models, low-lying Kaluza--Klein states of the bulk vector 
multiplets are observable in LHC experiments. 

Thus, the gauge symmetry, SO(10), SU(6), SO(12) or just SU(5), makes 
an observable difference. The above argument suggests, however, that 
it is unlikely to see the entire $E_8$ gauge group. 
There are also lots of phenomenological model buildings on the origin 
of fermion mass hierarchies on AdS background, and the determination 
of the matter curves provides restrictions on the degree of freedom 
available in such effective-field theory model buildings.

\section{Summary and Discussion}
\label{sec:Sum}

Gauge theories are often accompanied with 16 supersymmetry charges 
at the microscopic level, and when a large gauge symmetry $G$ is 
spontaneously broken down to a smaller group $H$, charged matter 
multiplets arise from the coset space of $\mathfrak{g}/\mathfrak{h}$. 
This is a salient feature of string theory, and this is a framework 
where we can expect to unify interactions of quark, lepton, neutrino 
and Higgs fields with gauge bosons. 

{\bf Physics Summary}

Requiring that the tri-linear Yukawa interactions be generated from 
the super Yang--Mills interactions, while the dimension-4 proton decay 
operators be not generated, we saw that $G=E_7$ is the minimal choice 
and $G=E_8$ is the only alternative, when we impose Georgi--Glashow 
SU(5) unification of quarks and leptons. There are four different 
patterns of how to break these symmetries to SU(5)$_{\rm GUT}$, 
namely, 
\begin{center}
\begin{tabular}{ll}
 $E_8 / \SU(5)_{\rm GUT} \times \vev{\SU(4)\times \U(1)_\chi}$ & A, \\
 $E_8 / \SU(5)_{\rm GUT} \times \vev{\SU(3)_2 \times \SU(2)_2 \times 
\U(1)_{\tilde{q}_7} }$ & A, B, \\
 $E_8 / \SU(5)_{\rm GUT} \times \vev{\SU(2)_1 \times \SU(2)_2 \times 
\U(1) \times \U(1)} $ & A, B, \\
 $E_7 / \SU(5)_{\rm GUT} \times \vev{\SU(2)_2 \times 
\U(1) \times \U(1)} $ & A, B. 
\end{tabular} 
\end{center}
We might call them, 41A-, 32A(B)-, 221A(B)- and 21A(B)-scenarios, 
for brevity.
The origins of all the low-energy multiplets are identified in the 
coset spaces, and labels A and B correspond to different particle 
identifications. A is when moduli multiplets are identified with 
right-handed neutrinos, and in case B, they are identified with a singlet 
chiral field that has $W \ni S H \bar{H}$ coupling.
In all these possibilities, all the Yukawa couplings for the quarks 
and 
charged leptons are generated from the super Yang--Mills interactions, 
and in possibilities A, Dirac Yukawa couplings of neutrinos also exist.
We have not discussed how to break SU(5)$_{\rm GUT}$ symmetry in this 
article. All U(1) vector fields that appear above have Green--Schwarz 
couplings, and are massive (\ref{eq:Amass-S}). Thus, they are not 
likely to be within the range of terrestrial experiments.

The dimension-4 proton decay operators are forbidden by U(1) global 
symmetries that correspond to the anomalous U(1) gauge symmetries.
When the Fayet--Iliopoulos parameters does not vanish, the U(1) symmetries 
can be spontaneously broken by fields of the same sign charges. 
When the sign of the Fayet--Iliopoulos parameter and the charges are 
appropriate, dimension-4 operators are not generated in any perturbative 
processes. In all four A-scenarios, either right-handed neutrinos or their 
anti-generation particles develop expectation values to absorb the 
Fayet--Iliopoulos parameters, and hence $\Z_2$ $B-L$ parity is not 
preserved anymore, yet the absence of dimension-4 proton decay is 
guaranteed. 

All possible terms in superpotential can be worked out, imposing all the 
underlying symmetry, and allowing to insert spurions that spontaneously 
break them. 
Using this approach, effective operators of some of the scenarios 
were studied in this article. See (\ref{eq:4N}), (\ref{eq:dim-7}), 
(\ref{eq:appPQ}), (\ref{eq:double}) and (\ref{eq:32B-dim5}) and discussion 
around these equations.
All the discussion so far are very robust, and does not depend on whether 
a vacuum is realized in Heterotic theory, M-theory or F-theory.

We also argued, in the context of M-theory compactification, that 
diagonal entries of up-type quark Yukawa matrix tend to be suppressed 
in some part of the moduli space.

In F-theory vacua, SU(5)$_{\rm GUT}$ gauge field propagates on an internal 
4-cycle, and the zero-mode wave functions are localized on internal 
2-cycles, in general. When the Fayet--Iliopoulos parameters vanish, 
we determined how the low-energy particles are localized, and found 
that the up-type Higgs multiplet propagates on the whole 4-cycle and 
gauge symmetry is locally SU(6) in 32A,B scenarios, 
both up-type and down-type Higgs multiplets propagate on the 4-cycle 
and the gauge symmetry is locally SO(12) in the 221A scenario. 
If the 4-cycle of the SU(5)$_{\rm GUT}$ is extended to a warped extra 
dimension explaining the hierarchy between the Planck scale and the 
electroweak scale, we can see at LHC the Kaluza--Klein vector fields 
of such an enhanced gauge symmetry.

{\bf String Theory Discussion}

The number of consistent vacua in String Theory continues to grow, 
while there is no prospect in near future that some non-perturbative 
dynamics is found and the over-degeneracy of vacua is lifted (see 
\cite{dd} for a full list of references). At least we may have to 
live with the multitude of vacua for a while, and there have been  
some attempts at finding constraints or predictions that hold for 
wide class of string vacua \cite{vafa,kachru}. There is another kind 
approach, where people try to find out dynamics that takes place 
over the landscape of vacua \cite{beauty} and to find out unknown 
``structure'' the true landscape has \cite{runaway}.

We found that the supersymmetric SU(5) landscape contains seven 
islands. Islands may be connected by continuous deformation of moduli 
parameters, but proton decays too fast in vacua between the islands, 
and ``lives'' may not exist in such vacua.
The minimal underlying symmetry of SU(5) unified theory is $E_7$, and 
D-brane Yang--Mills interactions in Type I, IIA or IIB string theory 
cannot provide the form of the up-type quark Yukawa couplings of 
SU(5) unified theories. 

The framework presented in this article does not have a massless 
U(1)$_{B\mbox{-}L}$ gauge boson, and yet, it is now clear
when dimension-4 proton decay operators are absent. 
In Heterotic language, a rank-5 holomorphic stable vector bundle $V_5$
has to have a certain sub-bundle, in order to prevent the dimension-4
proton decay. The condition will be easily taken into account
in search for a geometry describing our world. This condition
has also been translated into M-theory and F-theory compactification,
partially but not completely yet.

The Yukawa couplings of quarks and leptons come from super Yang--Mills 
interactions. The origin of low-energy particles has been identified 
in Heterotic, M and F-theory language. Thus, one can exploit various 
techniques in string theory to extract more information.

\appendix

\section{Note on Lie Algebra}
\label{sec:note}

\subsection{$\mathfrak{e}_6$ Lie algebra}

The $\mathfrak{e}_6$ Lie algebra contains 
$\mathfrak{su}(5)+\mathfrak{su}(2)_2+\mathfrak{u}(1)$ subalgebra. 
After removing the nodes for $\alpha_2$ and $-\theta$ from the 
extended Dynkin diagram of $E_6$ (see Fig.~\ref{fig:e6dynkin}), 
one node $\alpha_1$ for $\mathfrak{su}(2)_2$ and four 
$\alpha_{3,4,5,6}$ for $\mathfrak{su}(5)$ are left. The irreducible 
decomposition of the $\mathfrak{e}_6$-{\bf adj.} representation 
under the $\mathfrak{su}(5)+\mathfrak{su}(2)_2+\mathfrak{u}(1)$ 
subalgebra (\ref{eq:e6dcmp}) is known in the literature. We need 
to know, however, which roots of $\mathfrak{e}_6$ correspond to 
which irreducible components in order to provide full geometric 
description in M-theory, the equation (\ref{eq:e6dcmp}) is not enough.
We first follow an intuitively tractable way manipulating the extended 
Dynkin diagram to roughly determine the roots of the irreducible 
components and later provide the full results of explicit calculation.

When only the node of $\alpha_2$ is removed from the extended Dynkin 
diagram of $E_6$, the $\mathfrak{e}_6$-{\bf adj.} representation is 
decomposed into 
\begin{equation}
 {\rm Res}^{E_6}_{\SU(6) \times \SU(2)_2} 
      (\mathfrak{e}_6\mbox{-}{\rm adj.}) \simeq 
 ({\bf adj.},{\bf 1}) \oplus ({\bf 1},{\bf adj.}) \oplus 
 (\wedge^3 {\bf 6},{\bf 2}).
\end{equation} 
Thus the roots in the $(\wedge^3{\bf 6},{\bf 2})$ representation 
contain $\alpha_2$. Since this representation is a doublet of
SU(2)$_2$ generated by $\alpha_1$, half of roots in the representation 
contain the image of $\alpha_2$ under the Weyl reflection associated 
with $\alpha_1$, i.e., $\alpha_1+\alpha_2$.

When the node of $-\theta$ is further removed from the $A_5$ 
Dynkin diagram made of $\alpha_{3,4,5,6}$ and $-\theta$, breaking 
the SU(6) symmetry down to SU(5)$\times$U(1), 
the $(\wedge^3 {\bf 6},{\bf 2})$ representation splits up into 
$({\bf 10},{\bf 2})^{1}$+h.c. under the 
SU(5)$\times$SU(2)$_2\times$U(1), and SU(6)-{\bf adj.} representation 
into SU(5)-{\bf adj.}, -{\bf 5}$^{-2}$+h.c.. and a singlet. 
Thus, we have the irreducible decomposition (\ref{eq:e6dcmp}).
Roots in the SU(5)-{\bf 5} representation 
(and its Hermitian conjugate) contain $-\theta$. 

One can also carry out an explicit calculation, using a basis in the 
dual space of Cartan subalgebra. We adopt the basis in \cite{DKV} for 
$\mathfrak{e}_d$ ($d=6,7,8$) algebras, which can treat all three 
exceptional Lie algebra systematically and has also clear 
interpretation in del Pezzo surfaces. When the simple roots are chosen 
as 
\begin{eqnarray}
 \alpha_i & = & L_i - L_{i+1} \quad {\rm for}\quad 
  i=1,\cdots,(d-1), \quad {\rm and}  \nonumber \\
 \alpha_d & = & L_0-(L_1+L_2+L_3)
\label{eq:dPbasis}
\end{eqnarray} 
with $d=6$, all of positive roots are 
\begin{eqnarray}
 & & L_i - L_j \qquad (i<j), \\
 & & L_0 - (L_i + L_j + L_k) \qquad (i<j<k) \quad {\rm and} \\
 & & \theta \equiv 2L_0-(L_1+\cdots+L_6) = 
   \alpha_1+2\alpha_2+3\alpha_3+2\alpha_4+\alpha_5+2\alpha_6, 
\end{eqnarray}
where $\theta$ is the highest root. Under this convention, each 
irreducible component consists of roots shown as follows:
\begin{itemize}
\item ({\bf 1},{\bf adj.}): $\pm \alpha_1 = \pm(L_1-L_2).$
\item ({\bf adj.},{\bf 1}): $\pm(L_i-L_j)$ ($3 \leq i < j \leq 6$) and 
          $\pm(L_0-(L_1+L_2 + L_k))$ ($k=3,\cdots,6$).
\item $\overline{(\wedge^2{\bf 5},{\bf 2})}$: \\
$\qquad$ $L_0-(L_1 + L_2 + L_i + L_j) + L_a \quad$ 
 ($a=1,2$ and $3 \leq i < j \leq 6$) and \\
$\qquad$ $L_a - L_k$ ($a=1,2$ and $k=3,\cdots,6$). \\
$\qquad$ The lowest weights are a doublet 
$(\alpha_1+\alpha_2,\alpha_2)=(L_a-L_3)_{a=1,2}$.
\item $(\wedge^4 \bar{\bf 5},{\bf 1})\simeq ({\bf 5},{\bf 1})$: \\ 
$\qquad$ $L_0 - (L_i+L_j+L_k)$ ($3 \leq i < j < k \leq 6$) and \\
$\qquad$ $\theta = 2L_0 - (L_1+\cdots+L_6)$. \\
$\qquad$ The highest weight is $\theta = 
(\alpha_1+2\alpha_2)+(3\alpha_3+2\alpha_4+\alpha_5+2\alpha_6)$.
\end{itemize}
The roots of ($\wedge^2 {\bf 5}$,{\bf 2}) and $(\bar{\bf 5},{\bf 1})$ 
representations of SU(5)$\times$SU(2)$_2$ are obtained by just 
multiplying $(-1)$ to those of their Hermitian conjugate 
representations. In M-theory description, M2-branes wrapped on 
$\{-C_{-\theta},-(C_{-\theta}+C_6),-(C_{-\theta}+C_6+C_3),
-(C_{-\theta}+C_6+C_3+C_4),-(C_{-\theta}+C_6+C_3+C_4+C_5)\}$ provide 
the quintet $H({\bf 5})$ particles, where $C_i$'s are 2-cycles
corresponding to the simple roots $\alpha_i$'s.

\subsection{$\mathfrak{e}_7$ Lie algebra}
\label{ssec:AlgE7}

The $\mathfrak{e}_7$ Lie algebra contains an 
$\mathfrak{su}(6)_1+\mathfrak{su}_2+\mathfrak{u}(1)$ subalgebra, and its
subalgebra $\mathfrak{su}(5)+\mathfrak{su}_2+\mathfrak{u}(1)+\mathfrak{u}(1)$.
They are obtained by removing the nodes of 
$\alpha_2$, and $-\theta$ (and $\alpha_7$).\footnote{It is also fine 
to remove $\alpha_1$ instead of $-\theta$. The node for $\alpha_1$ 
is identified with the root of $\mathfrak{su}(2)_2$ subalgebra only 
because we want to use the same $\mathfrak{su}(2)_2$ in the 
$\mathfrak{e}_8$ algebra in section \ref{ssec:AlgE8}.} The 
$\mathfrak{su}(2)_2$ is generated by $\alpha_1$,  
$\mathfrak{su}(6)_1$ by $\alpha_{7,3,4,5,6}$, and $\mathfrak{su}(5)$ 
by $\alpha_{3,4,5,6}$ (see Fig.~\ref{fig:e7dynkinE}).

When only the node of $\alpha_2$ is removed from the extended Dynkin 
diagram Fig.~\ref{fig:e7dynkinE}, $E_7$ is reduced to 
SU(3)$\times$SU(6)$_1$, and 
\begin{equation}
 {\rm Res}^{E_7}_{\SU(6)_1 \times \SU(3)_1} (\mathfrak{e}\mbox{-}{\bf adj.})
   \simeq ({\bf adj.},{\bf 1})\oplus({\bf 1},{\bf adj.})\oplus
    (\wedge^2 {\bf 6},\bar{\bf 3})\oplus {\rm h.c.}.
\end{equation}
All the roots in the $(\wedge^2 {\bf 6},\bar{\bf 3})$+h.c. representation 
(and its Hermitian conjugate) should contain $\alpha_2$, and form a
triplet $\pm(-\theta+\alpha_1+\alpha_2,\alpha_1+\alpha_2,\alpha_2)$ 
under the SU(3)$_1$ symmetry; linear combinations of $\alpha_{3,\cdots,7}$ 
are just omitted here. Upon further removing the node of $-\theta$, 
the SU(3)$_1$ symmetry is reduced to SU(2)$_2\times$U(1)$_{q_6}$, and 
each component splits into irreducible pieces;
\begin{eqnarray}
 (\wedge^2{\bf 6},\bar{\bf 3}) & \rightarrow & 
  (\wedge^2{\bf 6},{\bf 2})^1 \oplus (\wedge^2{\bf 6},{\bf 1})^{-2}, \\
 ({\bf 1},{\bf adj.}) & \rightarrow & 
  ({\bf 1},{\bf adj.})^0 \oplus ({\bf 1},{\bf 1})^0 \oplus 
  ({\bf 1},{\bf 2})^3 \oplus {\rm h.c.}.
\end{eqnarray}
Thus, the irreducible decomposition (\ref{eq:e7dcmp}) is obtained.
Roots in the representation $({\bf 1},{\bf 2})^3+$h.c. should be 
a doublet $\pm(-\theta,-\theta+\alpha_1)$. 
The roots in $(\wedge^2 {\bf 6},{\bf 1})^{-2}+$h.c. are 
$\pm(-\theta+\alpha_1+\alpha_2)$, and those in 
$(\wedge^2 {\bf 6},{\bf 2})^1+$h.c. are a doublet 
$\pm (\alpha_1+\alpha_2, \alpha_2)$, 
up to some linear combinations of $\alpha_{3,\cdots,7}$.

When the node of $\alpha_7$ is further removed, the symmetry is 
reduced from SU(6)$_1$ to SU(5), and SU(6)-charged representations split 
into smaller irreducible pieces. In order to determine whether 
the roots of SU(5)$\times$SU(2)$_2$ irreducible representations contain 
$\alpha_7$ or not, let us consider another route of the irreducible 
decomposition. We could have remove $\alpha_7$ first, and $\alpha_2$ 
next, and $-\theta$ at the end; the symmetry is broken through $E_7$, 
SU(8), SU(5)$\times$SU(3)$_1\times$U(1), and finally 
to SU(5)$\times$SU(2)$_2\times$U(1)$\times$U(1).
In the first step in the new route, 
\begin{equation}
 {\rm Res}^{E_7}_{\SU(8)}(\mathfrak{e}_7\mbox{-}{\bf adj.})
  \simeq {\bf adj.} \oplus \wedge^4 {\bf 8}, 
\end{equation}
and $\wedge^4 {\bf 8}$ representation further reduces to 
$\left[({\bf 1},{\bf 5})\oplus({\bf 3},\overline{\bf 10})\right]
\oplus$h.c. when SU(8) is reduced to SU(5)$\times$SU(3)$_1 \times$U(1). 
Thus, among the irreducible components 
$({\bf 10},\bar{\bf 3})\oplus ({\bf 5},\bar{\bf 3})=
(\wedge^2 {\bf 6},\bar{\bf 3})$, the former 
comes from SU(8)-$\wedge^4 {\bf 8}$ representation, whereas the latter 
from SU(8)-{\bf adj.}. We see that the roots in the former contain 
$\pm\alpha_7$, and those in the latter do not. 

All the above argument on the roots of irreducible components can be 
confirmed explicitly, by adopting a basis of the space of root lattice. 
Simple roots are given by (\ref{eq:dPbasis}) with $d=7$.
Positive roots are 
\begin{eqnarray}
& & L_i - L_j \quad (i<j), \\
& & L_0- (L_i + L_j + L_k) \quad (i<j<k) \quad {\rm and} \\
& & 2L_0+L_k-(L_1+\cdots+L_7). 
\end{eqnarray}
The highest root is $\theta \equiv 2L_0 - (L_2+\cdots+L_7) = 
2\alpha_1+3\alpha_2+4\alpha_3+3\alpha_4+2\alpha_5+\alpha_6+2\alpha_7$.
Each irreducible components of $\mathfrak{e}_7$ under the subalgebra
$\mathfrak{su}(5)+\mathfrak{su}(2)_2+\mathfrak{u}(1)+\mathfrak{u}(1)$ 
consists of the following roots:
\begin{itemize}
\item ({\bf 1},{\bf adj.}): $\pm \alpha_1 = \pm(L_1-L_2)$.
\item (${\bf adj.}\oplus{\bf 5}\oplus \bar{\bf 5}$,{\bf 1}): 
 \begin{itemize}
  \item [$\diamond$] $\pm(L_i-L_j) \quad$ ($3 \leq i < j \leq 7$) 
  \item [$\diamond$] 
    $+ (L_0-(L_1+L_2) - L_k) \quad$ ($k=3,\cdots,7$),\\
    $\qquad$ The lowest weight is $\alpha_7 \quad$ ($k=3$), $\quad$ 
	[for $\bar{H}(\bar{\bf 5})$ in ID A].
  \item [$\diamond$] 
	$- (L_0-(L_1+L_2) - L_k) \quad$ ($k=3,\cdots,7$),$\quad$ 
	[for nothing].
 \end{itemize}
\item $\overline{(\wedge^2{\bf 5} \oplus {\bf 5},{\bf 2})}$: 
 \begin{itemize}
   \item [$\diamond$] $L_0-(L_a + L_i + L_j) \quad$ ($a=1,2$ and 
	 $3 \leq i < j \leq 7$). \\
	 $\qquad$ The lowest weights are a doublet
	 ($\alpha_7+\alpha_1+\alpha_2+\alpha_3$,
	 $\alpha_7+\alpha_2+\alpha_3$), \\
	 $\qquad$ [for Hermitian conjugate of $
	 {\bf 10}=(Q,\bar{U},\bar{E})$]. 
 \item [$\diamond$] $L_a - L_k$ for $k=3,...,7$, 
	 $\quad$ [for $\bar{\bf 5}=(\bar{D},L)$ in ID A]. \\
	 $\qquad$ The lowest weights are a doublet 
	 $(\alpha_1+\alpha_2,\alpha_2)$.
 \end{itemize}
\item $\overline{(\wedge^3 {\bf 5} \oplus \wedge^4{\bf 5},{\bf 1})}$: 
 \begin{itemize}
  \item [$\diamond$] 
	$L_0-(L_i+L_j+L_k) \quad$ ($3 \leq i < j < k \leq 7$), 
	$\quad$ [for nothing], \\
	$\qquad$ including 
	$\theta - (\alpha_7+\alpha_1+\alpha_2+\alpha_3) = 
	(\alpha_1+2\alpha_2+\alpha_7)+
	(3\alpha_3+3\alpha_4+2\alpha_5+\alpha_6)$. 
 \item [$\diamond$] $2L_0-(L_1+\cdots+L_7)+L_k \quad$ ($k=3,\cdots,7$),
	$\quad$ [for $H({\bf 5})$]. \\
	$\qquad$ The highest weight is $\theta-\alpha_1-\alpha_2 = 
	(\alpha_1+2\alpha_2+2\alpha_7)+
	(4\alpha_3+3\alpha_4+2\alpha_5+\alpha_6)$, \\
	$\qquad$ The lowest weight is $(\alpha_1+2\alpha_2+2\alpha_7)+
	(3\alpha_3+2\alpha_4+\alpha_5)=2L_0-(L_1+\cdots+L_6)$.
 \end{itemize}
\item $\overline{(\wedge^6{\bf 6},{\bf 2})}$: 
[for Hermitian conjugate of $\bar{N}$ in ID A] \\
$\qquad$ $2L_0 - (L_1+\cdots+L_7)+L_1=(2\alpha_1+3\alpha_2)+
(4\alpha_3+3\alpha_4+2\alpha_5+\alpha_6+2\alpha_7)=\theta, \; \qquad$ \\
$\qquad$ $2L_0 - (L_1+\cdots+L_7)+L_2=(\alpha_1+3\alpha_2)+
(4\alpha_3+3\alpha_4+2\alpha_5+\alpha_6+2\alpha_7)=\theta - \alpha_1$. 
\end{itemize}

\subsection{$\mathfrak{e}_8$ Lie algebra}
\label{ssec:AlgE8}

The simple roots of $\mathfrak{e}_8$ Lie algebra is given by 
(\ref{eq:dPbasis}) with $d=8$. Positive roots are 
\begin{eqnarray}
& & L_i-L_j \quad (i<j), \\
& & L_0-(L_i+L_j+L_k) \quad (i<j<k), \\
& & 2L_0+(L_i+L_j)-(L_1+\cdots+L_8) \quad (i<j) {\rm ~and} \\
& & 3L_0-L_k-(L_1+\cdots+L_8),
\end{eqnarray}
and the maximal root $\theta$ satisfies 
\begin{equation}
\theta =2\alpha_1+4\alpha_2+6\alpha_3+5\alpha_4+4\alpha_5+3\alpha_6
        +2\alpha_7+3\alpha_8=3L_0-(L_1+\cdots+L_8)-L_8.
\end{equation}

The $\mathfrak{su}(6)_1+\mathfrak{su}(2)_1+\mathfrak{su}(2)_2
+\mathfrak{u}(1)$ subalgebra of $\mathfrak{e}_8$ is obtained by 
removing the odes of $\alpha_2$ and $\alpha_7$ from the extended 
Dynkin diagram of $E_8$.  $\mathfrak{su}(2)_1$ is generated by 
the maximal root $\theta$, and its commutant $\mathfrak{e}_7$ 
by $\alpha_{1,\cdots,6}$ and $\alpha_8$.
$\mathfrak{e}_7$ contains the $\mathfrak{su}(6)_1+\mathfrak{su}(2)_2
+\mathfrak{u}(1)$ subalgebra.
Brute force calculation shows that the irreducible components of 
$\mathfrak{e}_8$ under the subalgebra 
$\mathfrak{su}(5)_{\rm GUT}+\mathfrak{su}(2)_1+\mathfrak{su}(2)_2$ are:
\begin{itemize}
 \item $\mathfrak{su}(5)_{\rm GUT}$-singlets:
\begin{eqnarray}
& & \!\!\!\!\!\!\!\!\!\!\!\!\!\! 
 \left({\bf 1}\oplus {\bf 2}_1 \oplus {\bf 2}_2\right)\otimes
 \left({\bf 1}\oplus {\bf 2}_1 \oplus {\bf 2}_2\right)^* 
  = \\ 
&&  
\left( \begin{array}{c|cc|cc}
     & \alpha^{''} & & & \\
\hline
 -\alpha^{''} & & -\theta & & \\
 L_0-(L_1+L_2+L_8) & \theta & & \alpha' & \alpha'+\alpha_1 \\	
\hline
\alpha^{'''}+\alpha_1 & L_1-L_8 & -\alpha' & & \alpha_1 \\ 
\alpha^{'''} & L_2-L_8 & -\alpha'-\alpha_1 & -\alpha_1 &
	   \end{array}\right),
\label{eq:root1inE8} \nonumber 
\end{eqnarray}
$\quad$ $\alpha^{''} \equiv 2L_0 - (L_3+\cdots+L_8) = 
  2\alpha_1 + 4\alpha_2 + \alpha_7 +2\alpha_8 + 
(5\alpha_3+4\alpha_4+3\alpha_3 + 2\alpha_6)$, \\ 
$\quad$ $\alpha' \equiv 3L_0 - (2L_1+L_2+\cdots+L_8) = 
  \alpha_1+3\alpha_2+\alpha_7+3\alpha_8+
(5\alpha_3+4\alpha_4+3\alpha_3 + 2\alpha_6)$, \\
$\quad$ $\alpha^{'''} \equiv -2L_0+(L_2+\cdots+L_7) = 
 - (2\alpha_1+3\alpha_2+2\alpha_8) + \cdots$, 
\item $\mathfrak{su}(5)_{\rm GUT}$-{\bf 10} representations:
\begin{equation}
 \left(\wedge^2 {\bf 10},
       \left({\bf 1}\oplus {\bf 2}_1 \oplus {\bf 2}_2\right)\right) 
= \left\{ L_i+L_j + \left(\begin{array}{c}
    L_0 - (L_3+\cdots+L_7) \\ 
\hline
    -L_0+L_8 \\
    2L_0-(L_1+\cdots+L_8) \\
\hline
    -L_0+L_1 \\
    -L_0+L_2
			  \end{array}\right)\right\}_{3\leq i<j\leq 7},
\label{eq:root10inE8}
\end{equation}
\item $\mathfrak{su}(5)_{\rm GUT}$-$\bar{\bf 5}$ representations:
\begin{eqnarray}
& & 
\left(\bar{\bf 5},\wedge^2 ({\bf 1}\oplus {\bf 2}_1 \oplus {\bf 2}_2)
\right) = \label{eq:root5inE8} \\
& & \left\{-L_k +\left(\begin{array}{c|cc|cc}
  & & & & \\
\hline
  L_8 & & L_0-(L_1+L_2) & & \\
  L_8+\theta &	& & & \\ 
\hline
  L_1 & L_8-\alpha' & L_0-(L_8+L_2) & & \alpha^{'''}+L_1 \\
  L_2 & L_8-\alpha'-\alpha_1 & L_0-(L_8+L_1) & & 
		       \end{array}\right)\right\}_{k=3,\cdots,7}.
\nonumber
\end{eqnarray}
\end{itemize}

Two different SU(6) subgroups, SU(6)$_1$ and SU(6)$_2$, appear in the 
main text. They are both subgroups of SO(12) in 
$E_8 \supset$SO(12)$\times$SU(2)$_1 \times$SU(2)$_2$. The irreducible 
components in 
\begin{eqnarray}
{\rm Res}^{E_8}_{\SO(12)\times \SU(2)_1 \times \SU(2)_2} 
(\mathfrak{e}_8\mbox{-}{\bf adj.}) & \simeq &
({\bf adj.},{\bf 1},{\bf 1}) \oplus ({\bf 1},{\bf adj.},{\bf 1})
\oplus ({\bf 1},{\bf 1},{\bf adj.}) \nonumber \\
 & & \oplus ({\bf 12},{\bf 2},{\bf 2}) \oplus 
({\bf 32}',{\bf 1},{\bf 2}) \oplus ({\bf 32},{\bf 2},{\bf 1})
\end{eqnarray}
reduces under SU(6)$_1 \times$U(1)$_{q_6} \subset$SO(12) to 
\begin{eqnarray}
{\rm Res}^{\SO(12)}_{\SU(6)_1 \times \U(1)_{q_6}} {\bf 32}' & \simeq & 
\left[{\bf 1}^3 \oplus \wedge^2 {\bf 6}^{\; 1} \right] 
\oplus {\rm h.c.}, \\
{\rm Res}^{\SO(12)}_{\SU(6)_1 \times \U(1)_{q_6}} {\bf 32} & \simeq & 
{\bf 6}^2 \oplus \wedge^3 {\bf 6}^{\; 0} \oplus \bar{\bf 6}^{-2}, 
\end{eqnarray}
and under SU(6)$_2 \times$U(1)$_{\tilde{q}_6}\subset$SO(12) to 
\begin{eqnarray}
{\rm Res}^{\SO(12)}_{\SU(6)_2 \times \U(1)_{\tilde{q}_6}} {\bf 32}' & \simeq & 
{\bf 6}^2 \oplus \wedge^3 {\bf 6}^{\; 0} \oplus \bar{\bf 6}^{-2}, \\
{\rm Res}^{\SO(12)}_{\SU(6)_2 \times \U(1)_{\tilde{q}_6}} {\bf 32} & \simeq & 
\left[{\bf 1}^3 \oplus \wedge^2 {\bf 6}^{\; 1} \right] 
\oplus {\rm h.c.}.
\end{eqnarray}
Two SU(3) subgroups, SU(3)$_1$ and SU(3)$_2$, contain 
SU(2)$_2 \times$U(1)$_{q_6}$ and SU(2)$_1 \times$U(1)$_{\tilde{q}_6}$, 
respectively, and 
\begin{eqnarray}
 \SU(6)_1 \times \SU(3)_1 \times \SU(2)_1 & \subset & E_8, \\
 \SU(6)_2 \times \SU(3)_2 \times \SU(2)_2 & \subset & E_8.
\end{eqnarray}

\section{ALE spaces}
\label{sec:ALE}

This section of the appendix is a quick summary of useful results. 

The ALE spaces of $A_r$, $D_r$ and $E_r$ types are constructed 
by hyper-K\"{a}hler quotient \cite{Kronheimer,DM} as vacuum moduli of 
quiver gauge theories associated with the extended Dynkin diagrams 
of the simply laced Lie algebras. Their moduli parameters are 
the Fayet--Iliopoulos parameters of the sigma model in physics 
terminology or the values of the moment maps in mathematical 
terminology. The Fayet--Iliopoulos parameters 
$\vec{\zeta}^i=(\zeta^i_1,\zeta^i_2,\zeta^i_3)$ for $i=0,\cdots,r$ 
have to satisfy a ``traceless condition''
\begin{equation}
 n_i \vec{\zeta}^i = 0, 
\label{eq:traceFI}
\end{equation}
where integers $n_i$ (sometimes denoted as $a_i$) are the labels of 
the extended Dynkin diagrams; for the simple roots $\alpha_i$ 
($i=1,\cdots,r$),
\begin{equation}
 n_i \alpha_i = 0
\end{equation}
where $\alpha_0 \equiv -\theta$ is the negative of the highest root 
and $n_0 \equiv 1$.

The metric is known for the $A_{n-1}$-type and $D_n$ type. The ALE 
spaces are described as $S^1$ fibration (coordinate $\tau$) over 
a three dimensional space (coordinates $\vec{x}=(x_1,x_2,x_3)$) 
and the metric of $A_{n-1}$ type and $D_n$ type is specified by 
$\vec{r}_i=(r_1,r_2,r_3)_i$ for $i=1,\cdots,n$. The $S^1$ fibre 
shrinks at $\vec{x}=\vec{r}_i$ in $A_{n-1}$ type, and 
at $\vec{x}=\vec{r}_i$ and $\vec{x}=-\vec{r}_i$ in $D_n$ type.  
Between two such points, there is a topological 2-cycle, 
given by $S^1$ fibration over the interval between the two points. 
The Fayet--Iliopoulos parameters $\vec{\zeta}^i$ describing the 
ALE spaces are given in terms of $\vec{r}_i$ by 
\begin{equation}
 \vec{\zeta}^i = \vec{r}_i - \vec{r}_{i+1} \qquad (i=0,\cdots,n-1),
\end{equation}
for $A_{n-1}$ case, where $i=0=n$ is understood, and 
\begin{eqnarray}
 \vec{\zeta}^0 = - \vec{r}_1 - \vec{r}_2, & 
 \vec{\zeta}^i = \vec{r}_i - \vec{r}_{i+1} & 
 \vec{\zeta}^{n} = \vec{r}_{n-1} + \vec{r}_n, \nonumber \\ 
 \vec{\zeta}^1 = \vec{r}_1 - \vec{r}_2, & 
 ({\rm for~}i=2,\cdots,n-2), &
 \vec{\zeta}^{n-1} = \vec{r}_{n-1} - \vec{r}_n, 
\end{eqnarray}
for $D_n$ case. It is easy to see that the constraint 
(\ref{eq:traceFI}) is satisfied for the $A_{n-1}$ and $D_n$ cases; 
note that $n_i = 1$ for ${}^\forall i$ in $A_{n-1}$, and $n_i=2$ 
in $D_{n}$ except $n_0=n_1=n_{n-1}=n_n=1$.
The data $\vec{r}_i$ ($i=1,\cdots,n$) correspond to the location of 
$n$ D6-branes in their transverse directions in the Type IIA
interpretation, if the ALE space of $A_{n-1}$ or $D_n$ type is used 
as the background geometry of the M-theory \cite{Sen-M}. 

When the 2-cycle associated with the root $-\theta$ shrinks, the 
first D6-brane comes on top the $n$-th D6-brane in the $A_{n-1}$ case, 
and an SU(2) symmetry is enhanced. In the case of $D_n$ type ALE 
space, the first D6-brane moves across an O6-plane and comes on 
top of the orientifold image of the second D6-brane, and an SU(2) 
symmetry is enhanced \cite{Sen-M}. 
Thus, 2-cycles involving the 2-cycle for $-\theta$ can be treated 
at the equal footing as other 2-cycles. In section \ref{sec:M}, we
see that massless matters arising from vanishing $C_{-\theta}$ cycle 
are identified with the $H({\bf 5})$ multiplet in the model with $E_6$ 
underlying symmetry, and with right-handed neutrinos $\bar{N}$ [or 
SU(5)$_{\rm GUT}$-singlet $S$ in $W \ni S H \bar{H}$] in the model 
with $E_7$.

\section*{Acknowledgements}
We thank Ralph~Blumenhagen, Alon~Faraggi, Yang-Hui~He, 
Karl~Landsteiner, Thomas~Mohaupt, Michael~Schulz, Eric~Sharpe and 
Timo~Weigand for fruitful discussion and correspondence.
This work was supported in part by a PPARC Advanced Fellowship (R.T.), 
and in part by the Director, Office of Science, 
Office of High Energy and Nuclear Physics, of the US Department of 
Energy under Contract DE-AC02-05CH11231, and in part 
by the Miller Intitute for Basic Research in Science (T.W.).


\begin{thebibliography}{99}
\bibitem{a1}
  A.~E.~Faraggi, D.~V.~Nanopoulos and K.~j.~Yuan,
  ``A Standard Like Model In The 4-D Free Fermionic String Formulation,''
  Nucl.\ Phys.\ B {\bf 335}, 347 (1990).
%
\bibitem{a2}
%
 A.~E.~Faraggi,
  ``Hierarchical top - bottom mass relation in a superstring derived standard -
  like model,'' Phys.\ Lett.\ B {\bf 274}, 47 (1992),~
%
  ``Construction of realistic standard - like models in the free fermionic
  superstring formulation,'' Nucl.\ Phys.\ B {\bf 387}, 239 (1992)
  [arXiv:hep-th/9208024],
%
~``A New standard - like model in the four-dimensional 
free fermionic string formulation,''Phys.\ Lett.\ B {\bf 278}, 131 (1992);\\
%
G.~B.~Cleaver, A.~E.~Faraggi and D.~V.~Nanopoulos,
``String derived MSSM and M-theory unification,''
Phys.\ Lett.\ B {\bf 455}, 135 (1999)
[arXiv:hep-ph/9811427].
%
\bibitem{cv0}
%
 M.~Cvetic, G.~Shiu and A.~M.~Uranga,
  ``Chiral four-dimensional N = 1 supersymmetric type IIA orientifolds from
  intersecting D6-branes,''
  Nucl.\ Phys.\ B {\bf 615}, 3 (2001)
  [arXiv:hep-th/0107166], 
%
  ``Chiral type II orientifold constructions as M theory on G(2) holonomy
  spaces,''
  arXiv:hep-th/0111179; \\
%
 M.~Cvetic, I.~Papadimitriou and G.~Shiu,
  ``Supersymmetric three family SU(5) grand unified models from type IIA
  orientifolds with intersecting D6-branes,''
  Nucl.\ Phys.\ B {\bf 659}, 193 (2003)
  [Erratum-ibid.\ B {\bf 696}, 298 (2004)]
  [arXiv:hep-th/0212177].
%
\bibitem{Penn4}
%
V.~Braun, B.~A.~Ovrut, T.~Pantev and R.~Reinbacher,
  ``Elliptic Calabi-Yau threefolds with Z(3) x Z(3) Wilson lines,''
  JHEP {\bf 0412} (2004) 062
  [arXiv:hep-th/0410055]; \\
%
V.~Braun, Y.~H.~He, B.~A.~Ovrut and T.~Pantev,
  ``The exact MSSM spectrum from string theory,''
  [arXiv:hep-th/0512177]; \\
%
V.~Braun, Y.~H.~He and B.~A.~Ovrut,
  ``Yukawa couplings in heterotic standard models,''
  [arXiv:hep-th/0601204].
%
\bibitem{cv1} 
%
R.~Blumenhagen, M.~Cvetic, P.~Langacker and G.~Shiu,
``Toward realistic intersecting D-brane models,''
[arXiv:hep-th/0502005].
%
\bibitem{DESY}
%
W.~Buchmuller, K.~Hamaguchi, O.~Lebedev and M.~Ratz,
  ``The supersymmetric standard model from the heterotic string,''
  arXiv:hep-ph/0511035.
%
\bibitem{dona1}
  V.~Bouchard and R.~Donagi,
  ``An SU(5) heterotic standard model,''
  Phys.\ Lett.\ B {\bf 633}, 783 (2006)
  [arXiv:hep-th/0512149].
%
\bibitem{dona2}
%
  V.~Bouchard, M.~Cvetic and R.~Donagi,
  ``Tri-linear Couplings in an Heterotic Minimal Supersymmetric Standard
  Model,''
  arXiv:hep-th/0602096.
%
\bibitem{gauge-higgs}
%
 N.~S.~Manton,
  ``A New Six-Dimensional Approach To The Weinberg-Salam Model,''
  Nucl.\ Phys.\ B {\bf 158}, 141 (1979); \\
%
 D.~B.~Fairlie,
  ``Higgs' Fields And The Determination Of The Weinberg Angle,''
  Phys.\ Lett.\ B {\bf 82}, 97 (1979), \\
%
  ``Two Consistent Calculations Of The Weinberg Angle,''
  J.\ Phys.\ G {\bf 5}, L55 (1979); \\
%
  P.~Forgacs and N.~S.~Manton,
  ``Space-Time Symmetries In Gauge Theories,''
  Commun.\ Math.\ Phys.\  {\bf 72}, 15 (1980); \\
%
 S.~Randjbar-Daemi, A.~Salam and J.~A.~Strathdee,
  ``Spontaneous Compactification In Six-Dimensional Einstein-Maxwell Theory,''
  Nucl.\ Phys.\ B {\bf 214}, 491 (1983).
%
\bibitem{coset-ql}
%
 W.~Buchmuller, S.~T.~Love, R.~D.~Peccei and T.~Yanagida,
  ``Quasi Goldstone Fermions,''
  Phys.\ Lett.\ B {\bf 115} (1982) 233; \\
%
C.~L.~Ong,
  ``Gauged Supersymmetric Generalized Nonlinear Sigma Models For Quarks And
  Leptons,''
  Phys.\ Rev.\ D {\bf 27}, 3044 (1983); \\
%
 T.~Kugo and T.~Yanagida,
  ``Unification Of Families Based On A Coset Space E7 / SU(5) X SU(3) X U(1),''
  Phys.\ Lett.\ B {\bf 134}, 313 (1984).
%
\bibitem{CHSW}
%
 P.~Candelas, G.~T.~Horowitz, A.~Strominger and E.~Witten,
  ``Vacuum Configurations For Superstrings,''
  Nucl.\ Phys.\ B {\bf 258}, 46 (1985).
%
\bibitem{non-parallel}
%
T.~Yanagida and J.~Sato,
  ``Large lepton mixing in seesaw models: Coset-space family unification,''
  Nucl.\ Phys.\ Proc.\ Suppl.\  {\bf 77}, 293 (1999)
  [arXiv:hep-ph/9809307]; \\
%
 P.~Ramond,
  ``Neutrinos: A glimpse beyond the standard model,''
  Nucl.\ Phys.\ Proc.\ Suppl.\  {\bf 77}, 3 (1999)
  [arXiv:hep-ph/9809401]. \\
%
See also  C.~H.~Albright and S.~M.~Barr,
  ``Fermion masses in SO(10) with a single adjoint Higgs field,''
  Phys.\ Rev.\ D {\bf 58}, 013002 (1998)
  [arXiv:hep-ph/9712488], 
%
  ``A minimality condition and atmospheric neutrino oscillations,''
  Phys.\ Rev.\ Lett.\  {\bf 81}, 1167 (1998)
  [arXiv:hep-ph/9802314], 
%
\bibitem{anarchy}
%
  L.~J.~Hall, H.~Murayama and N.~Weiner,
  ``Neutrino mass anarchy,''
  Phys.\ Rev.\ Lett.\  {\bf 84}, 2572 (2000)
  [arXiv:hep-ph/9911341]; \\
%
 N.~Haba and H.~Murayama,
  ``Anarchy and hierarchy,''
  Phys.\ Rev.\ D {\bf 63}, 053010 (2001)
  [arXiv:hep-ph/0009174].
%
\bibitem{see-saw}
%
 T.~Yanagida, {\it in} Proc. Workshop on the Unified Theory and the
   Baryon Number in the Universe, Tsukuba, 1979, eds. O.~Sawada
   and S.~Sugamoto, Report KEK-79-18 (1979); \\
%
 M.~Gell-mann, P.~Ramond and R.~Slansky, {\it in} ``Supergravity''
  North-Holland, Amsterdam, 1979, eds. D.Z.~Freedman and 
  P.~van~Nieuwenhuizen.
%
\bibitem{wittendual}
%
 E.~Witten,
 ``String theory dynamics in various dimensions,''
  Nucl.\ Phys.\ B {\bf 443}, 85 (1995)
  [arXiv:hep-th/9503124].
%
\bibitem{evidence}
%
 C.~Vafa,
  ``Evidence for F-Theory,''
  Nucl.\ Phys.\ B {\bf 469} (1996) 403
  [arXiv:hep-th/9602022].
%
\bibitem{MV1}
%
D.~R.~Morrison and C.~Vafa,
``Compactifications of F-Theory on Calabi--Yau Threefolds -- I,''
Nucl.\ Phys.\ B {\bf 473}, 74 (1996)
[arXiv:hep-th/9602114]
%
\bibitem{MV2}
%
  D.~R.~Morrison and C.~Vafa,
  ``Compactifications of F-Theory on Calabi--Yau Threefolds -- II,''
  Nucl.\ Phys.\ B {\bf 476} (1996) 437
  [arXiv:hep-th/9603161].
%
\bibitem{FMW}
%
 R.~Friedman, J.~Morgan and E.~Witten,
  ``Vector bundles and F theory,''
  Commun.\ Math.\ Phys.\  {\bf 187} (1997) 679
  [arXiv:hep-th/9701162].
%
\bibitem{Het-F-4D}
%
  M.~Bershadsky, A.~Johansen, T.~Pantev and V.~Sadov,
  ``On four-dimensional compactifications of F-theory,''
  Nucl.\ Phys.\ B {\bf 505} (1997) 165
  [arXiv:hep-th/9701165].
%
\bibitem{BSV}
%
M.~Bershadsky,V.~Sadov and C.~Vafa
``D-branes and Topological Field Thoeries''
Nucl.\ Phys.\ B{\bf 463}, 420(1996)
[arXiv:hep-th/9511222].
%
\bibitem{GZ}
%
 M.~R.~Gaberdiel and B.~Zwiebach,
  ``Exceptional groups from open strings,''
  Nucl.\ Phys.\ B {\bf 518} (1998) 151
  [arXiv:hep-th/9709013].
%
\bibitem{PDG}
%
[Particle Data Group] S. Eidelman et al., 
Phys. \ Lett. \ {\bf B592}, 1 (2004). 
%
\bibitem{AGW}
%
 N.~Arkani-Hamed, T.~Gregoire and J.~Wacker,
  ``Higher dimensional supersymmetry in 4D superspace,''
  JHEP {\bf 0203} (2002) 055
  [arXiv:hep-th/0101233].
%
\bibitem{KcVf}
%
  S.~Kachru and C.~Vafa,
  ``Exact results for N=2 compactifications of heterotic strings,''
  Nucl.\ Phys.\ B {\bf 450}, 69 (1995)
  [arXiv:hep-th/9505105].
%
\bibitem{6authors}
%
 M.~Bershadsky, K.~A.~Intriligator, S.~Kachru, D.~R.~Morrison, V.~Sadov and C.~Vafa,
  ``Geometric singularities and enhanced gauge symmetries,''
  Nucl.\ Phys.\ B {\bf 481} (1996) 215
  [arXiv:hep-th/9605200].
%
\bibitem{NMSSM}
%
 J.~R.~Ellis, J.~F.~Gunion, H.~E.~Haber, L.~Roszkowski and F.~Zwirner,
  ``Higgs Bosons In A Nonminimal Supersymmetric Model,''
  Phys.\ Rev.\ D {\bf 39} (1989) 844.
%
\bibitem{CPHW}
%
P.~Ciafaloni and A.~Pomarol,
  ``Dynamical determination of the supersymmetric Higgs mass,''
  Phys.\ Lett.\ B {\bf 404}, 83 (1997)
  [arXiv:hep-ph/9702410]; \\
%
 L.~J.~Hall and T.~Watari,
  ``Electroweak supersymmetry with an approximate U(1)(PQ),''
  Phys.\ Rev.\ D {\bf 70}, 115001 (2004)
  [arXiv:hep-ph/0405109]; \\
%
 D.~J.~Miller and R.~Nevzorov,
  ``The Peccei-Quinn axion in the next-to-minimal supersymmetric standard
  model,''
  arXiv:hep-ph/0309143; \\
%
B.~Feldstein, L.~J.~Hall and T.~Watari,
  ``Simultaneous solutions of the strong CP and mu problems,''
  Phys.\ Lett.\ B {\bf 607}, 155 (2005)
  [arXiv:hep-ph/0411013]; \\
%
 D.~J.~Miller, S.~Moretti and R.~Nevzorov,
  ``Higgs bosons in the NMSSM with exact and slightly broken PQ-symmetry,''
  arXiv:hep-ph/0501139.
%
\bibitem{LEP-limit}
%
 B.~A.~Dobrescu, G.~Landsberg and K.~T.~Matchev,
  ``Higgs boson decays to CP-odd scalars at the Tevatron and beyond,''
  Phys.\ Rev.\ D {\bf 63}, 075003 (2001)
  [arXiv:hep-ph/0005308]; \\
%
 B.~A.~Dobrescu and K.~T.~Matchev,
  ``Light axion within the next-to-minimal supersymmetric standard model,''
  JHEP {\bf 0009}, 031 (2000)
  [arXiv:hep-ph/0008192]; \\
%
R.~Dermisek and J.~F.~Gunion,
  ``Consistency of LEP event excesses with an h $\to$ a a decay scenario and
  low-fine-tuning NMSSM models,''
  arXiv:hep-ph/0510322; \\
%
 S.~Chang, P.~J.~Fox and N.~Weiner,
  ``Naturalness and Higgs decays in the MSSM with a singlet,''
  arXiv:hep-ph/0511250.
%
\bibitem{Witten23}
%
E.~Witten,
  ``Symmetry Breaking Patterns In Superstring Models,''
  Nucl.\ Phys.\ B {\bf 258}, 75 (1985).
%
\bibitem{missing}
%
A.~Masiero, D.~V.~Nanopoulos, K.~Tamvakis and T.~Yanagida,
  ``Naturally Massless Higgs Doublets In Supersymmetric SU(5),''
  Phys.\ Lett.\ B {\bf 115}, 380 (1982).
%
\bibitem{WY}
%
 T.~Watari and T.~Yanagida,
  ``Product-group unification in type IIB string thoery,''
  Phys.\ Rev.\ D {\bf 70}, 036009 (2004)
  [arXiv:hep-ph/0402160].
%
\bibitem{IY}
%
 K.~I.~Izawa and T.~Yanagida,
  ``R-invariant natural unification,''
  Prog.\ Theor.\ Phys.\  {\bf 97}, 913 (1997)
  [arXiv:hep-ph/9703350].
%
\bibitem{Penn5}
%
R.~Donagi, Y.~H.~He, B.~A.~Ovrut and R.~Reinbacher,
  ``The particle spectrum of heterotic compactifications,''
  JHEP {\bf 0412}, 054 (2004)
  [arXiv:hep-th/0405014].
%
\bibitem{Blumenhagen}
%
R.~Blumenhagen, G.~Honecker and T.~Weigand,
  ``Loop-corrected compactifications of the heterotic string with line
  bundles,''
  JHEP {\bf 0506} (2005) 020
  [arXiv:hep-th/0504232].
%
\bibitem{Hirzebruch}
%
F.~Hirzebruch, ``{\it Topological Methods in Algebraic Geometry,}'' 
Springer (1978).
%
\bibitem{Munich3}
%
R.~Blumenhagen, S.~Moster and T.~Weigand,
  ``Heterotic GUT and Standard Model vacua from simply connected Calabi-Yau
  manifolds,''
  arXiv:hep-th/0603015.
%
\bibitem{Witten-rk-red}
%
 E.~Witten,
  ``New Issues In Manifolds Of SU(3) Holonomy,''
  Nucl.\ Phys.\ B {\bf 268}, 79 (1986).
%
\bibitem{Witten-Stkbg}
%
E.~Witten,
  ``Some Properties Of O(32) Superstrings,''
  Phys.\ Lett.\ B {\bf 149}, 351 (1984).
%
\bibitem{FY}
%
 M.~Fukugita and T.~Yanagida,
  ``Baryogenesis Without Grand Unification,''
  Phys.\ Lett.\ B {\bf 174}, 45 (1986).
%
\bibitem{DiracLG}
%
 K.~Dick, M.~Lindner, M.~Ratz and D.~Wright,
  ``Leptogenesis with Dirac neutrinos,''
  Phys.\ Rev.\ Lett.\  {\bf 84}, 4039 (2000)
  [arXiv:hep-ph/9907562].
%
H.~Murayama and A.~Pierce,
  ``Realistic Dirac leptogenesis,''
  Phys.\ Rev.\ Lett.\  {\bf 89}, 271601 (2002)
  [arXiv:hep-ph/0206177].
%
\bibitem{DSWW}
%
 M.~Dine, N.~Seiberg, X.~G.~Wen and E.~Witten,
  ``Nonperturbative Effects On The String World Sheet,''
  Nucl.\ Phys.\ B {\bf 278}, 769 (1986)
%
\bibitem{DIN}
%
J.~P.~Derendinger, L.~E.~Ibanez and H.~P.~Nilles,
  ``On The Low-Energy Limit Of Superstring Theories,''
  Nucl.\ Phys.\ B {\bf 267}, 365 (1986).
%

%
\bibitem{local}
%
 G.~Aldazabal, L.~E.~Ibanez, F.~Quevedo and A.~M.~Uranga,
  ``D-branes at singularities: A bottom-up approach to the string  embedding of
  the standard model,''
  JHEP {\bf 0008}, 002 (2000)
  [arXiv:hep-th/0005067].
%
\bibitem{verlinde}
%
H.~Verlinde and M.~Wijnholt,
``Building the standard model on a D3-brane,''
[arXiv:hep-th/0508089].
\bibitem{WittenG2}
%
 E.~Witten,
  ``Deconstruction, G(2) holonomy, and doublet-triplet splitting,''
  arXiv:hep-ph/0201018; \\
%
T.~Friedmann and E.~Witten,
  ``Unification scale, proton decay, and manifolds of G(2) holonomy,''
  Adv.\ Theor.\ Math.\ Phys.\  {\bf 7}, 577 (2003)
  [arXiv:hep-th/0211269]; \\
%
I.~R.~Klebanov and E.~Witten,
  ``Proton decay in intersecting D-brane models,''
  Nucl.\ Phys.\ B {\bf 664}, 3 (2003)
  [arXiv:hep-th/0304079]; \\
%
M.~Axenides, E.~Floratos and C.~Kokorelis,
  ``SU(5) unified theories from intersecting branes,''
  JHEP {\bf 0310}, 006 (2003)
  [arXiv:hep-th/0307255].
%
B.~S.~Acharya and R.~Valandro,
  ``Suppressing proton decay in theories with localised fermions,''
  arXiv:hep-ph/0512144.
%
\bibitem{kv}
%
  S.~Katz and C.~Vafa,
  ``Matter from geometry,''
  Nucl.\ Phys.\ B {\bf 497}, 146 (1997)
  [arXiv:hep-th/9606086].
%
\bibitem{aw}
%
M. Atiyah and E.~Witten
``M-theory dynamics on a manifold of G(2) holonomy,''
Adv\ Theor.\ Math.\ Phys. {\bf 6}, 1 (2003)
[arXiv:hep-th/0107177]
%
\bibitem{acw}%
%
B.~Acharya and E.~Witten,
``Chiral Fermions from manifolds of G(2) holonomy,''
[arXiv:hep-th/0109152]
%
\bibitem{ag}
%
  B.~S.~Acharya and S.~Gukov,
  ``M theory and Singularities of Exceptional Holonomy Manifolds,''
  Phys.\ Rept.\  {\bf 392}, 121 (2004)
  [arXiv:hep-th/0409191].
%
\bibitem{Acharya}
%
B.~S.~Acharya,
  ``On realising N = 1 super Yang-Mills in M theory,''
  arXiv:hep-th/0011089.
%
\bibitem{Kronheimer}
%
P.~B.~Kronheimer,
``The Construction Of Ale Spaces As Hyperkahler Quotients,''
J.\ Diff.\ Geom.\  {\bf 29} (1989) 665.
%
\bibitem{BDL}
%
 M.~Berkooz, M.~R.~Douglas and R.~G.~Leigh,
  ``Branes intersecting at angles,''
  Nucl.\ Phys.\ B {\bf 480}, 265 (1996)
  [arXiv:hep-th/9606139].
%
\bibitem{CS}
%
 M.~Cvetic, P.~Langacker and G.~Shiu,
  ``A three-family standard-like orientifold model: Yukawa couplings and
  hierarchy,''
  Nucl.\ Phys.\ B {\bf 642}, 139 (2002)
  [arXiv:hep-th/0206115].
%
\bibitem{bb}
%
P.~Berglund and A.~Brandhuber,
``Matter from G(2) manifolds'',
Nucl.\ Phys.\ B {\bf 641}, 351 (2002)
[arXiv:hep-th/0205184].

%
\bibitem{BM}
%
 P.~Berglund and P.~Mayr,
  ``Heterotic string/F-theory duality from mirror symmetry,''
  Adv.\ Theor.\ Math.\ Phys.\  {\bf 2}, 1307 (1999)
  [arXiv:hep-th/9811217].
%
\bibitem{Rajesh}
G.~Rajesh,
``Toric geometry and F-theory/heterotic duality in four dimensions,''
JHEP {\bf 9812}, 018 (1998)
[arXiv:hep-th/9811240].
%
\bibitem{Hartshorne}
%
R.~Hartshorne,~ ``{\it Algebraic Geometry,}'' Springer (1977).
%
\bibitem{U(n)}
%
 B.~Andreas and D.~Hernandez Ruiperez,
  ``U(n) vector bundles on Calabi-Yau threefolds for string theory
  compactifications,''
  arXiv:hep-th/0410170.
%
\bibitem{ovrutpark}
 B.~A.~Ovrut, T.~Pantev and J.~Park,
 ``Small instanton transitions in heterotic M-theory,''
 JHEP {\bf 0005}, 045 (2000)
 [arXiv:hep-th/0001133].
%
\bibitem{DI}
%
G.~Curio,
``Chiral matter and transitions in heterotic string models,''
Phys.\ Lett.\ B {\bf 435}, 39 (1998)
[arXiv:hep-th/9803224]; \\
%
D.~E.~Diaconescu and G.~Ionesei,
``Spectral covers, charged matter and bundle cohomology,''
JHEP {\bf 9812}, 001 (1998)
[arXiv:hep-th/9811129].
%
\bibitem{Sethi}
S.~Sethi, C.~Vafa and E.~Witten,
``Constraints on low-dimensional string compactifications,''
Nucl.\ Phys.\ B {\bf 480}, 213 (1996)
[arXiv:hep-th/9606122].
%
\bibitem{Louis}
%
H.~Jockers and J.~Louis,
  ``The effective action of D7-branes in N = 1 Calabi-Yau orientifolds,''
  Nucl.\ Phys.\ B {\bf 705}, 167 (2005)
  [arXiv:hep-th/0409098].
%
\bibitem{Blumenhagen2}
%
R.~Blumenhagen, G.~Honecker and T.~Weigand,
  ``Non-abelian brane worlds: The heterotic string story,''
  JHEP {\bf 0510} (2005) 086
  [arXiv:hep-th/0510049].
%
\bibitem{FMW2}
%
R.~Friedman, J.~W.~Morgan and E.~Witten,
  ``Vector bundles over elliptic fibrations,''
  arXiv:alg-geom/9709029.
%
\bibitem{CD}
%
  G.~Curio and R.~Y.~Donagi,
  ``Moduli in N = 1 heterotic/F-theory duality,''
  Nucl.\ Phys.\ B {\bf 518} (1998) 603
  [arXiv:hep-th/9801057].
%
\bibitem{Sadov}
%
 V.~Sadov,
  ``Generalized Green-Schwarz mechanism in F theory,''
  Phys.\ Lett.\ B {\bf 388} (1996) 45
  [arXiv:hep-th/9606008].
%
\bibitem{CY4}
%
 A.~Klemm, B.~Lian, S.~S.~Roan and S.~T.~Yau,
  ``Calabi-Yau fourfolds for M- and F-theory compactifications,''
  Nucl.\ Phys.\ B {\bf 518}, 515 (1998)
  [arXiv:hep-th/9701023].
%
\bibitem{Nomura}
%
T.~Gherghetta and A.~Pomarol,
  ``Bulk fields and supersymmetry in a slice of AdS,''
  Nucl.\ Phys.\ B {\bf 586}, 141 (2000)
  [arXiv:hep-ph/0003129]; \\
%
 W.~D.~Goldberger, Y.~Nomura and D.~R.~Smith,
  ``Warped supersymmetric grand unification,''
  Phys.\ Rev.\ D {\bf 67}, 075021 (2003)
  [arXiv:hep-ph/0209158].
%
\bibitem{KS}
%
 I.~R.~Klebanov and M.~J.~Strassler,
  ``Supergravity and a confining gauge theory: Duality cascades and
  chiSB-resolution of naked singularities,''
  JHEP {\bf 0008}, 052 (2000)
  [arXiv:hep-th/0007191].
%
\bibitem{GKP}
%
S.~B.~Giddings, S.~Kachru and J.~Polchinski,
  ``Hierarchies from fluxes in string compactifications,''
  Phys.\ Rev.\ D {\bf 66}, 106006 (2002)
  [arXiv:hep-th/0105097].
%
\bibitem{radu}
%
K.~Dasgupta, K.~h.~Oh, J.~Park and R.~Tatar,
``Geometric transition versus cascading solution,''
JHEP {\bf 0201}, 031 (2002)
[arXiv:hep-th/0110050];~~M.~Becker, K.~Dasgupta, 
S.~Katz, A.~Knauf and R.~Tatar,
``Geometric transitions, flops and non-Kaehler manifolds. II,''
Nucl.\ Phys.\ B {\bf 738}, 124 (2006)
[arXiv:hep-th/0511099]
%
\bibitem{PzT}
%
 L.~A.~Pando Zayas and A.~A.~Tseytlin,
  ``3-branes on resolved conifold,''
  JHEP {\bf 0011}, 028 (2000)
  [arXiv:hep-th/0010088].
%
\bibitem{dd}
%
 F.~Denef and M.~R.~Douglas,
 ``Computational complexity of the landscape I,''
[arXiv:hep-th/0602072].
%
%
\bibitem{vafa}
%
  C.~Vafa,
  ``The string landscape and the swampland,''
  [arXiv:hep-th/0509212]; \\
%
N.~Arkani-Hamed, L.~Motl, A.~Nicolis and C.~Vafa,
  ``The string landscape, black holes and gravity as the weakest force,''
  [arXiv:hep-th/0601001].
%
\bibitem{kachru}
%
S.~Kachru, J.~McGreevy and P.~Svrcek,
``Bounds on masses of bulk fields in string compactifications,''
[arXiv:hep-th/0601111].
%
\bibitem{beauty}
%
L.~Kofman, A.~Linde, X.~Liu, A.~Maloney, L.~McAllister and E.~Silverstein,
  ``Beauty is attractive: Moduli trapping at enhanced symmetry points,''
  JHEP {\bf 0405}, 030 (2004)
  [arXiv:hep-th/0403001].
%
\bibitem{runaway}
%
 B.~Feldstein, L.~J.~Hall and T.~Watari,
  ``Density perturbations and the cosmological constant from inflationary
  landscapes,''
  Phys.\ Rev.\ D {\bf 72}, 123506 (2005)
  [arXiv:hep-th/0506235]; \\
%
J.~Garriga and A.~Vilenkin,
  ``Anthropic prediction for Lambda and the Q catastrophe,''
  arXiv:hep-th/0508005; \\
%
A.~Linde and V.~Mukhanov,
  ``The curvaton web,''
  arXiv:astro-ph/0511736; \\
%
L.~J.~Hall, T.~Watari and T.~T.~Yanagida,
  ``Taming the runaway problem of inflationary landscapes,''
  arXiv:hep-th/0601028.
%

%
\bibitem{DKV}
%
 M.~R.~Douglas, S.~Katz and C.~Vafa,
  ``Small instantons, del Pezzo surfaces and type I' theory,''
  Nucl.\ Phys.\ B {\bf 497}, 155 (1997)
  [arXiv:hep-th/9609071].
%
\bibitem{DM}
%
 M.~R.~Douglas and G.~W.~Moore,
  ``D-branes, Quivers, and ALE Instantons,''
  arXiv:hep-th/9603167. 
%
\bibitem{Sen-M}
%
 A.~Sen,
  ``A note on enhanced gauge symmetries in M- and string theory,''
  JHEP {\bf 9709} (1997) 001
  [arXiv:hep-th/9707123].
%
\end{thebibliography}
\end{document}